\documentclass[aps,prl,twocolumn, superscriptaddress,longbibliography]{revtex4-2}

\usepackage{graphicx,xcolor}
\usepackage{amsmath,amssymb,amsthm}
\usepackage[small,bf,justification=raggedright]{caption}
\usepackage{physics}
\usepackage[subrefformat=parens, labelformat=simple,singlelinecheck=false, justification=raggedright]{subcaption}
\usepackage[pdfencoding=auto]{hyperref}
\hypersetup{colorlinks=true,allcolors=black}
\usepackage[export]{adjustbox}
\usepackage{mathtools}
\usepackage{xr-hyper}

\makeatletter
\renewcommand*\p@subfigure{\thefigure}
\makeatother

\DeclareMathOperator{\diag}{diag}

\theoremstyle{definition}

\theoremstyle{plain}

\begin{document}

\title{Measurement-based quantum computation on weighted graph states \\ with arbitrarily small weight}

\author{Tomohiro Yamazaki}
\email{tomohiro.yamazaki@ntt.com}
\affiliation{Basic Research Laboratories, NTT Inc., Atsugi, Kanagawa 243-0198, Japan}
\affiliation{NTT Research Center for Theoretical Quantum Information, NTT Inc., Atsugi, Kanagawa 243-0198, Japan}
\author{Yuki Takeuchi}
\affiliation{Communication Science Laboratories, NTT Inc., Atsugi, Kanagawa 243-0198, Japan}
\affiliation{NTT Research Center for Theoretical Quantum Information, NTT Inc., Atsugi, Kanagawa 243-0198, Japan}
\affiliation{Information Technology R\&D Center, Mitsubishi Electric Corporation, Kamakura, Kanagawa 247-8501, Japan}
\begin{abstract}
    Weighted graph states are a natural generalization of graph states, which are generated by applying controlled-phase gates, instead of controlled-Z gates, to a separable state.
    In this paper, we show that \textit{uniformly} weighted graph states on a suitable planar graph constitute universal resources for measurement-based quantum computation for an arbitrary nonzero constant weight. To our knowledge, this is the first example of universal resources prepared with only non-maximally entangling gates and has potential applications to weakly interacting systems, such as photonic systems.
\end{abstract}

\maketitle

\textit{Introduction.}
Measurement-based quantum computation~\cite{Raussendorf_Phys.Rev.Lett._2001, Briegel_NaturePhys_2009} (MBQC) is a promising model of quantum computation, based on adaptive single-qubit measurements (SQMs) on a specific entangled state, called a universal resource.
To date, identifying universal resources has remained a central problem in MBQC, in which the class of universal resources has been extended from concrete examples~\cite{VandenNest_Phys.Rev.Lett._2006, Gross_Phys.Rev.Lett._2007, Gross_Phys.Rev.A_2007, Wei_Phys.Rev.Lett._2011, Wei_Phys.Rev.A_2012, VandenNest_Phys.Rev.Lett._2013,Wei_Phys.Rev.A_2013, Wei_Phys.Rev.A_2014, Wei_Phys.Rev.A_2015,Nautrup_Phys.Rev.A_2015, Chen_Phys.Rev.A_2018, Kissinger_Quantum_2019,Miller_npjQuantumInf_2016,Miller_Phys.Rev.Lett._2018,Takeuchi_SciRep_2019}, such as the 2D cluster state~\cite{Briegel_Phys.Rev.Lett._2001}, to phases of quantum states~\cite{Raussendorf_Phys.Rev.Lett._2019,Devakul_Phys.Rev.A_2018,Stephen_Quantum_2019,Daniel_Quantum_2020,Stephen_Phys.Rev.Lett._2024}, called computationally universal phases~\cite{Raussendorf_Phys.Rev.Lett._2019}. 
Finding new universal resources is not only of theoretical interest but also of practical importance because experimental constraints might prevent the direct generation of a 2D cluster state but allow the generation of other universal resources.

One such experimental constraint is that only controlled-phase (CP) gates with angle $\phi$ are available, instead of controlled-Z (CZ) gates ($\phi=\pi$).
Then, a generated state becomes a so-called weighted graph state~\cite{Dur_Phys.Rev.Lett._2005} (WGS), instead of a graph state~\cite{Raussendorf_Phys.Rev.A_2003,Hein__2006}.
A particularly relevant scenario in weakly interacting systems, such as photonic systems~\footnote{For example, CP gates are implemented with photon-photon interactions induced by matter systems such as single atoms~\cite{Turchette_Phys.Rev.Lett._1995,Duan_Phys.Rev.Lett._2004,Volz_NaturePhoton_2014,Hacker_Nature_2016,Beck_Proc.Natl.Acad.Sci._2016}, atomic ensembles~\cite{Sagona-Stophel_Phys.Rev.Lett._2020}, quantum dots~\cite{Fushman_Science_2008,Kim_NaturePhoton_2013,Sun_Science_2018}, nitrogen vacancy centers~\cite{Wang_Opt.ExpressOE_2013}, and Rydberg atoms~\cite{Gorshkov_Phys.Rev.Lett._2011,Firstenberg_Nature_2013,Tiarks_Sci.Adv._2016,Thompson_Nature_2017,Tiarks_NaturePhys_2019}.}, is when all edges are assigned the same weight $\phi$. However, despite extensive studies on WGSs~\cite{Calsamiglia_Phys.Rev.Lett._2005,Anders_Phys.Rev.Lett._2006,Hartmann_J.Phys.B:At.Mol.Opt.Phys._2007,Anders_NewJ.Phys._2007,Plato_Phys.Rev.A_2008,Xue_Phys.Rev.A_2012,Ghosh_Phys.Rev.A_2024,Ghosh__2025,Szymanski__2025}, such uniform WGSs have remained largely unexplored.
A notable recent exception is the scheme for extracting a GHZ state from a uniform WGS proposed in~\cite{Frantzeskakis_Phys.Rev.Res._2023}, although its success probability is exponentially small in the number of qubits.
Thus, a fundamental open question is whether there exists a uniform WGS from which an \textit{arbitrary state} can be extracted with (near-)\textit{unit probability}.
In fact, this question can be replaced by asking whether uniform WGSs can be universal resources for MBQC (in the sense of strict universality~\footnote{There are two notions of universality: strict universality and computational universality~\cite{Aharonov__2003,Takeuchi_Phys.Rev.Lett._2024}. The former means that any quantum circuit can be implemented, and the latter means that the output probability distribution of any quantum circuit can be generated. Most of the universal resources, including 2D cluster states, are strictly universal resources, while hypergraph states are known to be computationally universal resources~\cite{Miller_npjQuantumInf_2016,Miller_Phys.Rev.Lett._2018,Takeuchi_SciRep_2019}}) because MBQC can produce any quantum state with (near-)unit probability.
Although there have been several proposals for MBQC on WGSs~\cite{Gross_Phys.Rev.Lett._2007, Gross_Phys.Rev.A_2007,Kissinger_Quantum_2019}, all of them use WGSs whose weights are tuned artificially, where some edges are perfect ($\phi=\pi$), and the others are adjusted to be specific weights, such as $\phi=\pi/2$.
Thus, no example is known in which WGSs become universal resources without perfect edges.

\begin{figure}[tbp]
\centering
\includegraphics[scale=0.8, trim=0 0 0 0, clip]{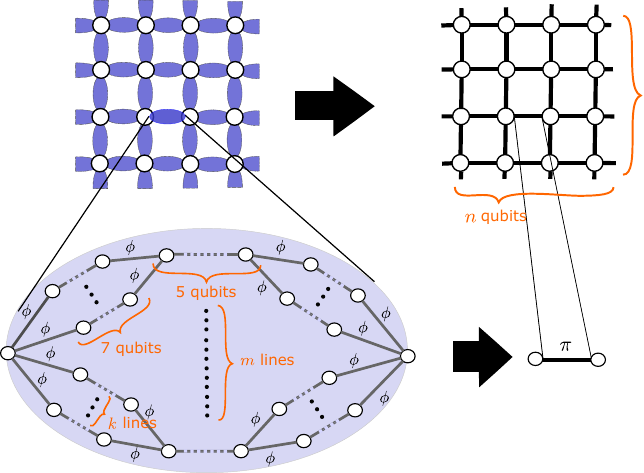}
\caption{Our weighted graph state serving as a universal resource and its transformation to a 2D cluster state. 
A concrete example with $k=m=3$ is shown in Fig.~\ref{fig:entire_transformation} in~\cite{Note99}.} 
\label{fig:entire_graph}
\end{figure}
In this paper, we show that the uniform WGS on the planar graph shown in Fig.~\ref{fig:entire_graph} becomes a universal resource for arbitrary nonzero weight $\phi$, providing an affirmative answer to the aforementioned question.
While the graph is based on a square lattice, each edge is replaced with a complicated structure inside the ellipse in Fig.~\ref{fig:entire_graph}.
We show that each of the structures can be probabilistically transformed to a perfect edge using SQMs, and its failure probability can be reduced exponentially in the number of consumed qubits.
After the transformations, the resulting state becomes a 2D cluster state; therefore, the WGS is shown to be a universal resource~\cite{Chen_Phys.Rev.Lett._2010}.

\begin{figure}[tpb]
\begin{minipage}[b]{0.48\linewidth}
\centering
\subcaption{}
\vspace{-5pt}
\includegraphics[scale=0.95, trim=0 0 0 0, clip]{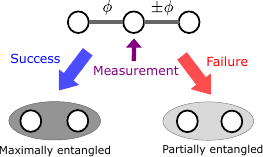}\label{fig:basic_transformation}
\end{minipage}
\begin{minipage}[b]{0.50\linewidth}
\centering
\subcaption{}
\vspace{0pt}
\includegraphics[scale=1.0, trim=0 0 0 0, clip]{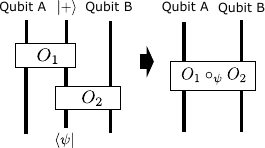}\label{fig:diagram_mbc}
\vspace{-15pt}
\end{minipage}
\caption{(a) Generation of a maximally or partially entangled state with a single-qubit measurement. (b) Concept of measurement-based composition of operators.}
\end{figure}

\textit{CZ-gate generation.}
We begin by considering the measurement of the middle qubit in a tripartite WGS, as in Fig.~\ref{fig:basic_transformation}.
Let the weight of an edge be $\phi$ and the weight of the other edge be $\phi$ or $-\phi$.
When the middle qubit is measured in the basis of $\{R_z(\phi)\ket{-},R_z(\phi)\ket{+}\}$ or $\{\ket{-},\ket{+}\}$, respectively,
the resulting state becomes a maximally or partially entangled state, depending on the measurement outcome. Here, $R_z(c)=\exp(-icZ/2)$ is a generalized Pauli-$Z$ rotation in which $c$ is allowed to be complex for later use.
This method of probabilistically generating a maximally entangled state from partially entangling gates is distinct from, but similar to, the entanglement concentration scheme~\cite{Bennett_Phys.Rev.A_1996,Bose_Phys.Rev.A_1999} especially on networks~\cite{Acin_NaturePhys_2007} and has (sometimes implicitly) been used in various contexts~\cite{Nemoto_Phys.Rev.Lett._2004,Barrett_Phys.Rev.A_2005,Spiller_NewJ.Phys._2006,HalilShah_LIPIcsVol.22TQC2013_2013,Frantzeskakis_Phys.Rev.Res._2023}.

To generalize the above process, we consider applying a two-qubit operator $O_1$ to qubit $A$ and an ancillary qubit, another two-qubit operator $O_2$ to the ancillary qubit and qubit $B$, and an SQM to the ancillary qubit, as in Fig.~\ref{fig:diagram_mbc}.
This process, which we call measurement-based composition (MBC) of $O_1$ and $O_2$, functions jointly as a two-qubit operator on qubits $A$ and $B$, depending on the measurement outcome on the ancillary qubit.
We let the initial and measured states of the ancillary qubit be $\ket{+}$ and $\ket{\psi}$ and represent the resulting operator as $O_1 \circ_\psi O_2 $.

This concept allows us to consider the process in Fig.~\ref{fig:basic_transformation} as a probabilistic generation of an entangling operator, rather than an entangled state.
The WGS is generated from the product state $\ket{+}^{\otimes n}$ by applying a CP gate $\mathrm{CP}(\phi)=\diag(1,1,1,e^{i\phi})$ for every edge with weight $\phi$.
Therefore, the above process corresponds to the MBC of $\mathrm{CP}(\phi)$ and $\mathrm{CP}(\pm \phi)$. 
For $\ket{\phi_\pm} \equiv R_z(\phi)\ket{\pm}$,
the generated operators are represented as $O^s_+ \equiv \mathrm{CP}(\phi) \circ_{\phi_-} \mathrm{CP}(\phi)$ and $O^s_- \equiv \mathrm{CP}(\phi) \circ_{-} \mathrm{CP}(-\phi)$ in the success events and $O^f_+ \equiv \mathrm{CP}(\phi) \circ_{\phi_+} \mathrm{CP}(\phi)$ and $O^f_- \equiv \mathrm{CP}(\phi) \circ_{+} \mathrm{CP}(-\phi)$ in the failure events, respectively.
They are calculated as
\begin{align}
    O^s_\pm &= (R_z(\phi/2)\otimes Z R_z(\pm\phi/2))(\sin(\phi/2)P_\pm), \label{eq:success_event} \\
    O^f_\pm &= (R_z(\phi/2)\otimes R_z(\pm\phi/2))(P_\mp + \cos(\phi/2)P_\pm)\label{eq:failure_event}
\end{align}
with projectors $P_+=\dyad{00}+\dyad{11}$ and $P_-=\dyad{01}+\dyad{10}$, up to the global phases.
In the success events in Eq.~\eqref{eq:success_event}, a maximally entangling projection (MEP) of $P_\pm$ has been applied to the remaining two qubits in $\ket{+}\ket{+}$, generating a maximally entangled state.
As used in~\cite{Frantzeskakis_Phys.Rev.Res._2023}, applying such an MEP multiple times is sufficient to generate a GHZ state; however, further strategies on how to apply SQMs are necessary to generate entangled states beyond GHZ states.

The CP gates are represented in the form of operator-Schmidt decomposition~\cite{Nielsen__2000,Nielsen_Phys.Rev.A_2003} as
\begin{align}\label{eq:Cphase-Schmidt_main}
    \mathrm{CP}(\phi) &= e^{i\phi/4}(R_z(\phi/2)\otimes R_z(\phi/2)) \notag  \\
    &\qquad (\cos(\phi/4) I \otimes I + i\sin(\phi/4) Z \otimes Z ).
\end{align}
This makes the role of measurement basis $\{R_z(\phi)\ket{\pm}\}$ or $\{\ket{\pm}\}$ clear in the above MBCs.
The prefactor $R_z(\phi)=R_z(\phi/2)R_z(\phi/2)$ or $I=R_z(\phi/2)R_z(-\phi/2)$ in the measurement basis cancels out the local operations applied to the ancillary qubit, and the following $X$-basis measurement measures the parity of the number of Pauli-$Z$ operations applied to the ancillary qubit.
As a result, the remaining two-qubit operator is again an entangling operator composed of two terms among $\{P_1 \otimes P_2 \mid P_1,P_2 \in \{I,Z\}\}$.
For example, in the case of $R_z(\phi)\ket{-}$, the resulting operator $O^s_+$ is
\begin{equation}\label{eq:maximally_entangling}
    \cos(\phi/4)\sin(\phi/4)(R_z(\phi/2)\otimes R_z(\phi/2))(I\otimes Z + Z \otimes I),
\end{equation}
which implies that the enhancement of the entangling power of $O^s_+$ is the result of the coefficients of $\cos(\phi/4)$ and $\sin(\phi/4)$ in Eq.~\eqref{eq:Cphase-Schmidt_main} being balanced to $\cos(\phi/4)\sin(\phi/4)$ in Eq.~\eqref{eq:maximally_entangling}.

We generalize the above MBCs as the \textit{weighted Pauli-X measurements}. 
Note that the SQMs in the above MBCs are identical to the Pauli-X measurement, except for the prefactors. Similarly, we will define an SQM for the MBC of two-qubit operators $O_1$ and $O_2$ in the form of 
\begin{align}\label{eq:general-Schmidt}
    O_i = (R_z(c'_i)\otimes R_z(c_i)) (\alpha_i I \otimes I + \beta_i Z \otimes Z )
\end{align}
for $i=1,2$ (where we assume that the right side of the tensor product acts on the ancillary qubit in both $O_1$ and $O_2$).
Here, an important generalization for later use is not assuming $R_z(c_i)$ to be unitary. Instead, we just require $R_z(c_i)$ to be invertible so that $R_z(c_i)$ can be canceled out by applying its inverse. 
Then, we define $X_w^{\pm}$-basis measurement as the measurement in the basis of $\{\ket{\psi_s},\ket{\psi_f}\}$, where
\begin{align}
    \ket{\psi_s} &= \mathcal{N}^{-1/2} R_z^\dag (-(c_1+c_2))\ket{\pm}, \label{eq:weightedPauliX_basis_success}\\ 
    \ket{\psi_f} &= \mathcal{N}^{-1/2} R_z (c_1+c_2)\ket{\mp}, \label{eq:weightedPauliX_basis_failure}
\end{align}
where $\mathcal{N}=\cosh(\Im(c_1+c_2))$. 
We let the measurements in $\ket{\psi_s}$ and $\ket{\psi_f}$ be the success and failure events, respectively. 
Note that, when $R_z(c_1)$ and $R_z(c_2)$ are unitary, $X_w^{+}$ and $X_w^{-}$ are the same except that the success and failure events are flipped. In general cases, $\{R_z^\dag (-(c_1+c_2))\ket{\pm}\}$ are neither normalized nor orthogonal to each other, requiring the normalization factor $\mathcal{N}$ and the complementary basis state $\ket{\psi_f}$ defined as in Eq.~\eqref{eq:weightedPauliX_basis_failure}.

\begin{figure}[tbp]
\begin{minipage}[b]{\linewidth}
    \centering
    \subcaption{}
    \vspace{-10pt}
    \includegraphics[scale=0.9, trim=0 0 0 0, clip]{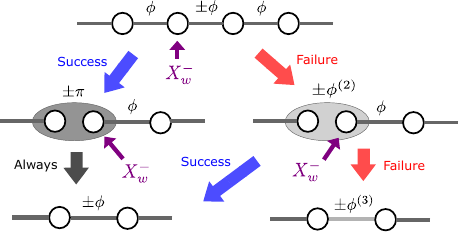}\label{fig:transformation_diagram}
    \vspace{10pt}
\end{minipage}\\
\begin{minipage}[b]{\linewidth}
    \centering
    \subcaption{}
    \vspace{-10pt}
    \includegraphics[scale=0.9, trim=0 0 0 0, clip]{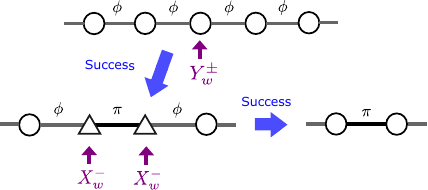}\label{fig:CZ_generation}
\end{minipage}
\caption{(a) State transformations under the weighted Pauli-X measurements on a 1D chain of WGS. Gray ellipses represent entangling projections being applied. (b) Probabilistic generation of a CZ gate. Triangles represent non-unitary single-qubit operations being applied. Throughout this paper, the multiple weighted Pauli measurements within one step are defined from the left.}
\end{figure}
Figure~\ref{fig:transformation_diagram} shows how the $X_w^{-}$-basis (or equivalently $X_w^{+}$-basis) measurement works on a 1D chain of the WGS.
Since each application of the $X_w^{-}$-basis measurement changes the relative phase between the terms $I\otimes I$ and $Z\otimes Z$ in Eq.~\eqref{eq:general-Schmidt} by $\pi/2$, CP gates and entangling projections are generated alternately.
On the other hand, let the weight $\phi$ for an entangling projection be the angle $-\pi \leq \phi \leq \pi$ such that $\tan(\phi/4) = \beta_i/\alpha_i$ in Eq.~\eqref{eq:general-Schmidt}; then, each $X_w^{-}$-basis measurement changes the weight of generated operators from $\phi^{(i)}$ to $\phi^{(i- 1)}$ ($\phi^{(i+1)}$) in the success (failure) event, where $\phi^{(i)}$ is the angle such that $\tan(\phi^{(i)}/4) = \tan^{i}(\phi/4)$.
These properties imply that the weight becomes $\phi^{(0)} = \pi$ only when the operation is an entangling projection, and thereby we cannot achieve a CZ gate only by applying $X_w^{\pm}$-basis measurements on the 1D chain.
 
To achieve a CZ gate, we introduce another type of SQMs, called \textit{weighted Pauli-Y measurements}. For the MBC of two identical CP gates, the $Y^{\pm}_w$ basis is defined as $\{R_z(\phi \pm \theta)\ket{-},R_z(\phi \pm \theta)\ket{+}\}$, where $0 < \theta < \pi$ is the angle satisfying
\begin{equation}\label{eq:condition_weightedY}
    \sqrt{2}\sin(\theta/2)=\abs{\sin(\phi/2)}.
\end{equation}
This angle $\theta$ is special in the sense that 
\begin{align}\label{eq:CZ_gate_with_nonunitaries}
    &\mathrm{CP}(\phi) \circ_{(\phi \pm \theta)_-} \mathrm{CP}(\phi) \notag \\ 
    &= (\sin(\phi/2)/\sqrt{2}) (R_z(\pm ir/2)\otimes R_z(\pm ir/2))\mathrm{CZ}
\end{align}
holds, up to local unitary operations, where $r=\log\abs{\sin((\phi+\theta)/2)/\sin((\phi-\theta)/2)}$. Therefore, the operator generated after the successful $Y^{\pm}_w$-basis measurement becomes a CZ gate, up to local non-unitary operations.
These local non-unitary operators can be removed by applying $X_w^{-}$-basis measurement twice, as shown in Fig.~\ref{fig:CZ_generation}. 
Therefore, we can probabilistically generate a CZ gate using three SQMs. See also~\footnote[99]{Supplemental Material.} for general definitions of weighted Pauli measurements and another CZ-gate generation method.

\textit{Near-deterministic implementations.}
Next, we consider increasing the success probability of the CZ-gate generation.
A naive approach to this end is to prepare multiple 1D chains connected to the same pair of target qubits in parallel.
Each chain can be consumed to implement a CZ gate on the target qubits with the procedure in Fig.~\ref{fig:CZ_generation}.
If we could ignore the effect of failure events on the CZ-gate generations and select one of the success events, the overall failure probability would be exponentially suppressed.
However, removing their effect is not straightforward because the failure operation corresponding to the failure event may act on the target qubits as a unitary operation and/or an entangling projection.
Note that, as long as an adjacent qubit of a target qubit remains unmeasured, any effect from the chain on the target qubit can be removed by measuring the adjacent qubit in the $Z$ basis. Therefore, we need to consider failure events only on qubits adjacent to the target qubits~\footnote{One might consider ``indirectly'' measuring the adjacent qubits in the $Z$ basis by attaching additional branches on the adjacent qubits and measuring the branches. However, it does not work because, if we measure these additional branches in the $Z$ basis to detach them from the main chain, these measurement outcomes can flip the success and failure events of the original CZ gate generation.}.

Let us first consider implementing MEPs, rather than CZ gates, in a near-deterministic manner.
An MEP can be implemented with the MBC of $\mathrm{CP}(\phi)$ and $\mathrm{CP}(\pm \phi)$ as in Eq.~\eqref{eq:success_event}.
An attempt to implement an MEP can be repeated by preparing multiple 1D chains of qubits connected to the target qubits in parallel, as explained above.
Here, in contrast to the case of CZ gates, failure events corresponding to Eq.~\eqref{eq:failure_event} have no effect once one of the events succeeds because of $P_\pm (P_\mp+\cos(\phi/2)P_\pm)\propto P_\pm$.
Therefore, after the $k$ repetitions of the MBC, only $(O^{f}_\pm)^k$ is considered to be a failure event.
To make the failure probability close to zero for an arbitrary input state, we need to combine the MBC of $\mathrm{CP}(\phi)$ and $\mathrm{CP}(\phi)$ and the MBC of $\mathrm{CP}(\phi)$ and $\mathrm{CP}(- \phi)$.
When both of the MBCs are applied $k$ times, the unique failure event is 
\begin{equation}\label{eq:failure_probability}
    (O^{f}_+ O^{f}_-)^k =\cos^k(\phi/2) (R_z(k\phi)\otimes I),
\end{equation}
which occurs with the probability of $\cos^{2k}(\phi/2)$.
Therefore, the MEP can be implemented with the success probability of $1-\cos^{2k}(\phi/2)$.

\begin{figure}[tbp]
\centering
\includegraphics[scale=0.9, trim=0 0 0 0, clip]{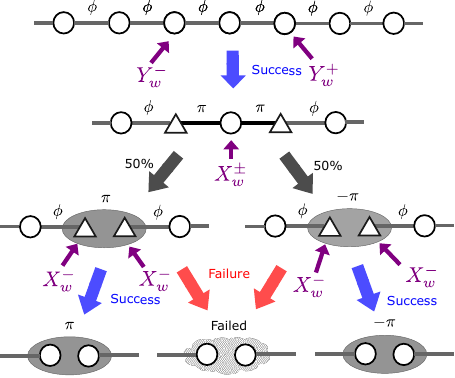}
\caption{Probabilistic generation of an entangling projection with weight $\pi$ or $-\pi$.}\label{fig:projection_generation}
\end{figure}

The preceding procedure requires not only $\mathrm{CP}(\phi)$ but also $\mathrm{CP}(-\phi)$. Thus, we need to prepare $\mathrm{CP}(-\phi)$ from $\mathrm{CP}(\phi)$ using SQMs.
To this end, we consider preparing the MEP with weight $-\pi$, that is, $P_-$.
Once we obtain such an MEP, we can deterministically prepare $\mathrm{CP}(-\phi)$ by applying an $X^-_w$-basis measurement to one of the projected qubits, as shown in Fig.~\ref{fig:transformation_diagram}.
Figure~\ref{fig:projection_generation} shows how to prepare the MEP with weight $-\pi$ from a 1D chain by applying five SQMs. First, we apply $Y_w^+$-basis and $Y_w^-$-basis measurements to two qubits that are connected to the same qubit, called the center qubit. If both of them succeed, the center qubit is connected to two other qubits with perfect edges. Note that the local non-unitary operations on the center qubit caused by $Y_w^+$-basis and $Y_w^-$-basis measurements cancel each other out, and thereby the center qubit is not affected by any non-unitary operation. Then, we apply an $X_w^-$-basis (or equivalently $X^+_w$-basis) measurement to the center qubit. Since it is the MBC of two CZ gates on the $X_w^-$ basis, the resulting operator becomes $O^s_+$ or $O^f_+$ with $\phi=\pi$, which correspond to an MEP with weight $\pi$ or $-\pi$, respectively.
Finally, we measure the remaining two qubits in the $X^-_w$ basis. Similar to the process in Fig.~\ref{fig:transformation_diagram}, the two successful applications of the $X^-_w$-basis measurement produce the same projection as one before their applications, except for the absence of non-unitary operations.
The resulting operator becomes a genuine MEP with weight $\pi$ or $-\pi$, depending on the measurement outcome on the center qubit, where both events occur with equal probability.

\begin{figure}[tbp]
\centering
\includegraphics[scale=0.9, trim=0 0 0 0, clip]{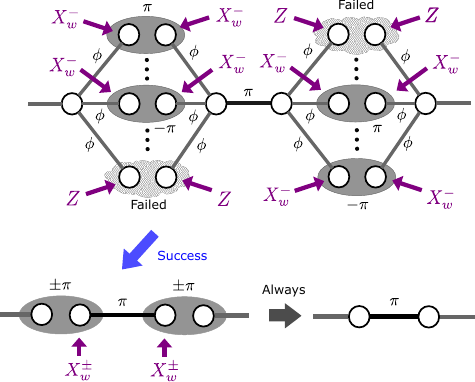}
\caption{Heralded generation of a CZ gate.}
\label{fig:gate_transportation}
\end{figure}

We now return to the ellipse in Fig.~\ref{fig:entire_graph} and focus on one of the lines, consisting of a five-qubit part and seven-qubit parts.
The middle three qubits in the five-qubit part and the middle five qubits in each seven-qubit part are used for the CZ-gate generation in Fig.~\ref{fig:CZ_generation} and the MEP generation in Fig.~\ref{fig:projection_generation}, respectively.
After executing the CZ-gate and MEP generations, if the CZ-gate generation succeeds, we obtain the state shown in Fig.~\ref{fig:gate_transportation}, where MEPs with weight $\pi$ or $-\pi$ or failure operations are randomly applied.
For a pair of qubits to which an MEP with weight $\pi$ or $-\pi$ is successfully applied, we apply two $X^-_w$-basis measurements to the qubits and generate $O^s_+$ or $O^f_+$ for weight $\pi$ and $O^s_-$ or $O^f_-$ for weight $-\pi$, depending on the measurement outcomes.
For a pair of qubits in which a failure operation is applied, we detach the qubits by applying the $Z$-basis measurements to them.
Except for the fact that the numbers of applied MEPs with weights $\pi$ and $-\pi$ are determined randomly, the situation on each side corresponds to the near-deterministic MEP generation discussed above.
As a result, MEPs are applied on both sides with failure probability exponentially small in $k$.
After the successful application of MEPs, two maximally entangled pairs are connected with a perfect edge, as in Fig.~\ref{fig:gate_transportation}.
Finally, we reduce each maximally entangled pair to a single qubit by applying an $X^\pm_w$-basis measurement and obtain a perfect edge between qubits. If the initial CZ-gate generation fails in this procedure, we can detach the line from the target qubits by measuring the adjacent qubits in the $Z$ basis, whereas this is impossible in the direct implementation of the CZ-gate generation.
Therefore, by preparing $m$ copies of the setup in parallel as in Fig.~\ref{fig:entire_graph}, we can implement a CZ-gate generation near-deterministically.

\textit{Entire performance.}
Here, we summarize the entire performance of our protocol. (See~\cite{Note99} for the detailed derivations.)
The SQMs in Figs.~\ref{fig:CZ_generation} and \ref{fig:projection_generation} and the first two weighted Pauli-X measurements in Fig.~\ref{fig:gate_transportation} succeed with probabilities $O(\phi^4)$, $O(\phi^6)$, and $O(\phi^2)$, respectively. These probabilities can be estimated from the fact that generations of a maximally entangling operation succeed with probabilities of $\sim \sin^2(\phi/2)$ as in Eqs.~\eqref{eq:success_event} and \eqref{eq:CZ_gate_with_nonunitaries}, while the others succeed with almost constant probabilities.
As a result, with $k$, $m$, and $n$ defined as in Fig.~\ref{fig:entire_graph}, $k$ and $m$ need to be of the order of $\phi^{-8}\log(\delta^{-1}n)$ and $\phi^{-4}\log(\delta^{-1}n)$, respectively, to obtain the overall success probability of $1-\delta$, where the factor of $\log{n}$ arises from the fact that CZ-gate generations on all the edges must succeed to ensure the success of the entire protocol.
The number of consumed qubits is $N=2n(n-1)(5m+14km)=O(n^2\phi^{-12}\log^2(\delta^{-1}n))$. 

The transformation from the uniform WGS to the 2D cluster state can be performed in a constant depth. More precisely, the CZ-gate generation in Fig.~\ref{fig:CZ_generation} and the MEP generation in Fig.~\ref{fig:projection_generation} take one and two steps, respectively, and the process in Fig.~\ref{fig:gate_transportation} takes two steps; thus, the total depth is four. (See \cite{Note99} for the concrete measurement bases.)
In principle, all the CP gates to generate a WGS can be performed simultaneously, for example, using a time evolution of Ising-type Hamiltonians~\cite{Dur_Phys.Rev.Lett._2005}, where the phase $\phi$ increases linearly with the interaction time. In such cases, our protocol implies that the preparation time of universal resources can be arbitrarily short, at the cost of additional constant-depth SQMs in MBQC.

Even when we are restricted to using a uniform WGS on a genuine square lattice, we can still perform MBQC with an exponentially small success probability~\footnote{By applying $Z$-basis measurements appropriately on the WGS on a square lattice, we can generate a WGS on a decorated square lattice in which each edge is replaced with a 1D chain of three qubits. Then, the WGS can be transformed to a 2D cluster state with the procedure in Fig.~\ref{fig:CZ_generation}.}.
According to the results in~\cite{Aaronson_Proc.R.Soc.Math.Phys.Eng.Sci._2005,Bremner_Proc.R.Soc.Math.Phys.Eng.Sci._2010}, this fact implies that SQMs even on the square-lattice WGS cannot be efficiently simulated on a classical computer within multiplicative error (under a plausible complexity assumption).
On the other hand, if we are allowed to choose each weight to be either $\phi$ or $-\phi$, the costly generation of $\mathrm{CP}(-\phi)$ becomes unnecessary.
Therefore, the set of CP gates with phases $\phi$ and $-\phi$ seems more effective for MBQC than each individual gate.

\textit{Conclusions.}
In this paper, we have shown that uniform WGSs can be universal resources for an arbitrary nonzero weight on a suitable planar graph.
To this end, we have introduced several new techniques, such as weighted Pauli measurements, which are expected to serve as fundamental tools for further investigation of quantum information processing using WGSs.
In particular, MBC is a concept that has likely not yet been studied systematically, despite the extensive research on entangling operations~\cite{Makhlin_QuantumInformationProcessing_2002,Zanardi_Phys.Rev.A_2000,Zanardi_Phys.Rev.A_2001,Cirac_Phys.Rev.Lett._2001,Kraus_Phys.Rev.A_2001,Dur_Phys.Rev.Lett._2002,Nielsen_Phys.Rev.A_2003}, and can be generalized in multiple ways, such as for more than two operators or for operators on qudits.
Another interesting direction is to examine whether our new universal resource can be interpreted as a symmetry-protected topological phase, as with cluster states and AKLT states, even though, unlike these existing examples, the WGS contains only arbitrarily small entanglement between neighboring qubits.

\textit{Acknowledgement.} We thank Yasuaki Nakayama, Seiseki Akibue, and Koji Azuma for fruitful discussion. YT is partially supported by the MEXT Quantum Leap Flagship Program (MEXT Q-LEAP) Grant Number JPMXS0120319794.

\bibliography{mbec.bib}

\begin{thebibliography}{76}%
\makeatletter
\providecommand \@ifxundefined [1]{%
 \@ifx{#1\undefined}
}%
\providecommand \@ifnum [1]{%
 \ifnum #1\expandafter \@firstoftwo
 \else \expandafter \@secondoftwo
 \fi
}%
\providecommand \@ifx [1]{%
 \ifx #1\expandafter \@firstoftwo
 \else \expandafter \@secondoftwo
 \fi
}%
\providecommand \natexlab [1]{#1}%
\providecommand \enquote  [1]{``#1''}%
\providecommand \bibnamefont  [1]{#1}%
\providecommand \bibfnamefont [1]{#1}%
\providecommand \citenamefont [1]{#1}%
\providecommand \href@noop [0]{\@secondoftwo}%
\providecommand \href [0]{\begingroup \@sanitize@url \@href}%
\providecommand \@href[1]{\@@startlink{#1}\@@href}%
\providecommand \@@href[1]{\endgroup#1\@@endlink}%
\providecommand \@sanitize@url [0]{\catcode `\\12\catcode `\$12\catcode `\&12\catcode `\#12\catcode `\^12\catcode `\_12\catcode `\%12\relax}%
\providecommand \@@startlink[1]{}%
\providecommand \@@endlink[0]{}%
\providecommand \url  [0]{\begingroup\@sanitize@url \@url }%
\providecommand \@url [1]{\endgroup\@href {#1}{\urlprefix }}%
\providecommand \urlprefix  [0]{URL }%
\providecommand \Eprint [0]{\href }%
\providecommand \doibase [0]{https://doi.org/}%
\providecommand \selectlanguage [0]{\@gobble}%
\providecommand \bibinfo  [0]{\@secondoftwo}%
\providecommand \bibfield  [0]{\@secondoftwo}%
\providecommand \translation [1]{[#1]}%
\providecommand \BibitemOpen [0]{}%
\providecommand \bibitemStop [0]{}%
\providecommand \bibitemNoStop [0]{.\EOS\space}%
\providecommand \EOS [0]{\spacefactor3000\relax}%
\providecommand \BibitemShut  [1]{\csname bibitem#1\endcsname}%
\let\auto@bib@innerbib\@empty
\bibitem [{\citenamefont {Raussendorf}\ and\ \citenamefont {Briegel}(2001)}]{Raussendorf_Phys.Rev.Lett._2001}%
  \BibitemOpen
  \bibfield  {author} {\bibinfo {author} {\bibfnamefont {R.}~\bibnamefont {Raussendorf}}\ and\ \bibinfo {author} {\bibfnamefont {H.~J.}\ \bibnamefont {Briegel}},\ }\bibfield  {title} {\bibinfo {title} {A {{One-Way Quantum Computer}}},\ }\href {https://doi.org/10.1103/PhysRevLett.86.5188} {\bibfield  {journal} {\bibinfo  {journal} {Physical Review Letters}\ }\textbf {\bibinfo {volume} {86}},\ \bibinfo {pages} {5188} (\bibinfo {year} {2001})}\BibitemShut {NoStop}%
\bibitem [{\citenamefont {Briegel}\ \emph {et~al.}(2009)\citenamefont {Briegel}, \citenamefont {Browne}, \citenamefont {D{\"u}r}, \citenamefont {Raussendorf},\ and\ \citenamefont {{Van den Nest}}}]{Briegel_NaturePhys_2009}%
  \BibitemOpen
  \bibfield  {author} {\bibinfo {author} {\bibfnamefont {H.~J.}\ \bibnamefont {Briegel}}, \bibinfo {author} {\bibfnamefont {D.~E.}\ \bibnamefont {Browne}}, \bibinfo {author} {\bibfnamefont {W.}~\bibnamefont {D{\"u}r}}, \bibinfo {author} {\bibfnamefont {R.}~\bibnamefont {Raussendorf}},\ and\ \bibinfo {author} {\bibfnamefont {M.}~\bibnamefont {{Van den Nest}}},\ }\bibfield  {title} {\bibinfo {title} {Measurement-based quantum computation},\ }\href {https://doi.org/10.1038/nphys1157} {\bibfield  {journal} {\bibinfo  {journal} {Nature Physics}\ }\textbf {\bibinfo {volume} {5}},\ \bibinfo {pages} {19} (\bibinfo {year} {2009})}\BibitemShut {NoStop}%
\bibitem [{\citenamefont {{Van den Nest}}\ \emph {et~al.}(2006)\citenamefont {{Van den Nest}}, \citenamefont {Miyake}, \citenamefont {D{\"u}r},\ and\ \citenamefont {Briegel}}]{VandenNest_Phys.Rev.Lett._2006}%
  \BibitemOpen
  \bibfield  {author} {\bibinfo {author} {\bibfnamefont {M.}~\bibnamefont {{Van den Nest}}}, \bibinfo {author} {\bibfnamefont {A.}~\bibnamefont {Miyake}}, \bibinfo {author} {\bibfnamefont {W.}~\bibnamefont {D{\"u}r}},\ and\ \bibinfo {author} {\bibfnamefont {H.~J.}\ \bibnamefont {Briegel}},\ }\bibfield  {title} {\bibinfo {title} {Universal {{Resources}} for {{Measurement-Based Quantum Computation}}},\ }\href {https://doi.org/10.1103/PhysRevLett.97.150504} {\bibfield  {journal} {\bibinfo  {journal} {Physical Review Letters}\ }\textbf {\bibinfo {volume} {97}},\ \bibinfo {pages} {150504} (\bibinfo {year} {2006})}\BibitemShut {NoStop}%
\bibitem [{\citenamefont {Gross}\ and\ \citenamefont {Eisert}(2007)}]{Gross_Phys.Rev.Lett._2007}%
  \BibitemOpen
  \bibfield  {author} {\bibinfo {author} {\bibfnamefont {D.}~\bibnamefont {Gross}}\ and\ \bibinfo {author} {\bibfnamefont {J.}~\bibnamefont {Eisert}},\ }\bibfield  {title} {\bibinfo {title} {Novel {{Schemes}} for {{Measurement-Based Quantum Computation}}},\ }\href {https://doi.org/10.1103/PhysRevLett.98.220503} {\bibfield  {journal} {\bibinfo  {journal} {Physical Review Letters}\ }\textbf {\bibinfo {volume} {98}},\ \bibinfo {pages} {220503} (\bibinfo {year} {2007})}\BibitemShut {NoStop}%
\bibitem [{\citenamefont {Gross}\ \emph {et~al.}(2007)\citenamefont {Gross}, \citenamefont {Eisert}, \citenamefont {Schuch},\ and\ \citenamefont {{Perez-Garcia}}}]{Gross_Phys.Rev.A_2007}%
  \BibitemOpen
  \bibfield  {author} {\bibinfo {author} {\bibfnamefont {D.}~\bibnamefont {Gross}}, \bibinfo {author} {\bibfnamefont {J.}~\bibnamefont {Eisert}}, \bibinfo {author} {\bibfnamefont {N.}~\bibnamefont {Schuch}},\ and\ \bibinfo {author} {\bibfnamefont {D.}~\bibnamefont {{Perez-Garcia}}},\ }\bibfield  {title} {\bibinfo {title} {Measurement-based quantum computation beyond the one-way model},\ }\href {https://doi.org/10.1103/PhysRevA.76.052315} {\bibfield  {journal} {\bibinfo  {journal} {Physical Review A}\ }\textbf {\bibinfo {volume} {76}},\ \bibinfo {pages} {052315} (\bibinfo {year} {2007})}\BibitemShut {NoStop}%
\bibitem [{\citenamefont {Wei}\ \emph {et~al.}(2011)\citenamefont {Wei}, \citenamefont {Affleck},\ and\ \citenamefont {Raussendorf}}]{Wei_Phys.Rev.Lett._2011}%
  \BibitemOpen
  \bibfield  {author} {\bibinfo {author} {\bibfnamefont {T.-C.}\ \bibnamefont {Wei}}, \bibinfo {author} {\bibfnamefont {I.}~\bibnamefont {Affleck}},\ and\ \bibinfo {author} {\bibfnamefont {R.}~\bibnamefont {Raussendorf}},\ }\bibfield  {title} {\bibinfo {title} {Affleck-{{Kennedy-Lieb-Tasaki State}} on a {{Honeycomb Lattice}} is a {{Universal Quantum Computational Resource}}},\ }\href {https://doi.org/10.1103/PhysRevLett.106.070501} {\bibfield  {journal} {\bibinfo  {journal} {Physical Review Letters}\ }\textbf {\bibinfo {volume} {106}},\ \bibinfo {pages} {070501} (\bibinfo {year} {2011})}\BibitemShut {NoStop}%
\bibitem [{\citenamefont {Wei}\ \emph {et~al.}(2012)\citenamefont {Wei}, \citenamefont {Affleck},\ and\ \citenamefont {Raussendorf}}]{Wei_Phys.Rev.A_2012}%
  \BibitemOpen
  \bibfield  {author} {\bibinfo {author} {\bibfnamefont {T.-C.}\ \bibnamefont {Wei}}, \bibinfo {author} {\bibfnamefont {I.}~\bibnamefont {Affleck}},\ and\ \bibinfo {author} {\bibfnamefont {R.}~\bibnamefont {Raussendorf}},\ }\bibfield  {title} {\bibinfo {title} {Two-dimensional {{Affleck-Kennedy-Lieb-Tasaki}} state on the honeycomb lattice is a universal resource for quantum computation},\ }\href {https://doi.org/10.1103/PhysRevA.86.032328} {\bibfield  {journal} {\bibinfo  {journal} {Physical Review A}\ }\textbf {\bibinfo {volume} {86}},\ \bibinfo {pages} {032328} (\bibinfo {year} {2012})}\BibitemShut {NoStop}%
\bibitem [{\citenamefont {{Van den Nest}}(2013)}]{VandenNest_Phys.Rev.Lett._2013}%
  \BibitemOpen
  \bibfield  {author} {\bibinfo {author} {\bibfnamefont {M.}~\bibnamefont {{Van den Nest}}},\ }\bibfield  {title} {\bibinfo {title} {Universal {{Quantum Computation}} with {{Little Entanglement}}},\ }\href {https://doi.org/10.1103/PhysRevLett.110.060504} {\bibfield  {journal} {\bibinfo  {journal} {Physical Review Letters}\ }\textbf {\bibinfo {volume} {110}},\ \bibinfo {pages} {060504} (\bibinfo {year} {2013})}\BibitemShut {NoStop}%
\bibitem [{\citenamefont {Wei}(2013)}]{Wei_Phys.Rev.A_2013}%
  \BibitemOpen
  \bibfield  {author} {\bibinfo {author} {\bibfnamefont {T.-C.}\ \bibnamefont {Wei}},\ }\bibfield  {title} {\bibinfo {title} {Quantum computational universality of {{Affleck-Kennedy-Lieb-Tasaki}} states beyond the honeycomb lattice},\ }\href {https://doi.org/10.1103/PhysRevA.88.062307} {\bibfield  {journal} {\bibinfo  {journal} {Physical Review A}\ }\textbf {\bibinfo {volume} {88}},\ \bibinfo {pages} {062307} (\bibinfo {year} {2013})}\BibitemShut {NoStop}%
\bibitem [{\citenamefont {Wei}\ \emph {et~al.}(2014)\citenamefont {Wei}, \citenamefont {Haghnegahdar},\ and\ \citenamefont {Raussendorf}}]{Wei_Phys.Rev.A_2014}%
  \BibitemOpen
  \bibfield  {author} {\bibinfo {author} {\bibfnamefont {T.-C.}\ \bibnamefont {Wei}}, \bibinfo {author} {\bibfnamefont {P.}~\bibnamefont {Haghnegahdar}},\ and\ \bibinfo {author} {\bibfnamefont {R.}~\bibnamefont {Raussendorf}},\ }\bibfield  {title} {\bibinfo {title} {Hybrid valence-bond states for universal quantum computation},\ }\href {https://doi.org/10.1103/PhysRevA.90.042333} {\bibfield  {journal} {\bibinfo  {journal} {Physical Review A}\ }\textbf {\bibinfo {volume} {90}},\ \bibinfo {pages} {042333} (\bibinfo {year} {2014})}\BibitemShut {NoStop}%
\bibitem [{\citenamefont {Wei}\ and\ \citenamefont {Raussendorf}(2015)}]{Wei_Phys.Rev.A_2015}%
  \BibitemOpen
  \bibfield  {author} {\bibinfo {author} {\bibfnamefont {T.-C.}\ \bibnamefont {Wei}}\ and\ \bibinfo {author} {\bibfnamefont {R.}~\bibnamefont {Raussendorf}},\ }\bibfield  {title} {\bibinfo {title} {Universal measurement-based quantum computation with spin-2 {{Affleck-Kennedy-Lieb-Tasaki}} states},\ }\href {https://doi.org/10.1103/PhysRevA.92.012310} {\bibfield  {journal} {\bibinfo  {journal} {Physical Review A}\ }\textbf {\bibinfo {volume} {92}},\ \bibinfo {pages} {012310} (\bibinfo {year} {2015})}\BibitemShut {NoStop}%
\bibitem [{\citenamefont {Nautrup}\ and\ \citenamefont {Wei}(2015)}]{Nautrup_Phys.Rev.A_2015}%
  \BibitemOpen
  \bibfield  {author} {\bibinfo {author} {\bibfnamefont {H.~P.}\ \bibnamefont {Nautrup}}\ and\ \bibinfo {author} {\bibfnamefont {T.-C.}\ \bibnamefont {Wei}},\ }\bibfield  {title} {\bibinfo {title} {Symmetry-protected topologically ordered states for universal quantum computation},\ }\href {https://doi.org/10.1103/PhysRevA.92.052309} {\bibfield  {journal} {\bibinfo  {journal} {Physical Review A}\ }\textbf {\bibinfo {volume} {92}},\ \bibinfo {pages} {052309} (\bibinfo {year} {2015})}\BibitemShut {NoStop}%
\bibitem [{\citenamefont {Chen}\ \emph {et~al.}(2018)\citenamefont {Chen}, \citenamefont {Prakash},\ and\ \citenamefont {Wei}}]{Chen_Phys.Rev.A_2018}%
  \BibitemOpen
  \bibfield  {author} {\bibinfo {author} {\bibfnamefont {Y.}~\bibnamefont {Chen}}, \bibinfo {author} {\bibfnamefont {A.}~\bibnamefont {Prakash}},\ and\ \bibinfo {author} {\bibfnamefont {T.-C.}\ \bibnamefont {Wei}},\ }\bibfield  {title} {\bibinfo {title} {Universal quantum computing using \$\textbraceleft (\textbraceleft\textbackslash mathbb\textbraceleft{{Z}}\textbraceright\textbraceright\_\textbraceleft d\textbraceright )\textbraceright\textasciicircum\textbraceleft 3\textbraceright\$ symmetry-protected topologically ordered states},\ }\href {https://doi.org/10.1103/PhysRevA.97.022305} {\bibfield  {journal} {\bibinfo  {journal} {Physical Review A}\ }\textbf {\bibinfo {volume} {97}},\ \bibinfo {pages} {022305} (\bibinfo {year} {2018})}\BibitemShut {NoStop}%
\bibitem [{\citenamefont {Kissinger}\ and\ \citenamefont {van~de Wetering}(2019)}]{Kissinger_Quantum_2019}%
  \BibitemOpen
  \bibfield  {author} {\bibinfo {author} {\bibfnamefont {A.}~\bibnamefont {Kissinger}}\ and\ \bibinfo {author} {\bibfnamefont {J.}~\bibnamefont {van~de Wetering}},\ }\bibfield  {title} {\bibinfo {title} {Universal {{MBQC}} with generalised parity-phase interactions and {{Pauli}} measurements},\ }\href {https://doi.org/10.22331/q-2019-04-26-134} {\bibfield  {journal} {\bibinfo  {journal} {Quantum}\ }\textbf {\bibinfo {volume} {3}},\ \bibinfo {pages} {134} (\bibinfo {year} {2019})}\BibitemShut {NoStop}%
\bibitem [{\citenamefont {Miller}\ and\ \citenamefont {Miyake}(2016)}]{Miller_npjQuantumInf_2016}%
  \BibitemOpen
  \bibfield  {author} {\bibinfo {author} {\bibfnamefont {J.}~\bibnamefont {Miller}}\ and\ \bibinfo {author} {\bibfnamefont {A.}~\bibnamefont {Miyake}},\ }\bibfield  {title} {\bibinfo {title} {Hierarchy of universal entanglement in {{2D}} measurement-based quantum computation},\ }\href {https://doi.org/10.1038/npjqi.2016.36} {\bibfield  {journal} {\bibinfo  {journal} {npj Quantum Information}\ }\textbf {\bibinfo {volume} {2}},\ \bibinfo {pages} {1} (\bibinfo {year} {2016})}\BibitemShut {NoStop}%
\bibitem [{\citenamefont {Miller}\ and\ \citenamefont {Miyake}(2018)}]{Miller_Phys.Rev.Lett._2018}%
  \BibitemOpen
  \bibfield  {author} {\bibinfo {author} {\bibfnamefont {J.}~\bibnamefont {Miller}}\ and\ \bibinfo {author} {\bibfnamefont {A.}~\bibnamefont {Miyake}},\ }\bibfield  {title} {\bibinfo {title} {Latent {{Computational Complexity}} of {{Symmetry-Protected Topological Order}} with {{Fractional Symmetry}}},\ }\href {https://doi.org/10.1103/PhysRevLett.120.170503} {\bibfield  {journal} {\bibinfo  {journal} {Physical Review Letters}\ }\textbf {\bibinfo {volume} {120}},\ \bibinfo {pages} {170503} (\bibinfo {year} {2018})}\BibitemShut {NoStop}%
\bibitem [{\citenamefont {Takeuchi}\ \emph {et~al.}(2019)\citenamefont {Takeuchi}, \citenamefont {Morimae},\ and\ \citenamefont {Hayashi}}]{Takeuchi_SciRep_2019}%
  \BibitemOpen
  \bibfield  {author} {\bibinfo {author} {\bibfnamefont {Y.}~\bibnamefont {Takeuchi}}, \bibinfo {author} {\bibfnamefont {T.}~\bibnamefont {Morimae}},\ and\ \bibinfo {author} {\bibfnamefont {M.}~\bibnamefont {Hayashi}},\ }\bibfield  {title} {\bibinfo {title} {Quantum computational universality of hypergraph states with {{Pauli-X}} and {{Z}} basis measurements},\ }\href {https://doi.org/10.1038/s41598-019-49968-3} {\bibfield  {journal} {\bibinfo  {journal} {Scientific Reports}\ }\textbf {\bibinfo {volume} {9}},\ \bibinfo {pages} {13585} (\bibinfo {year} {2019})}\BibitemShut {NoStop}%
\bibitem [{\citenamefont {Briegel}\ and\ \citenamefont {Raussendorf}(2001)}]{Briegel_Phys.Rev.Lett._2001}%
  \BibitemOpen
  \bibfield  {author} {\bibinfo {author} {\bibfnamefont {H.~J.}\ \bibnamefont {Briegel}}\ and\ \bibinfo {author} {\bibfnamefont {R.}~\bibnamefont {Raussendorf}},\ }\bibfield  {title} {\bibinfo {title} {Persistent {{Entanglement}} in {{Arrays}} of {{Interacting Particles}}},\ }\href {https://doi.org/10.1103/PhysRevLett.86.910} {\bibfield  {journal} {\bibinfo  {journal} {Physical Review Letters}\ }\textbf {\bibinfo {volume} {86}},\ \bibinfo {pages} {910} (\bibinfo {year} {2001})}\BibitemShut {NoStop}%
\bibitem [{\citenamefont {Raussendorf}\ \emph {et~al.}(2019)\citenamefont {Raussendorf}, \citenamefont {Okay}, \citenamefont {Wang}, \citenamefont {Stephen},\ and\ \citenamefont {Nautrup}}]{Raussendorf_Phys.Rev.Lett._2019}%
  \BibitemOpen
  \bibfield  {author} {\bibinfo {author} {\bibfnamefont {R.}~\bibnamefont {Raussendorf}}, \bibinfo {author} {\bibfnamefont {C.}~\bibnamefont {Okay}}, \bibinfo {author} {\bibfnamefont {D.-S.}\ \bibnamefont {Wang}}, \bibinfo {author} {\bibfnamefont {D.~T.}\ \bibnamefont {Stephen}},\ and\ \bibinfo {author} {\bibfnamefont {H.~P.}\ \bibnamefont {Nautrup}},\ }\bibfield  {title} {\bibinfo {title} {Computationally {{Universal Phase}} of {{Quantum Matter}}},\ }\href {https://doi.org/10.1103/PhysRevLett.122.090501} {\bibfield  {journal} {\bibinfo  {journal} {Physical Review Letters}\ }\textbf {\bibinfo {volume} {122}},\ \bibinfo {pages} {090501} (\bibinfo {year} {2019})}\BibitemShut {NoStop}%
\bibitem [{\citenamefont {Devakul}\ and\ \citenamefont {Williamson}(2018)}]{Devakul_Phys.Rev.A_2018}%
  \BibitemOpen
  \bibfield  {author} {\bibinfo {author} {\bibfnamefont {T.}~\bibnamefont {Devakul}}\ and\ \bibinfo {author} {\bibfnamefont {D.~J.}\ \bibnamefont {Williamson}},\ }\bibfield  {title} {\bibinfo {title} {Universal quantum computation using fractal symmetry-protected cluster phases},\ }\href {https://doi.org/10.1103/PhysRevA.98.022332} {\bibfield  {journal} {\bibinfo  {journal} {Physical Review A}\ }\textbf {\bibinfo {volume} {98}},\ \bibinfo {pages} {022332} (\bibinfo {year} {2018})}\BibitemShut {NoStop}%
\bibitem [{\citenamefont {Stephen}\ \emph {et~al.}(2019)\citenamefont {Stephen}, \citenamefont {Nautrup}, \citenamefont {{Bermejo-Vega}}, \citenamefont {Eisert},\ and\ \citenamefont {Raussendorf}}]{Stephen_Quantum_2019}%
  \BibitemOpen
  \bibfield  {author} {\bibinfo {author} {\bibfnamefont {D.~T.}\ \bibnamefont {Stephen}}, \bibinfo {author} {\bibfnamefont {H.~P.}\ \bibnamefont {Nautrup}}, \bibinfo {author} {\bibfnamefont {J.}~\bibnamefont {{Bermejo-Vega}}}, \bibinfo {author} {\bibfnamefont {J.}~\bibnamefont {Eisert}},\ and\ \bibinfo {author} {\bibfnamefont {R.}~\bibnamefont {Raussendorf}},\ }\bibfield  {title} {\bibinfo {title} {Subsystem symmetries, quantum cellular automata, and computational phases of quantum matter},\ }\href {https://doi.org/10.22331/q-2019-05-20-142} {\bibfield  {journal} {\bibinfo  {journal} {Quantum}\ }\textbf {\bibinfo {volume} {3}},\ \bibinfo {pages} {142} (\bibinfo {year} {2019})},\ \Eprint {https://arxiv.org/abs/1806.08780} {arXiv:1806.08780 [cond-mat, physics:quant-ph]} \BibitemShut {NoStop}%
\bibitem [{\citenamefont {Daniel}\ \emph {et~al.}(2020)\citenamefont {Daniel}, \citenamefont {Alexander},\ and\ \citenamefont {Miyake}}]{Daniel_Quantum_2020}%
  \BibitemOpen
  \bibfield  {author} {\bibinfo {author} {\bibfnamefont {A.~K.}\ \bibnamefont {Daniel}}, \bibinfo {author} {\bibfnamefont {R.~N.}\ \bibnamefont {Alexander}},\ and\ \bibinfo {author} {\bibfnamefont {A.}~\bibnamefont {Miyake}},\ }\bibfield  {title} {\bibinfo {title} {Computational universality of symmetry-protected topologically ordered cluster phases on {{2D Archimedean}} lattices},\ }\href {https://doi.org/10.22331/q-2020-02-10-228} {\bibfield  {journal} {\bibinfo  {journal} {Quantum}\ }\textbf {\bibinfo {volume} {4}},\ \bibinfo {pages} {228} (\bibinfo {year} {2020})}\BibitemShut {NoStop}%
\bibitem [{\citenamefont {Stephen}\ \emph {et~al.}(2024)\citenamefont {Stephen}, \citenamefont {Ho}, \citenamefont {Wei}, \citenamefont {Raussendorf},\ and\ \citenamefont {Verresen}}]{Stephen_Phys.Rev.Lett._2024}%
  \BibitemOpen
  \bibfield  {author} {\bibinfo {author} {\bibfnamefont {D.~T.}\ \bibnamefont {Stephen}}, \bibinfo {author} {\bibfnamefont {W.~W.}\ \bibnamefont {Ho}}, \bibinfo {author} {\bibfnamefont {T.-C.}\ \bibnamefont {Wei}}, \bibinfo {author} {\bibfnamefont {R.}~\bibnamefont {Raussendorf}},\ and\ \bibinfo {author} {\bibfnamefont {R.}~\bibnamefont {Verresen}},\ }\bibfield  {title} {\bibinfo {title} {Universal {{Measurement-Based Quantum Computation}} in a {{One-Dimensional Architecture Enabled}} by {{Dual-Unitary Circuits}}},\ }\href {https://doi.org/10.1103/PhysRevLett.132.250601} {\bibfield  {journal} {\bibinfo  {journal} {Physical Review Letters}\ }\textbf {\bibinfo {volume} {132}},\ \bibinfo {pages} {250601} (\bibinfo {year} {2024})}\BibitemShut {NoStop}%
\bibitem [{\citenamefont {D{\"u}r}\ \emph {et~al.}(2005)\citenamefont {D{\"u}r}, \citenamefont {Hartmann}, \citenamefont {Hein}, \citenamefont {Lewenstein},\ and\ \citenamefont {Briegel}}]{Dur_Phys.Rev.Lett._2005}%
  \BibitemOpen
  \bibfield  {author} {\bibinfo {author} {\bibfnamefont {W.}~\bibnamefont {D{\"u}r}}, \bibinfo {author} {\bibfnamefont {L.}~\bibnamefont {Hartmann}}, \bibinfo {author} {\bibfnamefont {M.}~\bibnamefont {Hein}}, \bibinfo {author} {\bibfnamefont {M.}~\bibnamefont {Lewenstein}},\ and\ \bibinfo {author} {\bibfnamefont {H.-J.}\ \bibnamefont {Briegel}},\ }\bibfield  {title} {\bibinfo {title} {Entanglement in {{Spin Chains}} and {{Lattices}} with {{Long-Range Ising-Type Interactions}}},\ }\href {https://doi.org/10.1103/PhysRevLett.94.097203} {\bibfield  {journal} {\bibinfo  {journal} {Physical Review Letters}\ }\textbf {\bibinfo {volume} {94}},\ \bibinfo {pages} {097203} (\bibinfo {year} {2005})}\BibitemShut {NoStop}%
\bibitem [{\citenamefont {Raussendorf}\ \emph {et~al.}(2003)\citenamefont {Raussendorf}, \citenamefont {Browne},\ and\ \citenamefont {Briegel}}]{Raussendorf_Phys.Rev.A_2003}%
  \BibitemOpen
  \bibfield  {author} {\bibinfo {author} {\bibfnamefont {R.}~\bibnamefont {Raussendorf}}, \bibinfo {author} {\bibfnamefont {D.~E.}\ \bibnamefont {Browne}},\ and\ \bibinfo {author} {\bibfnamefont {H.~J.}\ \bibnamefont {Briegel}},\ }\bibfield  {title} {\bibinfo {title} {Measurement-based quantum computation on cluster states},\ }\href {https://doi.org/10.1103/PhysRevA.68.022312} {\bibfield  {journal} {\bibinfo  {journal} {Physical Review A}\ }\textbf {\bibinfo {volume} {68}},\ \bibinfo {pages} {022312} (\bibinfo {year} {2003})}\BibitemShut {NoStop}%
\bibitem [{\citenamefont {Hein}\ \emph {et~al.}(2006)\citenamefont {Hein}, \citenamefont {D{\"u}r}, \citenamefont {Eisert}, \citenamefont {Raussendorf}, \citenamefont {den Nest},\ and\ \citenamefont {Briegel}}]{Hein__2006}%
  \BibitemOpen
  \bibfield  {author} {\bibinfo {author} {\bibfnamefont {M.}~\bibnamefont {Hein}}, \bibinfo {author} {\bibfnamefont {W.}~\bibnamefont {D{\"u}r}}, \bibinfo {author} {\bibfnamefont {J.}~\bibnamefont {Eisert}}, \bibinfo {author} {\bibfnamefont {R.}~\bibnamefont {Raussendorf}}, \bibinfo {author} {\bibfnamefont {M.~V.}\ \bibnamefont {den Nest}},\ and\ \bibinfo {author} {\bibfnamefont {H.-J.}\ \bibnamefont {Briegel}},\ }\href {https://doi.org/10.48550/arXiv.quant-ph/0602096} {\bibinfo {title} {Entanglement in {{Graph States}} and its {{Applications}}}} (\bibinfo {year} {2006}),\ \Eprint {https://arxiv.org/abs/quant-ph/0602096} {arXiv:quant-ph/0602096} \BibitemShut {NoStop}%
\bibitem [{Note1()}]{Note1}%
  \BibitemOpen
  \bibinfo {note} {For example, CP gates are implemented with photon-photon interactions induced by matter systems such as single atoms~\cite {Turchette_Phys.Rev.Lett._1995,Duan_Phys.Rev.Lett._2004,Volz_NaturePhoton_2014,Hacker_Nature_2016,Beck_Proc.Natl.Acad.Sci._2016}, atomic ensembles~\cite {Sagona-Stophel_Phys.Rev.Lett._2020}, quantum dots~\cite {Fushman_Science_2008,Kim_NaturePhoton_2013,Sun_Science_2018}, nitrogen vacancy centers~\cite {Wang_Opt.ExpressOE_2013}, and Rydberg atoms~\cite {Gorshkov_Phys.Rev.Lett._2011,Firstenberg_Nature_2013,Tiarks_Sci.Adv._2016,Thompson_Nature_2017,Tiarks_NaturePhys_2019}.}\BibitemShut {Stop}%
\bibitem [{\citenamefont {Calsamiglia}\ \emph {et~al.}(2005)\citenamefont {Calsamiglia}, \citenamefont {Hartmann}, \citenamefont {D{\"u}r},\ and\ \citenamefont {Briegel}}]{Calsamiglia_Phys.Rev.Lett._2005}%
  \BibitemOpen
  \bibfield  {author} {\bibinfo {author} {\bibfnamefont {J.}~\bibnamefont {Calsamiglia}}, \bibinfo {author} {\bibfnamefont {L.}~\bibnamefont {Hartmann}}, \bibinfo {author} {\bibfnamefont {W.}~\bibnamefont {D{\"u}r}},\ and\ \bibinfo {author} {\bibfnamefont {H.-J.}\ \bibnamefont {Briegel}},\ }\bibfield  {title} {\bibinfo {title} {Spin {{Gases}}: {{Quantum Entanglement Driven}} by {{Classical Kinematics}}},\ }\href {https://doi.org/10.1103/PhysRevLett.95.180502} {\bibfield  {journal} {\bibinfo  {journal} {Physical Review Letters}\ }\textbf {\bibinfo {volume} {95}},\ \bibinfo {pages} {180502} (\bibinfo {year} {2005})}\BibitemShut {NoStop}%
\bibitem [{\citenamefont {Anders}\ \emph {et~al.}(2006)\citenamefont {Anders}, \citenamefont {Plenio}, \citenamefont {D{\"u}r}, \citenamefont {Verstraete},\ and\ \citenamefont {Briegel}}]{Anders_Phys.Rev.Lett._2006}%
  \BibitemOpen
  \bibfield  {author} {\bibinfo {author} {\bibfnamefont {S.}~\bibnamefont {Anders}}, \bibinfo {author} {\bibfnamefont {M.~B.}\ \bibnamefont {Plenio}}, \bibinfo {author} {\bibfnamefont {W.}~\bibnamefont {D{\"u}r}}, \bibinfo {author} {\bibfnamefont {F.}~\bibnamefont {Verstraete}},\ and\ \bibinfo {author} {\bibfnamefont {H.-J.}\ \bibnamefont {Briegel}},\ }\bibfield  {title} {\bibinfo {title} {Ground-{{State Approximation}} for {{Strongly Interacting Spin Systems}} in {{Arbitrary Spatial Dimension}}},\ }\href {https://doi.org/10.1103/PhysRevLett.97.107206} {\bibfield  {journal} {\bibinfo  {journal} {Physical Review Letters}\ }\textbf {\bibinfo {volume} {97}},\ \bibinfo {pages} {107206} (\bibinfo {year} {2006})}\BibitemShut {NoStop}%
\bibitem [{\citenamefont {Hartmann}\ \emph {et~al.}(2007)\citenamefont {Hartmann}, \citenamefont {Calsamiglia}, \citenamefont {D{\"u}r},\ and\ \citenamefont {Briegel}}]{Hartmann_J.Phys.B:At.Mol.Opt.Phys._2007}%
  \BibitemOpen
  \bibfield  {author} {\bibinfo {author} {\bibfnamefont {L.}~\bibnamefont {Hartmann}}, \bibinfo {author} {\bibfnamefont {J.}~\bibnamefont {Calsamiglia}}, \bibinfo {author} {\bibfnamefont {W.}~\bibnamefont {D{\"u}r}},\ and\ \bibinfo {author} {\bibfnamefont {H.~J.}\ \bibnamefont {Briegel}},\ }\bibfield  {title} {\bibinfo {title} {Weighted graph states and applications to spin chains, lattices and gases},\ }\href {https://doi.org/10.1088/0953-4075/40/9/S01} {\bibfield  {journal} {\bibinfo  {journal} {Journal of Physics B: Atomic, Molecular and Optical Physics}\ }\textbf {\bibinfo {volume} {40}},\ \bibinfo {pages} {S1} (\bibinfo {year} {2007})}\BibitemShut {NoStop}%
\bibitem [{\citenamefont {Anders}\ \emph {et~al.}(2007)\citenamefont {Anders}, \citenamefont {Briegel},\ and\ \citenamefont {D{\"u}r}}]{Anders_NewJ.Phys._2007}%
  \BibitemOpen
  \bibfield  {author} {\bibinfo {author} {\bibfnamefont {S.}~\bibnamefont {Anders}}, \bibinfo {author} {\bibfnamefont {H.~J.}\ \bibnamefont {Briegel}},\ and\ \bibinfo {author} {\bibfnamefont {W.}~\bibnamefont {D{\"u}r}},\ }\bibfield  {title} {\bibinfo {title} {A variational method based on weighted graph states},\ }\href {https://doi.org/10.1088/1367-2630/9/10/361} {\bibfield  {journal} {\bibinfo  {journal} {New Journal of Physics}\ }\textbf {\bibinfo {volume} {9}},\ \bibinfo {pages} {361} (\bibinfo {year} {2007})}\BibitemShut {NoStop}%
\bibitem [{\citenamefont {Plato}\ \emph {et~al.}(2008)\citenamefont {Plato}, \citenamefont {Dahlsten},\ and\ \citenamefont {Plenio}}]{Plato_Phys.Rev.A_2008}%
  \BibitemOpen
  \bibfield  {author} {\bibinfo {author} {\bibfnamefont {A.~D.~K.}\ \bibnamefont {Plato}}, \bibinfo {author} {\bibfnamefont {O.~C.}\ \bibnamefont {Dahlsten}},\ and\ \bibinfo {author} {\bibfnamefont {M.~B.}\ \bibnamefont {Plenio}},\ }\bibfield  {title} {\bibinfo {title} {Random circuits by measurements on weighted graph states},\ }\href {https://doi.org/10.1103/PhysRevA.78.042332} {\bibfield  {journal} {\bibinfo  {journal} {Physical Review A}\ }\textbf {\bibinfo {volume} {78}},\ \bibinfo {pages} {042332} (\bibinfo {year} {2008})}\BibitemShut {NoStop}%
\bibitem [{\citenamefont {Xue}(2012)}]{Xue_Phys.Rev.A_2012}%
  \BibitemOpen
  \bibfield  {author} {\bibinfo {author} {\bibfnamefont {P.}~\bibnamefont {Xue}},\ }\bibfield  {title} {\bibinfo {title} {Spin-squeezing property of weighted graph states},\ }\href {https://doi.org/10.1103/PhysRevA.86.023812} {\bibfield  {journal} {\bibinfo  {journal} {Physical Review A}\ }\textbf {\bibinfo {volume} {86}},\ \bibinfo {pages} {023812} (\bibinfo {year} {2012})}\BibitemShut {NoStop}%
\bibitem [{\citenamefont {Ghosh}\ \emph {et~al.}(2024)\citenamefont {Ghosh}, \citenamefont {Das~Agarwal}, \citenamefont {Halder},\ and\ \citenamefont {Sen(De)}}]{Ghosh_Phys.Rev.A_2024}%
  \BibitemOpen
  \bibfield  {author} {\bibinfo {author} {\bibfnamefont {D.}~\bibnamefont {Ghosh}}, \bibinfo {author} {\bibfnamefont {K.}~\bibnamefont {Das~Agarwal}}, \bibinfo {author} {\bibfnamefont {P.}~\bibnamefont {Halder}},\ and\ \bibinfo {author} {\bibfnamefont {A.}~\bibnamefont {Sen(De)}},\ }\bibfield  {title} {\bibinfo {title} {Entanglement of weighted graphs uncovering transitions in variable-range interacting models},\ }\href {https://doi.org/10.1103/PhysRevA.110.022431} {\bibfield  {journal} {\bibinfo  {journal} {Physical Review A}\ }\textbf {\bibinfo {volume} {110}},\ \bibinfo {pages} {022431} (\bibinfo {year} {2024})}\BibitemShut {NoStop}%
\bibitem [{\citenamefont {Ghosh}\ \emph {et~al.}(2025)\citenamefont {Ghosh}, \citenamefont {Agarwal}, \citenamefont {Halder},\ and\ \citenamefont {De}}]{Ghosh__2025}%
  \BibitemOpen
  \bibfield  {author} {\bibinfo {author} {\bibfnamefont {D.}~\bibnamefont {Ghosh}}, \bibinfo {author} {\bibfnamefont {K.~D.}\ \bibnamefont {Agarwal}}, \bibinfo {author} {\bibfnamefont {P.}~\bibnamefont {Halder}},\ and\ \bibinfo {author} {\bibfnamefont {A.~S.}\ \bibnamefont {De}},\ }\href {https://doi.org/10.48550/arXiv.2506.11909} {\bibinfo {title} {Measurement-based quantum computation with variable-range interacting systems}} (\bibinfo {year} {2025}),\ \Eprint {https://arxiv.org/abs/2506.11909} {arXiv:2506.11909 [quant-ph]} \BibitemShut {NoStop}%
\bibitem [{\citenamefont {Szyma{\'n}ski}\ \emph {et~al.}(2025)\citenamefont {Szyma{\'n}ski}, \citenamefont {Vandr{\'e}},\ and\ \citenamefont {G{\"u}hne}}]{Szymanski__2025}%
  \BibitemOpen
  \bibfield  {author} {\bibinfo {author} {\bibfnamefont {K.}~\bibnamefont {Szyma{\'n}ski}}, \bibinfo {author} {\bibfnamefont {L.}~\bibnamefont {Vandr{\'e}}},\ and\ \bibinfo {author} {\bibfnamefont {O.}~\bibnamefont {G{\"u}hne}},\ }\href {https://doi.org/10.48550/arXiv.2402.00937} {\bibinfo {title} {Useful entanglement can be extracted from noisy graph states}} (\bibinfo {year} {2025}),\ \Eprint {https://arxiv.org/abs/2402.00937} {arXiv:2402.00937 [quant-ph]} \BibitemShut {NoStop}%
\bibitem [{\citenamefont {Frantzeskakis}\ \emph {et~al.}(2023)\citenamefont {Frantzeskakis}, \citenamefont {Liu}, \citenamefont {Raissi}, \citenamefont {Barnes},\ and\ \citenamefont {Economou}}]{Frantzeskakis_Phys.Rev.Res._2023}%
  \BibitemOpen
  \bibfield  {author} {\bibinfo {author} {\bibfnamefont {R.}~\bibnamefont {Frantzeskakis}}, \bibinfo {author} {\bibfnamefont {C.}~\bibnamefont {Liu}}, \bibinfo {author} {\bibfnamefont {Z.}~\bibnamefont {Raissi}}, \bibinfo {author} {\bibfnamefont {E.}~\bibnamefont {Barnes}},\ and\ \bibinfo {author} {\bibfnamefont {S.~E.}\ \bibnamefont {Economou}},\ }\bibfield  {title} {\bibinfo {title} {Extracting perfect {{GHZ}} states from imperfect weighted graph states via entanglement concentration},\ }\href {https://doi.org/10.1103/PhysRevResearch.5.023124} {\bibfield  {journal} {\bibinfo  {journal} {Physical Review Research}\ }\textbf {\bibinfo {volume} {5}},\ \bibinfo {pages} {023124} (\bibinfo {year} {2023})}\BibitemShut {NoStop}%
\bibitem [{Note2()}]{Note2}%
  \BibitemOpen
  \bibinfo {note} {There are two notions of universality: strict universality and computational universality~\cite {Aharonov__2003,Takeuchi_Phys.Rev.Lett._2024}. The former means that any quantum circuit can be implemented, and the latter means that the output probability distribution of any quantum circuit can be generated. Most of the universal resources, including 2D cluster states, are strictly universal resources, while hypergraph states are known to be computationally universal resources~\cite {Miller_npjQuantumInf_2016,Miller_Phys.Rev.Lett._2018,Takeuchi_SciRep_2019}}\BibitemShut {NoStop}%
\bibitem [{Note99()}]{Note99}%
  \BibitemOpen
  \bibinfo {note} {Supplemental Material.}\BibitemShut {Stop}%
\bibitem [{\citenamefont {Chen}\ \emph {et~al.}(2010)\citenamefont {Chen}, \citenamefont {Duan}, \citenamefont {Ji},\ and\ \citenamefont {Zeng}}]{Chen_Phys.Rev.Lett._2010}%
  \BibitemOpen
  \bibfield  {author} {\bibinfo {author} {\bibfnamefont {X.}~\bibnamefont {Chen}}, \bibinfo {author} {\bibfnamefont {R.}~\bibnamefont {Duan}}, \bibinfo {author} {\bibfnamefont {Z.}~\bibnamefont {Ji}},\ and\ \bibinfo {author} {\bibfnamefont {B.}~\bibnamefont {Zeng}},\ }\bibfield  {title} {\bibinfo {title} {Quantum {{State Reduction}} for {{Universal Measurement Based Computation}}},\ }\href {https://doi.org/10.1103/PhysRevLett.105.020502} {\bibfield  {journal} {\bibinfo  {journal} {Physical Review Letters}\ }\textbf {\bibinfo {volume} {105}},\ \bibinfo {pages} {020502} (\bibinfo {year} {2010})}\BibitemShut {NoStop}%
\bibitem [{\citenamefont {Bennett}\ \emph {et~al.}(1996)\citenamefont {Bennett}, \citenamefont {Bernstein}, \citenamefont {Popescu},\ and\ \citenamefont {Schumacher}}]{Bennett_Phys.Rev.A_1996}%
  \BibitemOpen
  \bibfield  {author} {\bibinfo {author} {\bibfnamefont {C.~H.}\ \bibnamefont {Bennett}}, \bibinfo {author} {\bibfnamefont {H.~J.}\ \bibnamefont {Bernstein}}, \bibinfo {author} {\bibfnamefont {S.}~\bibnamefont {Popescu}},\ and\ \bibinfo {author} {\bibfnamefont {B.}~\bibnamefont {Schumacher}},\ }\bibfield  {title} {\bibinfo {title} {Concentrating partial entanglement by local operations},\ }\href {https://doi.org/10.1103/PhysRevA.53.2046} {\bibfield  {journal} {\bibinfo  {journal} {Physical Review A}\ }\textbf {\bibinfo {volume} {53}},\ \bibinfo {pages} {2046} (\bibinfo {year} {1996})}\BibitemShut {NoStop}%
\bibitem [{\citenamefont {Bose}\ \emph {et~al.}(1999)\citenamefont {Bose}, \citenamefont {Vedral},\ and\ \citenamefont {Knight}}]{Bose_Phys.Rev.A_1999}%
  \BibitemOpen
  \bibfield  {author} {\bibinfo {author} {\bibfnamefont {S.}~\bibnamefont {Bose}}, \bibinfo {author} {\bibfnamefont {V.}~\bibnamefont {Vedral}},\ and\ \bibinfo {author} {\bibfnamefont {P.~L.}\ \bibnamefont {Knight}},\ }\bibfield  {title} {\bibinfo {title} {Purification via entanglement swapping and conserved entanglement},\ }\href {https://doi.org/10.1103/PhysRevA.60.194} {\bibfield  {journal} {\bibinfo  {journal} {Physical Review A}\ }\textbf {\bibinfo {volume} {60}},\ \bibinfo {pages} {194} (\bibinfo {year} {1999})}\BibitemShut {NoStop}%
\bibitem [{\citenamefont {Ac{\'i}n}\ \emph {et~al.}(2007)\citenamefont {Ac{\'i}n}, \citenamefont {Cirac},\ and\ \citenamefont {Lewenstein}}]{Acin_NaturePhys_2007}%
  \BibitemOpen
  \bibfield  {author} {\bibinfo {author} {\bibfnamefont {A.}~\bibnamefont {Ac{\'i}n}}, \bibinfo {author} {\bibfnamefont {J.~I.}\ \bibnamefont {Cirac}},\ and\ \bibinfo {author} {\bibfnamefont {M.}~\bibnamefont {Lewenstein}},\ }\bibfield  {title} {\bibinfo {title} {Entanglement percolation in quantum networks},\ }\href {https://doi.org/10.1038/nphys549} {\bibfield  {journal} {\bibinfo  {journal} {Nature Physics}\ }\textbf {\bibinfo {volume} {3}},\ \bibinfo {pages} {256} (\bibinfo {year} {2007})}\BibitemShut {NoStop}%
\bibitem [{\citenamefont {Nemoto}\ and\ \citenamefont {Munro}(2004)}]{Nemoto_Phys.Rev.Lett._2004}%
  \BibitemOpen
  \bibfield  {author} {\bibinfo {author} {\bibfnamefont {K.}~\bibnamefont {Nemoto}}\ and\ \bibinfo {author} {\bibfnamefont {W.~J.}\ \bibnamefont {Munro}},\ }\bibfield  {title} {\bibinfo {title} {Nearly {{Deterministic Linear Optical Controlled-NOT Gate}}},\ }\href {https://doi.org/10.1103/PhysRevLett.93.250502} {\bibfield  {journal} {\bibinfo  {journal} {Physical Review Letters}\ }\textbf {\bibinfo {volume} {93}},\ \bibinfo {pages} {250502} (\bibinfo {year} {2004})}\BibitemShut {NoStop}%
\bibitem [{\citenamefont {Barrett}\ \emph {et~al.}(2005)\citenamefont {Barrett}, \citenamefont {Kok}, \citenamefont {Nemoto}, \citenamefont {Beausoleil}, \citenamefont {Munro},\ and\ \citenamefont {Spiller}}]{Barrett_Phys.Rev.A_2005}%
  \BibitemOpen
  \bibfield  {author} {\bibinfo {author} {\bibfnamefont {S.~D.}\ \bibnamefont {Barrett}}, \bibinfo {author} {\bibfnamefont {P.}~\bibnamefont {Kok}}, \bibinfo {author} {\bibfnamefont {K.}~\bibnamefont {Nemoto}}, \bibinfo {author} {\bibfnamefont {R.~G.}\ \bibnamefont {Beausoleil}}, \bibinfo {author} {\bibfnamefont {W.~J.}\ \bibnamefont {Munro}},\ and\ \bibinfo {author} {\bibfnamefont {T.~P.}\ \bibnamefont {Spiller}},\ }\bibfield  {title} {\bibinfo {title} {Symmetry analyzer for nondestructive {{Bell-state}} detection using weak nonlinearities},\ }\href {https://doi.org/10.1103/PhysRevA.71.060302} {\bibfield  {journal} {\bibinfo  {journal} {Physical Review A}\ }\textbf {\bibinfo {volume} {71}},\ \bibinfo {pages} {060302} (\bibinfo {year} {2005})}\BibitemShut {NoStop}%
\bibitem [{\citenamefont {Spiller}\ \emph {et~al.}(2006)\citenamefont {Spiller}, \citenamefont {Nemoto}, \citenamefont {Braunstein}, \citenamefont {Munro}, \citenamefont {{van Loock}},\ and\ \citenamefont {Milburn}}]{Spiller_NewJ.Phys._2006}%
  \BibitemOpen
  \bibfield  {author} {\bibinfo {author} {\bibfnamefont {T.~P.}\ \bibnamefont {Spiller}}, \bibinfo {author} {\bibfnamefont {K.}~\bibnamefont {Nemoto}}, \bibinfo {author} {\bibfnamefont {S.~L.}\ \bibnamefont {Braunstein}}, \bibinfo {author} {\bibfnamefont {W.~J.}\ \bibnamefont {Munro}}, \bibinfo {author} {\bibfnamefont {P.}~\bibnamefont {{van Loock}}},\ and\ \bibinfo {author} {\bibfnamefont {G.~J.}\ \bibnamefont {Milburn}},\ }\bibfield  {title} {\bibinfo {title} {Quantum computation by communication},\ }\href {https://doi.org/10.1088/1367-2630/8/2/030} {\bibfield  {journal} {\bibinfo  {journal} {New Journal of Physics}\ }\textbf {\bibinfo {volume} {8}},\ \bibinfo {pages} {30} (\bibinfo {year} {2006})}\BibitemShut {NoStop}%
\bibitem [{\citenamefont {Halil~Shah}\ and\ \citenamefont {Oi}(2013)}]{HalilShah_LIPIcsVol.22TQC2013_2013}%
  \BibitemOpen
  \bibfield  {author} {\bibinfo {author} {\bibfnamefont {K.}~\bibnamefont {Halil~Shah}}\ and\ \bibinfo {author} {\bibfnamefont {D.~K.}\ \bibnamefont {Oi}},\ }\bibfield  {title} {\bibinfo {title} {Ancilla {{Driven Quantum Computation}} with {{Arbitrary Entangling Strength}}},\ }\href {https://doi.org/10.4230/LIPICS.TQC.2013.1} {\bibfield  {journal} {\bibinfo  {journal} {LIPIcs, Volume 22, TQC 2013}\ }\textbf {\bibinfo {volume} {22}},\ \bibinfo {pages} {1} (\bibinfo {year} {2013})}\BibitemShut {NoStop}%
\bibitem [{\citenamefont {Nielsen}(2000)}]{Nielsen__2000}%
  \BibitemOpen
  \bibfield  {author} {\bibinfo {author} {\bibfnamefont {M.~A.}\ \bibnamefont {Nielsen}},\ }\href {https://doi.org/10.48550/arXiv.quant-ph/0011036} {\bibinfo {title} {Quantum information theory}} (\bibinfo {year} {2000}),\ \Eprint {https://arxiv.org/abs/quant-ph/0011036} {arXiv:quant-ph/0011036} \BibitemShut {NoStop}%
\bibitem [{\citenamefont {Nielsen}\ \emph {et~al.}(2003)\citenamefont {Nielsen}, \citenamefont {Dawson}, \citenamefont {Dodd}, \citenamefont {Gilchrist}, \citenamefont {Mortimer}, \citenamefont {Osborne}, \citenamefont {Bremner}, \citenamefont {Harrow},\ and\ \citenamefont {Hines}}]{Nielsen_Phys.Rev.A_2003}%
  \BibitemOpen
  \bibfield  {author} {\bibinfo {author} {\bibfnamefont {M.~A.}\ \bibnamefont {Nielsen}}, \bibinfo {author} {\bibfnamefont {C.~M.}\ \bibnamefont {Dawson}}, \bibinfo {author} {\bibfnamefont {J.~L.}\ \bibnamefont {Dodd}}, \bibinfo {author} {\bibfnamefont {A.}~\bibnamefont {Gilchrist}}, \bibinfo {author} {\bibfnamefont {D.}~\bibnamefont {Mortimer}}, \bibinfo {author} {\bibfnamefont {T.~J.}\ \bibnamefont {Osborne}}, \bibinfo {author} {\bibfnamefont {M.~J.}\ \bibnamefont {Bremner}}, \bibinfo {author} {\bibfnamefont {A.~W.}\ \bibnamefont {Harrow}},\ and\ \bibinfo {author} {\bibfnamefont {A.}~\bibnamefont {Hines}},\ }\bibfield  {title} {\bibinfo {title} {Quantum dynamics as a physical resource},\ }\href {https://doi.org/10.1103/PhysRevA.67.052301} {\bibfield  {journal} {\bibinfo  {journal} {Physical Review A}\ }\textbf {\bibinfo {volume} {67}},\ \bibinfo {pages} {052301} (\bibinfo {year} {2003})}\BibitemShut {NoStop}%
\bibitem [{Note3()}]{Note3}%
  \BibitemOpen
  \bibinfo {note} {One might consider ``indirectly'' measuring the adjacent qubits in the $Z$ basis by attaching additional branches on the adjacent qubits and measuring the branches. However, it does not work because, if we measure these additional branches in the $Z$ basis to detach them from the main chain, these measurement outcomes can flip the success and failure events of the original CZ gate generation.}\BibitemShut {Stop}%
\bibitem [{Note4()}]{Note4}%
  \BibitemOpen
  \bibinfo {note} {By applying $Z$-basis measurements appropriately on the WGS on a square lattice, we can generate a WGS on a decorated square lattice in which each edge is replaced with a 1D chain of three qubits. Then, the WGS can be transformed to a 2D cluster state with the procedure in Fig.~\ref {fig:CZ_generation}.}\BibitemShut {Stop}%
\bibitem [{\citenamefont {Aaronson}(2005)}]{Aaronson_Proc.R.Soc.Math.Phys.Eng.Sci._2005}%
  \BibitemOpen
  \bibfield  {author} {\bibinfo {author} {\bibfnamefont {S.}~\bibnamefont {Aaronson}},\ }\bibfield  {title} {\bibinfo {title} {Quantum computing, postselection, and probabilistic polynomial-time},\ }\href {https://doi.org/10.1098/rspa.2005.1546} {\bibfield  {journal} {\bibinfo  {journal} {Proceedings of the Royal Society A: Mathematical, Physical and Engineering Sciences}\ }\textbf {\bibinfo {volume} {461}},\ \bibinfo {pages} {3473} (\bibinfo {year} {2005})}\BibitemShut {NoStop}%
\bibitem [{\citenamefont {Bremner}\ \emph {et~al.}(2010)\citenamefont {Bremner}, \citenamefont {Jozsa},\ and\ \citenamefont {Shepherd}}]{Bremner_Proc.R.Soc.Math.Phys.Eng.Sci._2010}%
  \BibitemOpen
  \bibfield  {author} {\bibinfo {author} {\bibfnamefont {M.~J.}\ \bibnamefont {Bremner}}, \bibinfo {author} {\bibfnamefont {R.}~\bibnamefont {Jozsa}},\ and\ \bibinfo {author} {\bibfnamefont {D.~J.}\ \bibnamefont {Shepherd}},\ }\bibfield  {title} {\bibinfo {title} {Classical simulation of commuting quantum computations implies collapse of the polynomial hierarchy},\ }\href {https://doi.org/10.1098/rspa.2010.0301} {\bibfield  {journal} {\bibinfo  {journal} {Proceedings of the Royal Society A: Mathematical, Physical and Engineering Sciences}\ }\textbf {\bibinfo {volume} {467}},\ \bibinfo {pages} {459} (\bibinfo {year} {2010})}\BibitemShut {NoStop}%
\bibitem [{\citenamefont {Makhlin}(2002)}]{Makhlin_QuantumInformationProcessing_2002}%
  \BibitemOpen
  \bibfield  {author} {\bibinfo {author} {\bibfnamefont {Y.}~\bibnamefont {Makhlin}},\ }\bibfield  {title} {\bibinfo {title} {Nonlocal {{Properties}} of {{Two-Qubit Gates}} and {{Mixed States}}, and the {{Optimization}} of {{Quantum Computations}}},\ }\href {https://doi.org/10.1023/A:1022144002391} {\bibfield  {journal} {\bibinfo  {journal} {Quantum Information Processing}\ }\textbf {\bibinfo {volume} {1}},\ \bibinfo {pages} {243} (\bibinfo {year} {2002})}\BibitemShut {NoStop}%
\bibitem [{\citenamefont {Zanardi}\ \emph {et~al.}(2000)\citenamefont {Zanardi}, \citenamefont {Zalka},\ and\ \citenamefont {Faoro}}]{Zanardi_Phys.Rev.A_2000}%
  \BibitemOpen
  \bibfield  {author} {\bibinfo {author} {\bibfnamefont {P.}~\bibnamefont {Zanardi}}, \bibinfo {author} {\bibfnamefont {C.}~\bibnamefont {Zalka}},\ and\ \bibinfo {author} {\bibfnamefont {L.}~\bibnamefont {Faoro}},\ }\bibfield  {title} {\bibinfo {title} {Entangling power of quantum evolutions},\ }\href {https://doi.org/10.1103/PhysRevA.62.030301} {\bibfield  {journal} {\bibinfo  {journal} {Physical Review A}\ }\textbf {\bibinfo {volume} {62}},\ \bibinfo {pages} {030301} (\bibinfo {year} {2000})}\BibitemShut {NoStop}%
\bibitem [{\citenamefont {Zanardi}(2001)}]{Zanardi_Phys.Rev.A_2001}%
  \BibitemOpen
  \bibfield  {author} {\bibinfo {author} {\bibfnamefont {P.}~\bibnamefont {Zanardi}},\ }\bibfield  {title} {\bibinfo {title} {Entanglement of quantum evolutions},\ }\href {https://doi.org/10.1103/PhysRevA.63.040304} {\bibfield  {journal} {\bibinfo  {journal} {Physical Review A}\ }\textbf {\bibinfo {volume} {63}},\ \bibinfo {pages} {040304} (\bibinfo {year} {2001})}\BibitemShut {NoStop}%
\bibitem [{\citenamefont {Cirac}\ \emph {et~al.}(2001)\citenamefont {Cirac}, \citenamefont {D{\"u}r}, \citenamefont {Kraus},\ and\ \citenamefont {Lewenstein}}]{Cirac_Phys.Rev.Lett._2001}%
  \BibitemOpen
  \bibfield  {author} {\bibinfo {author} {\bibfnamefont {J.~I.}\ \bibnamefont {Cirac}}, \bibinfo {author} {\bibfnamefont {W.}~\bibnamefont {D{\"u}r}}, \bibinfo {author} {\bibfnamefont {B.}~\bibnamefont {Kraus}},\ and\ \bibinfo {author} {\bibfnamefont {M.}~\bibnamefont {Lewenstein}},\ }\bibfield  {title} {\bibinfo {title} {Entangling {{Operations}} and {{Their Implementation Using}} a {{Small Amount}} of {{Entanglement}}},\ }\href {https://doi.org/10.1103/PhysRevLett.86.544} {\bibfield  {journal} {\bibinfo  {journal} {Physical Review Letters}\ }\textbf {\bibinfo {volume} {86}},\ \bibinfo {pages} {544} (\bibinfo {year} {2001})}\BibitemShut {NoStop}%
\bibitem [{\citenamefont {Kraus}\ and\ \citenamefont {Cirac}(2001)}]{Kraus_Phys.Rev.A_2001}%
  \BibitemOpen
  \bibfield  {author} {\bibinfo {author} {\bibfnamefont {B.}~\bibnamefont {Kraus}}\ and\ \bibinfo {author} {\bibfnamefont {J.~I.}\ \bibnamefont {Cirac}},\ }\bibfield  {title} {\bibinfo {title} {Optimal creation of entanglement using a two-qubit gate},\ }\href {https://doi.org/10.1103/PhysRevA.63.062309} {\bibfield  {journal} {\bibinfo  {journal} {Physical Review A}\ }\textbf {\bibinfo {volume} {63}},\ \bibinfo {pages} {062309} (\bibinfo {year} {2001})}\BibitemShut {NoStop}%
\bibitem [{\citenamefont {D{\"u}r}\ \emph {et~al.}(2002)\citenamefont {D{\"u}r}, \citenamefont {Vidal},\ and\ \citenamefont {Cirac}}]{Dur_Phys.Rev.Lett._2002}%
  \BibitemOpen
  \bibfield  {author} {\bibinfo {author} {\bibfnamefont {W.}~\bibnamefont {D{\"u}r}}, \bibinfo {author} {\bibfnamefont {G.}~\bibnamefont {Vidal}},\ and\ \bibinfo {author} {\bibfnamefont {J.~I.}\ \bibnamefont {Cirac}},\ }\bibfield  {title} {\bibinfo {title} {Optimal {{Conversion}} of {{Nonlocal Unitary Operations}}},\ }\href {https://doi.org/10.1103/PhysRevLett.89.057901} {\bibfield  {journal} {\bibinfo  {journal} {Physical Review Letters}\ }\textbf {\bibinfo {volume} {89}},\ \bibinfo {pages} {057901} (\bibinfo {year} {2002})}\BibitemShut {NoStop}%
\bibitem [{\citenamefont {Turchette}\ \emph {et~al.}(1995)\citenamefont {Turchette}, \citenamefont {Hood}, \citenamefont {Lange}, \citenamefont {Mabuchi},\ and\ \citenamefont {Kimble}}]{Turchette_Phys.Rev.Lett._1995}%
  \BibitemOpen
  \bibfield  {author} {\bibinfo {author} {\bibfnamefont {Q.~A.}\ \bibnamefont {Turchette}}, \bibinfo {author} {\bibfnamefont {C.~J.}\ \bibnamefont {Hood}}, \bibinfo {author} {\bibfnamefont {W.}~\bibnamefont {Lange}}, \bibinfo {author} {\bibfnamefont {H.}~\bibnamefont {Mabuchi}},\ and\ \bibinfo {author} {\bibfnamefont {H.~J.}\ \bibnamefont {Kimble}},\ }\bibfield  {title} {\bibinfo {title} {Measurement of {{Conditional Phase Shifts}} for {{Quantum Logic}}},\ }\href {https://doi.org/10.1103/PhysRevLett.75.4710} {\bibfield  {journal} {\bibinfo  {journal} {Physical Review Letters}\ }\textbf {\bibinfo {volume} {75}},\ \bibinfo {pages} {4710} (\bibinfo {year} {1995})}\BibitemShut {NoStop}%
\bibitem [{\citenamefont {Duan}\ and\ \citenamefont {Kimble}(2004)}]{Duan_Phys.Rev.Lett._2004}%
  \BibitemOpen
  \bibfield  {author} {\bibinfo {author} {\bibfnamefont {L.-M.}\ \bibnamefont {Duan}}\ and\ \bibinfo {author} {\bibfnamefont {H.~J.}\ \bibnamefont {Kimble}},\ }\bibfield  {title} {\bibinfo {title} {Scalable {{Photonic Quantum Computation}} through {{Cavity-Assisted Interactions}}},\ }\href {https://doi.org/10.1103/PhysRevLett.92.127902} {\bibfield  {journal} {\bibinfo  {journal} {Physical Review Letters}\ }\textbf {\bibinfo {volume} {92}},\ \bibinfo {pages} {127902} (\bibinfo {year} {2004})}\BibitemShut {NoStop}%
\bibitem [{\citenamefont {Volz}\ \emph {et~al.}(2014)\citenamefont {Volz}, \citenamefont {Scheucher}, \citenamefont {Junge},\ and\ \citenamefont {Rauschenbeutel}}]{Volz_NaturePhoton_2014}%
  \BibitemOpen
  \bibfield  {author} {\bibinfo {author} {\bibfnamefont {J.}~\bibnamefont {Volz}}, \bibinfo {author} {\bibfnamefont {M.}~\bibnamefont {Scheucher}}, \bibinfo {author} {\bibfnamefont {C.}~\bibnamefont {Junge}},\ and\ \bibinfo {author} {\bibfnamefont {A.}~\bibnamefont {Rauschenbeutel}},\ }\bibfield  {title} {\bibinfo {title} {Nonlinear {$\pi$} phase shift for single fibre-guided photons interacting with a single resonator-enhanced atom},\ }\href {https://doi.org/10.1038/nphoton.2014.253} {\bibfield  {journal} {\bibinfo  {journal} {Nature Photonics}\ }\textbf {\bibinfo {volume} {8}},\ \bibinfo {pages} {965} (\bibinfo {year} {2014})}\BibitemShut {NoStop}%
\bibitem [{\citenamefont {Hacker}\ \emph {et~al.}(2016)\citenamefont {Hacker}, \citenamefont {Welte}, \citenamefont {Rempe},\ and\ \citenamefont {Ritter}}]{Hacker_Nature_2016}%
  \BibitemOpen
  \bibfield  {author} {\bibinfo {author} {\bibfnamefont {B.}~\bibnamefont {Hacker}}, \bibinfo {author} {\bibfnamefont {S.}~\bibnamefont {Welte}}, \bibinfo {author} {\bibfnamefont {G.}~\bibnamefont {Rempe}},\ and\ \bibinfo {author} {\bibfnamefont {S.}~\bibnamefont {Ritter}},\ }\bibfield  {title} {\bibinfo {title} {A photon--photon quantum gate based on a single atom in an optical resonator},\ }\href {https://doi.org/10.1038/nature18592} {\bibfield  {journal} {\bibinfo  {journal} {Nature}\ }\textbf {\bibinfo {volume} {536}},\ \bibinfo {pages} {193} (\bibinfo {year} {2016})}\BibitemShut {NoStop}%
\bibitem [{\citenamefont {Beck}\ \emph {et~al.}(2016)\citenamefont {Beck}, \citenamefont {Hosseini}, \citenamefont {Duan},\ and\ \citenamefont {Vuleti{\'c}}}]{Beck_Proc.Natl.Acad.Sci._2016}%
  \BibitemOpen
  \bibfield  {author} {\bibinfo {author} {\bibfnamefont {K.~M.}\ \bibnamefont {Beck}}, \bibinfo {author} {\bibfnamefont {M.}~\bibnamefont {Hosseini}}, \bibinfo {author} {\bibfnamefont {Y.}~\bibnamefont {Duan}},\ and\ \bibinfo {author} {\bibfnamefont {V.}~\bibnamefont {Vuleti{\'c}}},\ }\bibfield  {title} {\bibinfo {title} {Large conditional single-photon cross-phase modulation},\ }\href {https://doi.org/10.1073/pnas.1524117113} {\bibfield  {journal} {\bibinfo  {journal} {Proceedings of the National Academy of Sciences}\ }\textbf {\bibinfo {volume} {113}},\ \bibinfo {pages} {9740} (\bibinfo {year} {2016})}\BibitemShut {NoStop}%
\bibitem [{\citenamefont {{Sagona-Stophel}}\ \emph {et~al.}(2020)\citenamefont {{Sagona-Stophel}}, \citenamefont {Shahrokhshahi}, \citenamefont {Jordaan}, \citenamefont {Namazi},\ and\ \citenamefont {Figueroa}}]{Sagona-Stophel_Phys.Rev.Lett._2020}%
  \BibitemOpen
  \bibfield  {author} {\bibinfo {author} {\bibfnamefont {S.}~\bibnamefont {{Sagona-Stophel}}}, \bibinfo {author} {\bibfnamefont {R.}~\bibnamefont {Shahrokhshahi}}, \bibinfo {author} {\bibfnamefont {B.}~\bibnamefont {Jordaan}}, \bibinfo {author} {\bibfnamefont {M.}~\bibnamefont {Namazi}},\ and\ \bibinfo {author} {\bibfnamefont {E.}~\bibnamefont {Figueroa}},\ }\bibfield  {title} {\bibinfo {title} {Conditional {$\pi$} -{{Phase Shift}} of {{Single-Photon-Level Pulses}} at {{Room Temperature}}},\ }\href {https://doi.org/10.1103/PhysRevLett.125.243601} {\bibfield  {journal} {\bibinfo  {journal} {Physical Review Letters}\ }\textbf {\bibinfo {volume} {125}},\ \bibinfo {pages} {243601} (\bibinfo {year} {2020})}\BibitemShut {NoStop}%
\bibitem [{\citenamefont {Fushman}\ \emph {et~al.}(2008)\citenamefont {Fushman}, \citenamefont {Englund}, \citenamefont {Faraon}, \citenamefont {Stoltz}, \citenamefont {Petroff},\ and\ \citenamefont {Vu{\v c}kovi{\'c}}}]{Fushman_Science_2008}%
  \BibitemOpen
  \bibfield  {author} {\bibinfo {author} {\bibfnamefont {I.}~\bibnamefont {Fushman}}, \bibinfo {author} {\bibfnamefont {D.}~\bibnamefont {Englund}}, \bibinfo {author} {\bibfnamefont {A.}~\bibnamefont {Faraon}}, \bibinfo {author} {\bibfnamefont {N.}~\bibnamefont {Stoltz}}, \bibinfo {author} {\bibfnamefont {P.}~\bibnamefont {Petroff}},\ and\ \bibinfo {author} {\bibfnamefont {J.}~\bibnamefont {Vu{\v c}kovi{\'c}}},\ }\bibfield  {title} {\bibinfo {title} {Controlled {{Phase Shifts}} with a {{Single Quantum Dot}}},\ }\href {https://doi.org/10.1126/science.1154643} {\bibfield  {journal} {\bibinfo  {journal} {Science}\ }\textbf {\bibinfo {volume} {320}},\ \bibinfo {pages} {769} (\bibinfo {year} {2008})}\BibitemShut {NoStop}%
\bibitem [{\citenamefont {Kim}\ \emph {et~al.}(2013)\citenamefont {Kim}, \citenamefont {Bose}, \citenamefont {Shen}, \citenamefont {Solomon},\ and\ \citenamefont {Waks}}]{Kim_NaturePhoton_2013}%
  \BibitemOpen
  \bibfield  {author} {\bibinfo {author} {\bibfnamefont {H.}~\bibnamefont {Kim}}, \bibinfo {author} {\bibfnamefont {R.}~\bibnamefont {Bose}}, \bibinfo {author} {\bibfnamefont {T.~C.}\ \bibnamefont {Shen}}, \bibinfo {author} {\bibfnamefont {G.~S.}\ \bibnamefont {Solomon}},\ and\ \bibinfo {author} {\bibfnamefont {E.}~\bibnamefont {Waks}},\ }\bibfield  {title} {\bibinfo {title} {A quantum logic gate between a solid-state quantum bit and a photon},\ }\href {https://doi.org/10.1038/nphoton.2013.48} {\bibfield  {journal} {\bibinfo  {journal} {Nature Photonics}\ }\textbf {\bibinfo {volume} {7}},\ \bibinfo {pages} {373} (\bibinfo {year} {2013})}\BibitemShut {NoStop}%
\bibitem [{\citenamefont {Sun}\ \emph {et~al.}(2018)\citenamefont {Sun}, \citenamefont {Kim}, \citenamefont {Luo}, \citenamefont {Solomon},\ and\ \citenamefont {Waks}}]{Sun_Science_2018}%
  \BibitemOpen
  \bibfield  {author} {\bibinfo {author} {\bibfnamefont {S.}~\bibnamefont {Sun}}, \bibinfo {author} {\bibfnamefont {H.}~\bibnamefont {Kim}}, \bibinfo {author} {\bibfnamefont {Z.}~\bibnamefont {Luo}}, \bibinfo {author} {\bibfnamefont {G.~S.}\ \bibnamefont {Solomon}},\ and\ \bibinfo {author} {\bibfnamefont {E.}~\bibnamefont {Waks}},\ }\bibfield  {title} {\bibinfo {title} {A single-photon switch and transistor enabled by a solid-state quantum memory},\ }\href {https://doi.org/10.1126/science.aat3581} {\bibfield  {journal} {\bibinfo  {journal} {Science}\ }\textbf {\bibinfo {volume} {361}},\ \bibinfo {pages} {57} (\bibinfo {year} {2018})}\BibitemShut {NoStop}%
\bibitem [{\citenamefont {Wang}\ \emph {et~al.}(2013)\citenamefont {Wang}, \citenamefont {Zhang}, \citenamefont {Jiao},\ and\ \citenamefont {Jin}}]{Wang_Opt.ExpressOE_2013}%
  \BibitemOpen
  \bibfield  {author} {\bibinfo {author} {\bibfnamefont {C.}~\bibnamefont {Wang}}, \bibinfo {author} {\bibfnamefont {Y.}~\bibnamefont {Zhang}}, \bibinfo {author} {\bibfnamefont {R.-z.}\ \bibnamefont {Jiao}},\ and\ \bibinfo {author} {\bibfnamefont {G.-s.}\ \bibnamefont {Jin}},\ }\bibfield  {title} {\bibinfo {title} {Universal quantum controlled phase gate on photonic qubits based on nitrogen vacancy centers and microcavity resonators},\ }\href {https://doi.org/10.1364/OE.21.019252} {\bibfield  {journal} {\bibinfo  {journal} {Optics Express}\ }\textbf {\bibinfo {volume} {21}},\ \bibinfo {pages} {19252} (\bibinfo {year} {2013})}\BibitemShut {NoStop}%
\bibitem [{\citenamefont {Gorshkov}\ \emph {et~al.}(2011)\citenamefont {Gorshkov}, \citenamefont {Otterbach}, \citenamefont {Fleischhauer}, \citenamefont {Pohl},\ and\ \citenamefont {Lukin}}]{Gorshkov_Phys.Rev.Lett._2011}%
  \BibitemOpen
  \bibfield  {author} {\bibinfo {author} {\bibfnamefont {A.~V.}\ \bibnamefont {Gorshkov}}, \bibinfo {author} {\bibfnamefont {J.}~\bibnamefont {Otterbach}}, \bibinfo {author} {\bibfnamefont {M.}~\bibnamefont {Fleischhauer}}, \bibinfo {author} {\bibfnamefont {T.}~\bibnamefont {Pohl}},\ and\ \bibinfo {author} {\bibfnamefont {M.~D.}\ \bibnamefont {Lukin}},\ }\bibfield  {title} {\bibinfo {title} {Photon-{{Photon Interactions}} via {{Rydberg Blockade}}},\ }\href {https://doi.org/10.1103/PhysRevLett.107.133602} {\bibfield  {journal} {\bibinfo  {journal} {Physical Review Letters}\ }\textbf {\bibinfo {volume} {107}},\ \bibinfo {pages} {133602} (\bibinfo {year} {2011})}\BibitemShut {NoStop}%
\bibitem [{\citenamefont {Firstenberg}\ \emph {et~al.}(2013)\citenamefont {Firstenberg}, \citenamefont {Peyronel}, \citenamefont {Liang}, \citenamefont {Gorshkov}, \citenamefont {Lukin},\ and\ \citenamefont {Vuleti{\'c}}}]{Firstenberg_Nature_2013}%
  \BibitemOpen
  \bibfield  {author} {\bibinfo {author} {\bibfnamefont {O.}~\bibnamefont {Firstenberg}}, \bibinfo {author} {\bibfnamefont {T.}~\bibnamefont {Peyronel}}, \bibinfo {author} {\bibfnamefont {Q.-Y.}\ \bibnamefont {Liang}}, \bibinfo {author} {\bibfnamefont {A.~V.}\ \bibnamefont {Gorshkov}}, \bibinfo {author} {\bibfnamefont {M.~D.}\ \bibnamefont {Lukin}},\ and\ \bibinfo {author} {\bibfnamefont {V.}~\bibnamefont {Vuleti{\'c}}},\ }\bibfield  {title} {\bibinfo {title} {Attractive photons in a quantum nonlinear medium},\ }\href {https://doi.org/10.1038/nature12512} {\bibfield  {journal} {\bibinfo  {journal} {Nature}\ }\textbf {\bibinfo {volume} {502}},\ \bibinfo {pages} {71} (\bibinfo {year} {2013})}\BibitemShut {NoStop}%
\bibitem [{\citenamefont {Tiarks}\ \emph {et~al.}(2016)\citenamefont {Tiarks}, \citenamefont {Schmidt}, \citenamefont {Rempe},\ and\ \citenamefont {D{\"u}rr}}]{Tiarks_Sci.Adv._2016}%
  \BibitemOpen
  \bibfield  {author} {\bibinfo {author} {\bibfnamefont {D.}~\bibnamefont {Tiarks}}, \bibinfo {author} {\bibfnamefont {S.}~\bibnamefont {Schmidt}}, \bibinfo {author} {\bibfnamefont {G.}~\bibnamefont {Rempe}},\ and\ \bibinfo {author} {\bibfnamefont {S.}~\bibnamefont {D{\"u}rr}},\ }\bibfield  {title} {\bibinfo {title} {Optical {$\pi$} phase shift created with a single-photon pulse},\ }\href {https://doi.org/10.1126/sciadv.1600036} {\bibfield  {journal} {\bibinfo  {journal} {Science Advances}\ }\textbf {\bibinfo {volume} {2}},\ \bibinfo {pages} {e1600036} (\bibinfo {year} {2016})}\BibitemShut {NoStop}%
\bibitem [{\citenamefont {Thompson}\ \emph {et~al.}(2017)\citenamefont {Thompson}, \citenamefont {Nicholson}, \citenamefont {Liang}, \citenamefont {Cantu}, \citenamefont {Venkatramani}, \citenamefont {Choi}, \citenamefont {Fedorov}, \citenamefont {Viscor}, \citenamefont {Pohl}, \citenamefont {Lukin},\ and\ \citenamefont {Vuleti{\'c}}}]{Thompson_Nature_2017}%
  \BibitemOpen
  \bibfield  {author} {\bibinfo {author} {\bibfnamefont {J.~D.}\ \bibnamefont {Thompson}}, \bibinfo {author} {\bibfnamefont {T.~L.}\ \bibnamefont {Nicholson}}, \bibinfo {author} {\bibfnamefont {Q.-Y.}\ \bibnamefont {Liang}}, \bibinfo {author} {\bibfnamefont {S.~H.}\ \bibnamefont {Cantu}}, \bibinfo {author} {\bibfnamefont {A.~V.}\ \bibnamefont {Venkatramani}}, \bibinfo {author} {\bibfnamefont {S.}~\bibnamefont {Choi}}, \bibinfo {author} {\bibfnamefont {I.~A.}\ \bibnamefont {Fedorov}}, \bibinfo {author} {\bibfnamefont {D.}~\bibnamefont {Viscor}}, \bibinfo {author} {\bibfnamefont {T.}~\bibnamefont {Pohl}}, \bibinfo {author} {\bibfnamefont {M.~D.}\ \bibnamefont {Lukin}},\ and\ \bibinfo {author} {\bibfnamefont {V.}~\bibnamefont {Vuleti{\'c}}},\ }\bibfield  {title} {\bibinfo {title} {Symmetry-protected collisions between strongly interacting photons},\ }\href {https://doi.org/10.1038/nature20823} {\bibfield  {journal} {\bibinfo  {journal} {Nature}\ }\textbf {\bibinfo {volume} {542}},\ \bibinfo {pages} {206} (\bibinfo {year} {2017})}\BibitemShut {NoStop}%
\bibitem [{\citenamefont {Tiarks}\ \emph {et~al.}(2019)\citenamefont {Tiarks}, \citenamefont {{Schmidt-Eberle}}, \citenamefont {Stolz}, \citenamefont {Rempe},\ and\ \citenamefont {D{\"u}rr}}]{Tiarks_NaturePhys_2019}%
  \BibitemOpen
  \bibfield  {author} {\bibinfo {author} {\bibfnamefont {D.}~\bibnamefont {Tiarks}}, \bibinfo {author} {\bibfnamefont {S.}~\bibnamefont {{Schmidt-Eberle}}}, \bibinfo {author} {\bibfnamefont {T.}~\bibnamefont {Stolz}}, \bibinfo {author} {\bibfnamefont {G.}~\bibnamefont {Rempe}},\ and\ \bibinfo {author} {\bibfnamefont {S.}~\bibnamefont {D{\"u}rr}},\ }\bibfield  {title} {\bibinfo {title} {A photon--photon quantum gate based on {{Rydberg}} interactions},\ }\href {https://doi.org/10.1038/s41567-018-0313-7} {\bibfield  {journal} {\bibinfo  {journal} {Nature Physics}\ }\textbf {\bibinfo {volume} {15}},\ \bibinfo {pages} {124} (\bibinfo {year} {2019})}\BibitemShut {NoStop}%
\bibitem [{\citenamefont {Aharonov}(2003)}]{Aharonov__2003}%
  \BibitemOpen
  \bibfield  {author} {\bibinfo {author} {\bibfnamefont {D.}~\bibnamefont {Aharonov}},\ }\href {https://doi.org/10.48550/arXiv.quant-ph/0301040} {\bibinfo {title} {A {{Simple Proof}} that {{Toffoli}} and {{Hadamard}} are {{Quantum Universal}}}} (\bibinfo {year} {2003}),\ \Eprint {https://arxiv.org/abs/quant-ph/0301040} {arXiv:quant-ph/0301040} \BibitemShut {NoStop}%
\bibitem [{\citenamefont {Takeuchi}(2024)}]{Takeuchi_Phys.Rev.Lett._2024}%
  \BibitemOpen
  \bibfield  {author} {\bibinfo {author} {\bibfnamefont {Y.}~\bibnamefont {Takeuchi}},\ }\bibfield  {title} {\bibinfo {title} {Catalytic {{Transformation}} from {{Computationally Universal}} to {{Strictly Universal Measurement-Based Quantum Computation}}},\ }\href {https://doi.org/10.1103/PhysRevLett.133.050601} {\bibfield  {journal} {\bibinfo  {journal} {Physical Review Letters}\ }\textbf {\bibinfo {volume} {133}},\ \bibinfo {pages} {050601} (\bibinfo {year} {2024})}\BibitemShut {NoStop}%
\end{thebibliography}%


\begin{thebibliography}{3}%
\makeatletter
\providecommand \@ifxundefined [1]{%
 \@ifx{#1\undefined}
}%
\providecommand \@ifnum [1]{%
 \ifnum #1\expandafter \@firstoftwo
 \else \expandafter \@secondoftwo
 \fi
}%
\providecommand \@ifx [1]{%
 \ifx #1\expandafter \@firstoftwo
 \else \expandafter \@secondoftwo
 \fi
}%
\providecommand \natexlab [1]{#1}%
\providecommand \enquote  [1]{``#1''}%
\providecommand \bibnamefont  [1]{#1}%
\providecommand \bibfnamefont [1]{#1}%
\providecommand \citenamefont [1]{#1}%
\providecommand \href@noop [0]{\@secondoftwo}%
\providecommand \href [0]{\begingroup \@sanitize@url \@href}%
\providecommand \@href[1]{\@@startlink{#1}\@@href}%
\providecommand \@@href[1]{\endgroup#1\@@endlink}%
\providecommand \@sanitize@url [0]{\catcode `\\12\catcode `\$12\catcode `\&12\catcode `\#12\catcode `\^12\catcode `\_12\catcode `\%12\relax}%
\providecommand \@@startlink[1]{}%
\providecommand \@@endlink[0]{}%
\providecommand \url  [0]{\begingroup\@sanitize@url \@url }%
\providecommand \@url [1]{\endgroup\@href {#1}{\urlprefix }}%
\providecommand \urlprefix  [0]{URL }%
\providecommand \Eprint [0]{\href }%
\providecommand \doibase [0]{https://doi.org/}%
\providecommand \selectlanguage [0]{\@gobble}%
\providecommand \bibinfo  [0]{\@secondoftwo}%
\providecommand \bibfield  [0]{\@secondoftwo}%
\providecommand \translation [1]{[#1]}%
\providecommand \BibitemOpen [0]{}%
\providecommand \bibitemStop [0]{}%
\providecommand \bibitemNoStop [0]{.\EOS\space}%
\providecommand \EOS [0]{\spacefactor3000\relax}%
\providecommand \BibitemShut  [1]{\csname bibitem#1\endcsname}%
\let\auto@bib@innerbib\@empty
\bibitem [{\citenamefont {Nielsen}(2000)}]{Nielsen__2000}%
  \BibitemOpen
  \bibfield  {author} {\bibinfo {author} {\bibfnamefont {M.~A.}\ \bibnamefont {Nielsen}},\ }\href {https://doi.org/10.48550/arXiv.quant-ph/0011036} {\bibinfo {title} {Quantum information theory}} (\bibinfo {year} {2000}),\ \Eprint {https://arxiv.org/abs/quant-ph/0011036} {arXiv:quant-ph/0011036} \BibitemShut {NoStop}%
\bibitem [{\citenamefont {Nielsen}\ \emph {et~al.}(2003)\citenamefont {Nielsen}, \citenamefont {Dawson}, \citenamefont {Dodd}, \citenamefont {Gilchrist}, \citenamefont {Mortimer}, \citenamefont {Osborne}, \citenamefont {Bremner}, \citenamefont {Harrow},\ and\ \citenamefont {Hines}}]{Nielsen_Phys.Rev.A_2003}%
  \BibitemOpen
  \bibfield  {author} {\bibinfo {author} {\bibfnamefont {M.~A.}\ \bibnamefont {Nielsen}}, \bibinfo {author} {\bibfnamefont {C.~M.}\ \bibnamefont {Dawson}}, \bibinfo {author} {\bibfnamefont {J.~L.}\ \bibnamefont {Dodd}}, \bibinfo {author} {\bibfnamefont {A.}~\bibnamefont {Gilchrist}}, \bibinfo {author} {\bibfnamefont {D.}~\bibnamefont {Mortimer}}, \bibinfo {author} {\bibfnamefont {T.~J.}\ \bibnamefont {Osborne}}, \bibinfo {author} {\bibfnamefont {M.~J.}\ \bibnamefont {Bremner}}, \bibinfo {author} {\bibfnamefont {A.~W.}\ \bibnamefont {Harrow}},\ and\ \bibinfo {author} {\bibfnamefont {A.}~\bibnamefont {Hines}},\ }\bibfield  {title} {\bibinfo {title} {Quantum dynamics as a physical resource},\ }\href {https://doi.org/10.1103/PhysRevA.67.052301} {\bibfield  {journal} {\bibinfo  {journal} {Physical Review A}\ }\textbf {\bibinfo {volume} {67}},\ \bibinfo {pages} {052301} (\bibinfo {year} {2003})}\BibitemShut {NoStop}%
\bibitem [{\citenamefont {D{\"u}r}\ \emph {et~al.}(2002)\citenamefont {D{\"u}r}, \citenamefont {Vidal},\ and\ \citenamefont {Cirac}}]{Dur_Phys.Rev.Lett._2002}%
  \BibitemOpen
  \bibfield  {author} {\bibinfo {author} {\bibfnamefont {W.}~\bibnamefont {D{\"u}r}}, \bibinfo {author} {\bibfnamefont {G.}~\bibnamefont {Vidal}},\ and\ \bibinfo {author} {\bibfnamefont {J.~I.}\ \bibnamefont {Cirac}},\ }\bibfield  {title} {\bibinfo {title} {Optimal {{Conversion}} of {{Nonlocal Unitary Operations}}},\ }\href {https://doi.org/10.1103/PhysRevLett.89.057901} {\bibfield  {journal} {\bibinfo  {journal} {Physical Review Letters}\ }\textbf {\bibinfo {volume} {89}},\ \bibinfo {pages} {057901} (\bibinfo {year} {2002})}\BibitemShut {NoStop}%
\end{thebibliography}%
\end{document}


\section{Supplemental Material}

\subsection{Measurement-based composition (MBC) of operators and weighted Pauli measurements}
Here, we introduce the definitions of MBC and weighted Pauli measurements with their motivations and derive their properties.

\subsubsection{Definition and properties of MBC}
Let us first define the MBC of operators in a general way as follows:
\begin{definition}[measurement-based composition]\label{def:mbo}
    Let $O_i$ be an operator acting on two systems $A_i$ and $B$ for $i=1,2,\dots,n$. Then, the MBC operator of $O_1, O_2, \dots, O_n$ from $\ket{\psi_\text{in}}$ to $\ket{\psi_\text{out}}$ is defined as the operator $\bra{\psi_\text{out}}_B (O_1)_{A_1B} (O_2)_{A_2B} \cdots (O_n)_{A_nB} \ket{\psi_\text{in}}_B$ acting on $n$ systems $A_1$, $A_2$, \dots, and $A_n$. In particular, we denote $\bra{\psi}_B (O_1)_{AB} (O_2)_{BC} \ket{+}_B$ as $(O_1 \circ_\psi O_2)_{AC}$.
\end{definition}
We can consider a sequence of MBCs by taking the resulting operator of an MBC as an input operator of another MBC.
If we decompose a resource state into a separable state and multi-qubit operations applied to the state, measurement-based quantum computation itself is essentially a sequence of MBCs of the multi-qubit operations from the initial separable state to the measurement-basis states of single-qubit measurements.
In this study, we focus on the weighted graph state (WGS) as a resource state.
Since the WGS is generated by applying controlled-phase (CP) gates to the product state $\ket{+}^{\otimes n}$, it is sufficient to consider a sequence of MBCs of CP gates from $\ket{+}$.
Then, the problem of showing the universality of the WGS reduces to generating a CZ gate via the sequence of MBCs because any graph state including the 2D cluster state can be generated by applying CZ gates to the state $\ket{+}^{\otimes n}$.
In fact, it is sufficient to consider MBCs of two input operations, where both input and output operators of MBCs are always two-qubit operators, for generating a CZ gate. A sequence of MBCs of two two-qubit operators satisfies the associative law: $(O_1 \circ_{\psi_1} O_2) \circ_{\psi_2} O_3= O_1 \circ_{\psi_1} (O_2 \circ_{\psi_2} O_3)$ because the single-qubit measurements commute with each other as long as the measurement bases are independently determined. Therefore, we use the notation such as $O_1 \circ_{\psi_1} O_2 \circ_{\psi_2} O_3$.

Any linear operator $Q$ acting on systems $A$ and $B$ can be represented in the following operator-Schmidt decomposition~\cite{Nielsen__2000,Nielsen_Phys.Rev.A_2003}:
\begin{equation}\label{eq:operator-Schmidt}
    Q=\sum_i c_i Q^A_i \otimes Q^B_i,
\end{equation}
where $c_i\geq 0$ and $\{Q^A_i\}$ and $\{Q^B_i\}$ are operator bases for $A$ and $B$. According to the number of terms in Eq.~\eqref{eq:operator-Schmidt}, called the Schmidt number, two-qubit unitary operators can be classified into three classes whose Schmidt rank is one, two, and four, which respectively include the identity, CNOT, and SWAP operations~\cite{Dur_Phys.Rev.Lett._2002}.
In particular, the CP gate is represented as
\begin{align}
    \mathrm{CP}(\phi) &= \diag(e^{-i\phi/4}, e^{i\phi/4}, e^{i\phi/4}, e^{3i\phi/4})\diag(e^{i\phi/4}, e^{-i\phi/4}, e^{-i\phi/4}, e^{i\phi/4}) \\
    &= e^{i\phi/4}(R_z(\phi/2)\otimes R_z(\phi/2)) (\cos(\phi/4) I \otimes I + i\sin(\phi/4) Z \otimes Z )\label{eq:Cphase-Schmidt},
\end{align}
and its Schmidt rank is two.
Let us consider operators $O$ acting on qubits $A$ and $B$ with the following operator-Schmidt decomposition:
\begin{equation}\label{eq:general_diagonal operators}
    O =  (R_z(c^A) \otimes R_z(c^B)) (\cos(\phi/4)I\otimes I+e^{i\theta} \sin(\phi/4)Z \otimes Z),
\end{equation}
where $-\pi<\phi,\theta\leq \pi$, and $c^A, c^B \in \mathbb{C}$.
It is clear that the CP gate is one of such operators. 

\begin{definition}\label{def:measurement_basis}
    Let $O_1$ and $O_2$ be two-qubit operators acting on qubits $A$ and $C$ and qubits $B$ and $C$, respectively, and decomposed in the form of Eq.~\eqref{eq:general_diagonal operators} as
    \begin{align}
        (O_1)_{AB} &= C_1(R_z(c^A_{1}) \otimes R_z(c^B_{1}))(\cos(\phi_1/4) I \otimes I + e^{i\theta_1} \sin(\phi_1/4)  (Z \otimes Z)), \label{eq:general_O1} \\
        (O_2)_{BC} &= C_2(R_z(c^B_{2}) \otimes R_z(c^C_{2}))(\cos(\phi_2/4) I \otimes I + e^{i\theta_2} \sin(\phi_2/4)  (Z \otimes Z)) \label{eq:general_O2}
    \end{align}
    for nonzero complex numbers $c_{1}^A,c_{1}^B,c_{2}^B,c_{2}^C,C_1,C_2 \in \mathbb{C}\setminus \{0\}$ and real numbers $-\pi < \phi_1, \phi_2, \theta_1, \theta_2 \leq \pi$.
    Then, we define the success state $\ket{\psi_s(\theta)}$ and the failure state $\ket{\psi_f(\theta)}$ on qubit $B$ as
    \begin{equation}\label{eq:def_measurement_basis}
        \ket{\psi_s(\theta)}_B = \mathcal{N}^{-1/2} R^\dag_z(-(c_{1}^B+c_{2}^B+\theta))\ket{-}_B, \quad 
        \ket{\psi_f(\theta)}_B = \mathcal{N}^{-1/2} R_z(c_{1}^B+c_{2}^B+\theta)\ket{+}_B,
    \end{equation}
    for a real number $-\pi < \theta \leq \pi$ and $\mathcal{N}=\cosh(\Im(c_1^B+c_2^B))$.
\end{definition}
\begin{proposition}\label{prop:basis_properties}\ 
    \begin{itemize}
        \item $\{\ket{\psi_s(\theta)}, \ket{\psi_f(\theta)}\}$ is an orthonormal basis.
        \item If $\Im(c_1^B+c_2^B)=0$, the MBC of $O_1$ and $O_2$ to $\ket{\psi_f(\theta)}$ is equivalent to one to $\ket{\psi_s(\theta+\pi)}$.
        \item For the MBC of $O_1$ and $O_2$ to $\ket{\psi_s(\theta)}$, local operations on the qubit $B$ can be absorbed in the definition of $\ket{\psi_s(\theta)}$ as 
        \begin{equation}\label{eq:local_invariance}
            (O_1 \circ_{\psi_s(\theta)} O_2) =  C_1C_2 \mathcal{N}^{-1/2}(R_z(c_1^A) \otimes R_z(c_{2}^C)) (\widetilde{O}_1 \circ_{\tilde{\psi}_s(\theta)}\widetilde{O}_2),
        \end{equation}
        where 
        \begin{equation}
            \widetilde{O}_1 = \cos(\phi_1/4) I \otimes I + e^{i\theta_1} \sin(\phi_1/4)  (Z \otimes Z), \quad \widetilde{O}_2 = \cos(\phi_2/4) I \otimes I + e^{i\theta_2} \sin(\phi_2/4)  (Z \otimes Z)
        \end{equation}
        and $\ket*{\tilde{\psi}_s(\theta)}$ is the success state defined for operators $\widetilde{O}_1$ and $\widetilde{O}_2$.
    \end{itemize}
\end{proposition}
\begin{proof}
    The first statement follows from
    \begin{equation}
    \norm{R^\dag_z(-(c^B_{1}+c^B_{2}+\theta))\ket{-}}^2=\norm{R_z(c^B_{1}+c^B_{2}+\theta)\ket{+}}^2 =\frac{1}{2}(e^{\Im(c^B_{1}+c^B_{2})}+e^{-\Im(c^B_{1}+c^B_{2})})=\cosh(\Im(c_1^B+c_2^B))=\mathcal{N}
    \end{equation}
    and $\bra{\psi_s(\theta)}\ket{\psi_f(\theta)}\propto \bra{-}R_z(-(c_{1}^B+c_{2}^B+\theta))R_z(c_{1}^B+c_{2}^B+\theta)\ket{+} =0$.
    The second statement follows from 
    \begin{equation}
        \ket{\psi_f(\theta)} = e^{i\pi/2}\mathcal{N}^{-1/2} R^\dag_z(-(c_{1}^B+c_{2}^B+\theta+\pi)+2i\Im(c_{1}^B+c_{2}^B))\ket{-} = e^{i\pi/2}\ket{\psi_s(\theta+\pi-2i\Im(c_{1}^B+c_{2}^B))}.
    \end{equation}
    The third statement follows from the definition as 
    \begin{align}
        (O_1 \circ_{\psi_s(\theta)} O_2) &= C_1 C_2 (R_z(c_1^A) \otimes R_z(c_{2}^C)) \mathcal{N}^{-1/2} \bra{-}_B R_z(-(c_{1}^B+c_{2}^B+\theta)) R_z(c_{1}^B+c_{2}^B)\widetilde{O}_1 \widetilde{O}_2 \ket{+}_B \notag \\
        &= C_1C_2 \mathcal{N}^{-1/2}(R_z(c_1^A) \otimes R_z(c_{2}^C)) (\widetilde{O}_1 \circ_{\tilde{\psi}_s(\theta)}\widetilde{O}_2).
    \end{align}
\end{proof}

From Eq.~\eqref{eq:local_invariance}, we can ignore constant factors and local operations of input operators on the calculation of MBC operators to $\ket{\psi_s(\theta)}$ without loss of generality. 
In particular, we define the following operators:
\begin{align}
    \widetilde{\mathrm{CP}}(\phi) &= \cos(\phi/4) I \otimes I + i \sin(\phi/4)  (Z \otimes Z), \\
    \widetilde{\mathrm{EP}}(\phi) &= \cos(\phi/4) I \otimes I + \sin(\phi/4)  (Z \otimes Z)
\end{align}
for $-\pi \leq \phi \leq \pi$. We call $\phi$ the weight of these operators.
Then, the following Proposition~\ref{prop:fundamental_formula} concretely calculates the MBC operator of a pair of $\widetilde{\mathrm{CP}}$ to $\ket{\psi_s(\theta)}$, which motivates us to define weighted Pauli measurements.

\begin{proposition}\label{prop:fundamental_formula}
    For $-\pi \leq \phi_1,  \phi_2, \theta \leq \pi$ with $\phi_1,\phi_2 \neq 0$ and $\abs{\theta} \neq \abs{\phi_\pm}$, where $\phi_\pm = (\phi_1\pm \phi_2)/2$, it holds that
    \begin{equation}\label{eq:success_MBCoperator} 
        (\widetilde{\mathrm{CP}}(\phi_1) \circ
        _{\psi_s(\theta)} \widetilde{\mathrm{CP}}(\phi_2)) = iC (R_z(i(r_++r_-)/2)\otimes R_z(i(r_+-r_-)/2))\widetilde{\mathrm{EP}}(\chi) U_\mathrm{sign},
    \end{equation}
    where 
    \begin{equation}\label{eq:def_r_m}
        r_\pm = \log{\abs{\frac{\sin((\phi_\pm +\theta)/2)}{\sin((\phi_\pm -\theta)/2)}}}, \quad C=\sqrt{(m_+^2+m_-^2)/2}, \quad m_\pm =\sqrt{\abs{\sin^2(\phi_\pm/2)-\sin^2(\theta/2)}},
    \end{equation}
    \begin{equation}
        U_\mathrm{sign} = \diag(\sgn(\theta+\phi_+),\sgn(\theta+\phi_-),\sgn(\theta-\phi_-),\sgn(\theta-\phi_+)),
    \end{equation}
    and weight $-\pi \leq \chi \leq \pi$ is the angle such that
    \begin{equation}\label{eq:entangling_power_measure}
    \tan(\chi/4) = \frac{m_+ - m_-}{m_+ + m_-}.
    \end{equation}
    In particular, for case (i) $\abs{\theta} < \phi_l$ or $ \phi_u < \abs{\theta}$ and case (ii) $\phi_l < \abs{\theta} < \phi_u$ with $\phi_u = \max(\abs{\phi_+}, \abs{\phi_-})$ and $\phi_l = \min(\abs{\phi_+}, \abs{\phi_-})$, it holds that 
    \begin{align}
        \sin(\theta/2) &= \begin{cases} \pm 
        \sqrt{S_+ - S_- /\sin(\chi/2)} & \text{ for case (i)} \\
            \pm  \sqrt{S_+ - S_- \sin(\chi/2)} & \text{ for case (ii)}
        \end{cases}, \label{eq:MBC_angle} \\ 
        C &= \begin{cases}
            \sqrt{\abs{S_+ -\sin^2(\theta/2)}} =  \sqrt{\abs{S_-/\sin(\chi/2)}} & \text{ for case (i)}\\
            \sqrt{\abs{S_-}} & \text{ for case (ii)}
        \end{cases}, \label{eq:MBC_coefficient}
    \end{align}
    where $S_\pm = (\sin^2(\phi_+/2)\pm \sin^2(\phi_-/2))/2$. Furthermore, $U_\mathrm{sign}$ is the identity operator for case (i) and the CZ gate for case (ii), up to local unitary operations.
    Letting $\mathcal{N}_1 = \cosh((r_+ + r_-) /2)$ and $\mathcal{N}_2 = \cosh((r_+ - r_-) /2)$, it holds that
    \begin{equation}\label{eq:MBC_noninitary_amplitude}
        \mathcal{N}_i
        =\begin{dcases}
            \abs{\frac{\tan(\chi/2)\qty{\cos(\phi_{j}/2)-\cos(\theta)\cos(\phi_i/2) }}{2S_-}}
             & \text{ for case (i)}\\
            \abs{\frac{\sin(\phi_i /2)\sin(\theta)}{2S_-\cos(\chi/2)}} & \text{ for case (ii)}
        \end{dcases},
    \end{equation}
    for $(i,j)=(1,2),(2,1)$.
\end{proposition}
\begin{proof}
    From the definition, Eq.~\eqref{eq:success_MBCoperator} follows as
    \begin{align}
        &(\widetilde{\mathrm{CP}}(\phi_1) \circ_{\psi_s(\theta)} \widetilde{\mathrm{CP}}(\phi_2)) \notag \\
        &= \qty{\cos(\phi_1/4)\cos(\phi_2/4) (I\otimes I) -\sin(\phi_1/4)\sin(\phi_2/4) (Z \otimes Z)}\bra{-} R_z (-\theta) \ket{+} \notag \\  
        &\quad +i\qty{\cos(\phi_1/4)\sin(\phi_2/4) (I\otimes Z) +\sin(\phi_1/4)\cos(\phi_2/4) (Z \otimes I)}\bra{-} R_z (-\theta) \ket{-}
        \label{eq:straightforward_calculation}
        \\
        &=\diag(\cos(\phi_+/2),\cos(\phi_-/2),\cos(\phi_-/2),\cos(\phi_+/2))i \sin(\theta/2) \notag \\ 
        &\quad +i\diag(\sin(\phi_+/2),\sin(\phi_-/2),-\sin(\phi_-/2),-\sin(\phi_+/2))\cos(\theta/2) \\
        &=i\diag(\sin((\theta + \phi_+)/2),\sin((\theta+\phi_- )/2),\sin((\theta-\phi_-)/2),\sin((\theta- \phi_+)/2)) \\
        &=i\diag(\abs{\sin((\theta + \phi_+)/2)},\abs{\sin((\theta+\phi_-)/2)},\abs{\sin((\theta -\phi_- )/2)},\abs{\sin((\theta-\phi_+)/2)})U_\mathrm{sign} 
        \label{eq:hold_without_assumption}\\
        &=i(R_z(i(r_++r_-)/2)\otimes R_z(i(r_+-r_-)/2)) \qty(\frac{m_+ + m_-}{2} I\otimes I + \frac{m_+ - m_-}{2} Z \otimes Z)U_\mathrm{sign},
        \label{eq:hold_with_assumption}
    \end{align}
    where we have used 
    \begin{equation}
        e^{ r_\pm/2} = \sqrt{\abs{\frac{\sin((\phi_\pm + \theta)/2)}{\sin((\phi_\pm - \theta)/2)}}}, \quad m_\pm =\sqrt{\abs{\sin((\phi_\pm + \theta)/2)\sin((\phi_\pm - \theta)/2)}}
    \end{equation}
    in the last line.
    Observing that both $\abs{\sin(\phi_+/2)}$ and $\abs{\sin(\phi_-/2)}$ are greater or smaller than $\abs{\sin(\theta/2)}$ for case (i) while $\abs{\sin(\theta/2)}$ lies between $\abs{\sin(\phi_+/2)}$ and $\abs{\sin(\phi_-/2)}$ for case (ii), Eqs.~\eqref{eq:MBC_coefficient} and \eqref{eq:MBC_angle} are obtained from straightforward calculations. 
    Since $U_\mathrm{sign}(\theta)$ is a function of the signs of $\theta + \phi_\pm$ and $\theta - \phi_\pm$, it changes discontinuously four times depending on $\theta$.
    It is summarized as
    \begin{align}\label{eq:summary_Usign}
        U_\mathrm{sign}&=\begin{cases}
            I \otimes I & (\phi_u <  \theta \leq \pi) \\
            U_{\sigma(1)} \mathrm{CZ}  & (\phi_l <  \theta< \phi_u) \\
            U_{\sigma(1)} U_{\sigma(2)}  & (-\phi_l < \theta< \phi_l) \\
            -U_{\sigma(4)} \mathrm{CZ} & (-\phi_u <  \theta< -\phi_l) \\
            -I \otimes I & (-\pi \leq \theta< -\phi_u) \\
        \end{cases},
    \end{align}
    where 
    \begin{equation}
        U_1=I\otimes I, \quad U_2=Z \otimes I,\quad U_3=I\otimes Z, \quad U_4 = -Z \otimes Z,
    \end{equation}
    and $\sigma$ is a permutation of $\{1,2,3,4\}$ determined from the signs of $\phi_+$, $\phi_-$, and $\abs{\phi_+}-\abs{\phi_-}$, which is summarized in Table~\ref{table:values_U_sign}.
    Equation~\eqref{eq:MBC_noninitary_amplitude} is also derived from calculations as
    \begin{align}
        \cosh((r_+ \pm r_-) /2) & 
        = \frac{1}{2}\qty(\sqrt{\abs{\frac{\sin((\phi_+ + \theta)/2)\sin((\phi_- \pm \theta)/2)}{\sin((\phi_+ - \theta)/2)\sin((\phi_- \mp \theta)/2)}}}+\sqrt{\abs{\frac{\sin((\phi_+ - \theta)/2)\sin((\phi_- \mp \theta)/2)}{\sin((\phi_+ + \theta)/2)\sin((\phi_- \pm \theta)/2)}}})\\
        &= \frac{\abs{\sin((\phi_+ + \theta)/2)\sin((\phi_- \pm \theta)/2)}+\abs{\sin((\phi_+ - \theta)/2)\sin((\phi_- \mp \theta)/2)}}{2m_+m_-} \\
        &=\begin{dcases}
            \frac{\abs{\sin(\phi_+ /2)\sin(\phi_- /2)\cos^2(\theta/2)\pm \cos(\phi_+ /2)\cos(\phi_- /2)\sin^2(\theta/2)}}{m_+m_-} & \text{ for case (i)}\\
            \frac{\abs{(\sin(\phi_+ /2)\cos(\phi_- /2)\pm\cos(\phi_+ /2)\sin(\phi_- /2))\sin(\theta/2)\cos(\theta/2)}}{m_+m_-}& \text{ for case (ii)}
        \end{dcases} \\
        &=\begin{dcases}
            \abs{\frac{\tan(\chi/2)\qty{\cos((\phi_+ \mp \phi_-)/2)-\cos(\theta)\cos((\phi_+ \pm \phi_-)/2) }}{2S_-}}
             & \text{ for case (i)}\\
            \abs{\frac{\sin((\phi_+ \pm \phi_-) /2)\sin(\theta)}{2S_-\cos(\chi/2)}} & \text{ for case (ii)}
        \end{dcases},
    \end{align}
    where we have used $m_+m_-=\abs{S_-}\cos(\chi/2)$.
\end{proof}

\begin{table}[tbp]
    \centering
    \caption{Summary of the permutation $\sigma$ determined from the signs of $\phi_+$, $\phi_-$, and $\abs{\phi_+}-\abs{\phi_-}$. We let $\sgn(0)=+1$ for convenience.}
    \begin{tabular}{|c|c|c||c|}
        \hline
        $\phi_+$ & $\phi_-$ & $\abs{\phi_+}-\abs{\phi_-}$ & $\sigma$ \\ 
        \hline \hline
        $+$ & $+$ & $+$ & $e = (1)$ \\ \hline
        $-$ & $+$ & $+$ & $(14)$ \\ \hline
        $+$ & $-$ & $+$ & $(23)$ \\ \hline
        $+$ & $+$ & $-$ & $(12)(34)$ \\ \hline
        $-$ & $-$ & $+$ & $(14)(23)$ \\ \hline
        $-$ & $+$ & $-$ & $(14)(12)(34) = (1243)$ \\ \hline
        $+$ & $-$ & $-$ & $(23)(12)(34)=(1342)$ \\ \hline
        $-$ & $-$ & $-$ & $(14)(23)(12)(34) = (13)(24)$ \\ \hline
    \end{tabular}
    \label{table:values_U_sign}
\end{table}

Proposition~\ref{prop:fundamental_formula} has two important implications. First, there are two regions of parameter $\theta$: case (i) $\abs{\theta} < \phi_l$ or $ \phi_u < \abs{\theta}$ and case (ii) $\phi_l < \abs{\theta} < \phi_u$, and the resulting MBC operator obtains the additional CZ gate only for case (ii). Second, the entangling power characterized by weight $\chi$ of the entangling projection part $\widetilde{\mathrm{EP}}(\chi)$ of the MBC operator is tunable within the range in which a solution of Eq.~\eqref{eq:MBC_angle} exists.
For case (i), there is a solution of Eq.~\eqref{eq:MBC_angle} only when
\begin{equation}\label{eq:condition_PauliX}
    \begin{cases}
        -1 < \sin(\chi/2) \leq (C_-/C_+),\  (S_-/S_+) \leq \sin(\chi/2)< 1 & (\abs{\phi_+} > \abs{\phi_-}) \\
        -1 < \sin(\chi/2) \leq (S_-/S_+),\  (C_-/C_+) \leq \sin(\chi/2)< 1 & (\abs{\phi_-} > \abs{\phi_+})
    \end{cases},
\end{equation}
where $C_\pm = (\cos^2(\phi_+/2)\pm\cos^2(\phi_-/2))/2$.
On the other hand, a solution of Eq.~\eqref{eq:MBC_angle} always exists for case (ii) unless $\chi = \pm \pi$.
Therefore, the entangling power can be arbitrarily close to the maximum value ($\chi = \pm \pi$) for both cases, at the expense of amplifying the effect of the local non-unitary operators.

\subsubsection{Definitions and properties of weighted Pauli measurements 1}
Motivated by the implications of Proposition~\ref{prop:fundamental_formula}, we define the measurements in the basis of $\{\ket{\psi_s(\theta)}, \ket{\psi_f(\theta)}\}$ with particularly useful choices of $\theta$ as weighted Pauli measurements.
The simplest case is when $\theta = 0$ or $\pi$, where the local non-unitary operator parts vanish because of $r_\pm=0$.
In addition, this case corresponds to the cases of $\sin(\chi/2)= (S_-/S_+), (C_-/C_+)$ in Eq.~\eqref{eq:condition_PauliX}.
We call them the weighted Pauli-X measurements.
Another important special case is when $m_+=m_-$, where the part of an entangling projection becomes the identity.
Since it corresponds to $\chi = 0$, such solutions can exist only for case (ii) from Eq.~\eqref{eq:condition_PauliX}; therefore the resulting MBC operator is purely the CZ gate up to local (non-unitary) operations.
We call them weighted Pauli-Y measurements, because they are reduced to the Pauli-Y measurement when $\phi_1=\phi_2=\pi$, and its action on a pair of CP gates is similar to the action of the Pauli-Y measurement on a pair of CZ gates.
We can consider introducing other types of measurement bases (see Def.~\ref{def:weight_Pauli2} for additional measurement bases), but these two types of measurement bases are sufficient to achieve universal MBQC with the WGS, as shown in the manuscript. 
See also Fig.~\ref{fig:weighted_Pauli} for the summary of the defined weighted Pauli measurements and the corresponding measurement angles $\theta$ and weights $\chi$.

\begin{definition}[Weighted Pauli measurements 1]\label{def:weight_Pauli}\ 
    \begin{itemize}
        \item Weighted Pauli-X measurements are the measurements in the $X_w^{+}$ basis of $\{\ket{\psi_s(\pi)}, \ket{\psi_f(\pi)}\}$ and the $X_w^{-}$ basis of $\{\ket{\psi_s(0)}, \ket{\psi_f(0)}\}$. 
        \item Weighted Pauli-Y measurements are the measurements in the $Y_w^{+}$ basis of $\{\ket{\psi_s(\theta_{*})}, \ket{\psi_f(\theta_{*})}\}$ and the $Y_w^{-}$ basis of $\{\ket{\psi_s(-\theta_{*})}, \ket{\psi_f(-\theta_{*})}\}$, where $0 < \theta_{*}< \pi$ satisfies 
        \begin{equation}
            \sin(\theta_*/2)=\sqrt{(\sin^2(\phi_+/2)+\sin^2(\phi_-/2))/2}=\sqrt{S_+}.
        \end{equation}
    \end{itemize}
    In particular, we use the following notations for the MBC of $O_1$ and $O_2$:
    \begin{align}
        (O_1 \circ_{\psi_s(\pi)} O_2) = X_w^{+} (O_1,O_2), &\quad  (O_1 \circ_{\psi_s(0)} O_2) = X_w^{-} (O_1,O_2), \\
        (O_1 \circ_{\psi_s(\theta_{*})} O_2) = Y_w^{+} (O_1,O_2), &\quad  (O_1 \circ_{\psi_s(-\theta_{*})} O_2) = Y_w^{-} (O_1,O_2).
    \end{align}
\end{definition}

Since $\phi_\pm$ is never equal to $\pm\theta_*$ unless $\phi_1=0$ or $\phi_2=0$, we obtain the following as a corollary of Proposition~\ref{prop:fundamental_formula}.
\begin{corollary}\label{cor:weighted_PauliY_MBC}
    For $-\pi \leq \phi_1, \phi_2 \leq \pi$ with $\phi_1, \phi_2 \neq 0$, it holds that
    \begin{align}
        Y_w^+ (\widetilde{\mathrm{CP}}(\phi_1), \widetilde{\mathrm{CP}}(\phi_2)) &= i \sqrt{\abs{S_-}}U_{\sigma(1)}(R_z(i(r^*_++r^*_-)/2)\otimes R_z(i(r_+^*-r_-^*)/2))   \mathrm{CZ} \label{eq:CPCP_pauliY+}\\
        Y_w^- (\widetilde{\mathrm{CP}}(\phi_1), \widetilde{\mathrm{CP}}(\phi_2)) &= - i \sqrt{\abs{S_-}}U_{\sigma(4)}(R_z(-i(r^*_++r^*_-)/2)\otimes R_z(-i(r^*_+-r^*_-)/2)) \mathrm{CZ} \label{eq:CPCP_pauliY-},
    \end{align}
    where $r^*_\pm = \log{\abs{{\sin((\phi_\pm +\theta_*)/2)}/{\sin((\phi_\pm -\theta_*)/2)}}}$  satisfies 
    \begin{equation}\label{eq:def_r*_m*}
        \mathcal{N}^*_1 = \cosh((r^*_+ + r^*_-)/2)= \frac{\sqrt{S_+C_+}\abs{\sin(\phi_1/2)}}{\abs{S_-}},\quad \mathcal{N}^*_2 = \cosh((r^*_+-r^*_-)/2)= \frac{\sqrt{S_+C_+}\abs{\sin(\phi_2/2)}}{\abs{S_-}}.
    \end{equation}
\end{corollary}

For weighted Pauli-X measurements, we can extend the results of Proposition~\ref{prop:fundamental_formula} further. We can remove the assumption of $\abs{\phi_\pm} \neq \theta = 0,\pi$ because of $r_\pm = 0$ and obtain similar results not only for the MBC of two CP gates, but also for the MBCs of two EP gates and a pair of a CP gate and an EP gate. 

\begin{proposition}\label{prop:weighted_PauliX_MBC}
    For $-\pi \leq \phi_1, \phi_2 \leq \pi$ with $\phi_1, \phi_2 \neq 0$, it holds that
    \begin{align}
        X_w^+ (\widetilde{\mathrm{CP}}(\phi_1), \widetilde{\mathrm{CP}}(\phi_2)) &=  i\sqrt{C_+} \widetilde{\mathrm{EP}}(-\chi_c), \quad X_w^+ (\widetilde{\mathrm{EP}}(\phi_1), \widetilde{\mathrm{EP}}(\phi_2)) = i\sqrt{C_+} \widetilde{\mathrm{EP}}(\chi_c), \notag \\
        X_w^+ (\widetilde{\mathrm{CP}}(\phi_1), \widetilde{\mathrm{EP}}(\phi_2)) &= X_w^+ (\widetilde{\mathrm{EP}}(\phi_1), \widetilde{\mathrm{CP}}(\phi_2)) = i\sqrt{C_+} \widetilde{\mathrm{CP}}(\chi_c) \label{eq:CPCP_pauliX+},\\
        X_w^- (\widetilde{\mathrm{CP}}(\phi_1), \widetilde{\mathrm{CP}}(\phi_2)) &= i\sqrt{S_+}(I\otimes Z) \widetilde{\mathrm{EP}}(\chi_s), \quad 
        X_w^- (\widetilde{\mathrm{EP}}(\phi_1), \widetilde{\mathrm{EP}}(\phi_2))
        =  \sqrt{S_+}(I\otimes Z) \widetilde{\mathrm{EP}}(\chi_s), \notag \\
        X_w^- (\widetilde{\mathrm{CP}}(\phi_1), \widetilde{\mathrm{EP}}(\phi_2)) &=\sqrt{S_+} (I\otimes Z)\widetilde{\mathrm{CP}}(\chi_s),\quad  X_w^- (\widetilde{\mathrm{EP}}(\phi_1), \widetilde{\mathrm{CP}}(\phi_2)) = i\sqrt{S_+} (I\otimes Z)\widetilde{\mathrm{CP}}(-\chi_s) \label{eq:CPCP_pauliX-}
    \end{align}
    where $\chi_c=-2\arcsin(C_-/C_+)$ and $\chi_s=2\arcsin(S_-/S_+)$ satisfy 
    \begin{equation}
        \sqrt{C_+} \widetilde{\mathrm{EP}}(-\chi_c)=\cos(\phi_+/2)P_+ +  \cos(\phi_-/2) P_- , \quad   \sqrt{S_+} \widetilde{\mathrm{EP}}(\chi_s) = \sin(\phi_+/2)P_+ -  \sin(\phi_-/2) P_-,
    \end{equation}
    where $P_+=\dyad{00}+\dyad{11}=\sqrt{2^{-1}}\widetilde{\mathrm{EP}}( \pi)$ and $P_-=\dyad{01}+\dyad{10}=\sqrt{2^{-1}}\widetilde{\mathrm{EP}}(- \pi)$.
    In particular, letting
    \begin{equation}
        \widetilde{\mathrm{CP}}_{(0)}(\phi) = \widetilde{\mathrm{EP}}(\phi) ,\quad \widetilde{\mathrm{CP}}_{(1)}(\phi)  = \widetilde{\mathrm{CP}}(\phi), \quad \widetilde{\mathrm{CP}}_{(2)}(\phi) = \widetilde{\mathrm{EP}}(-\phi), \quad \widetilde{\mathrm{CP}}_{(3)}(\phi) = \widetilde{\mathrm{CP}}(-\phi),
    \end{equation}
    these results are summarized as
    \begin{align}
        X_w^+ (\widetilde{\mathrm{CP}}_{(i)}(\phi_1), \widetilde{\mathrm{CP}}_{(j)}(\phi_2)) &=  \sqrt{C_+} \widetilde{\mathrm{CP}}_{(i+j)}(\chi_c) \\ 
        X_w^- (\widetilde{\mathrm{CP}}_{(i)}(\phi_1), \widetilde{\mathrm{CP}}_{(j)}(\phi_2))
        &=  \sqrt{S_+}(I\otimes Z) \widetilde{\mathrm{CP}}_{(i-j)}(\chi_s),
    \end{align}
    up to the global phases, where $i,j \in \mathbb{Z}_4 =\{0,1,2,3\}$ in which the addition is modulo $4$.
\end{proposition}
\begin{proof}
    Note that it holds that
    \begin{equation}
        \tan(\chi_c/4) = \tan(\phi_1/4)\tan(\phi_2/4), \quad \tan(\chi_s/4) = \tan(\phi_1/4)/\tan(\phi_2/4)
    \end{equation}
    from 
    \begin{equation}
        \tan(\chi/4)=\frac{1-\sqrt{1-\sin^2(\chi/2)}}{\sin(\chi/2)}
    \end{equation}
    for $0<\abs{\chi}\leq \pi$.
    Then, all the equations are derived from straightforward calculations similar to Eq.~\eqref{eq:straightforward_calculation}.
\end{proof}

As a corollary of Proposition~\ref{prop:weighted_PauliX_MBC}, we can see what happens when the weighted Pauli-X measurements are repeatedly applied to a 1D chain of uniform WGS as in Fig.~\ref{fig:transformation_diagram}.
As mentioned in the manuscript, none of the resulting MBC operators becomes a CZ gate as follows.
\begin{corollary}\label{cor:MBC_operator_weighted_PauliX}
    The operators generated from a sequence of MBCs of CP gates, $\mathrm{CP}(\phi)$ or $\mathrm{CP}(-\phi)$, with the weighted Pauli-X measurements are represented as 
    \begin{equation}
        X_w^\pm(\cdots (X_w^\pm(X_w^\pm(\mathrm{CP}(\pm \phi),\mathrm{CP}(\pm \phi)),\mathrm{CP}(\pm \phi)))\cdots,\mathrm{CP}(\pm \phi)) \propto \mathrm{CP}_{(i-j+1 \ \mathrm{mod}\ {4})}((-1)^m \phi^{(i-j+1)}),
    \end{equation}
    up to local unitary operations, where all the signs in the left-hand side are independently chosen. Here, $i$ and $j$ are the numbers of measurements with the measurement basis states $\ket{\psi_s(\pi)} (=\ket{\psi_f(0)})$ and $\ket{\psi_s(0)} (=\ket{\psi_f(\pi)})$, respectively, $m$ is the number of $\mathrm{CP}(-\phi)$, and $-\pi \leq \phi^{(i)} \leq \pi $ is the angle satisfying
    \begin{align}
        \tan(\phi^{(i)}/4) = \tan^i(\phi/4).
    \end{align}
\end{corollary}

\subsection{Detailed descriptions and proofs of the protocol}
Here, we describe the protocol presented in the manuscript in detail, which includes the success probabilities, resulting operators, and concrete measurement bases of all the sequences of MBCs used in the manuscript.
At the end, we derive the analytical expression of the success probability of the entire protocol and its lower bound.

\subsubsection{Probabilistic generations of CZ gates and MEPs}
Let us first introduce the notation to describe measurement bases briefly as 
\begin{equation}\label{eq:notation_measurement_basis}
    \ket{(v,w)} \equiv R_z(w)R_y(v)\ket{-}=e^{-i w/2}\cos(\frac{v}{2}-\frac{\pi}{4})\ket{0}+e^{i w/2}\sin(\frac{v}{2}-\frac{\pi}{4})\ket{1}
\end{equation}
for $-\pi < v, w \leq \pi$.
For a real number $r$, $R_z(ir)\ket{-}\propto R_y(v)\ket{-}$ holds when $v$ satisfies $\tan(v/2)=\tanh(r/2)$.
Thus, as seen from Eq.~\eqref{eq:def_measurement_basis} in Def.~\ref{def:measurement_basis}, we often choose such a value of $v$ to represent the measurement basis of a weighted Pauli measurement, where $\{\ket{(v,w)},\ket{(v+\pi,w)}\}$ becomes the concrete measurement basis.
Note that $-iZ\ket{(v,w)} = \ket{(v,w+\pi)}$ is equal to $\ket{(v + \pi,w)}$ only when $\theta=0$. 
That is, the prefactor of $Z$ changes the measurement basis unless $\theta=0$.

In the following Proposition~\ref{prop:list_MBCoperator}, we derive three sequences of MBCs. One is the conventional scheme for probabilistic MEP generation, which is shown in Fig.~\ref{fig:basic_transformation} and described as Eqs.~\eqref{eq:success_event} and \eqref{eq:failure_event} in the manuscript.
Here, Eqs.~\eqref{eq:CP_CP_success} and \eqref{eq:CP_CP_failure} and Eqs.~\eqref{eq:CP_ICP_success} and \eqref{eq:CP_ICP_failure} are pairs of the possible measurement outcomes on the same measurements, respectively.
This part is somewhat redundant, as this conventional scheme is not used in our protocol in its original form.
The second one is the probabilistic CZ-gate generation shown in Fig.~\ref{fig:CZ_generation} in the manuscript, where there exist two independent variations, $\mathrm{CZ}_+$ and $\mathrm{CZ}_-$ in Eq.~\eqref{eq:CZgate_generation}, depending on which of $Y_w^+$ and $Y_w^-$ is used as the weighted Pauli-Y measurement.
The third one is the probabilistic MEP generation shown in Fig.~\ref{fig:projection_generation} in the manuscript.
There also exist two variations, $\mathrm{MP}_+$ and $\mathrm{MP}_-$ in Eq.~\eqref{eq:MEP_generation}, but these are the same sequence of MBCs until the center qubit is measured and then split to different sequences of MBCs depending on the measurement outcome on the center qubit.
In the probabilistic CZ-gate generation, the success event is unique because, if one of the weighted Pauli measurements fails, the probabilistic CZ-gate generation also fails.
Therefore, the measurement bases are independent of each other, and we can implement all the measurements at once.
On the other hand, in the probabilistic MEP generation, both $\mathrm{MP}_+$ and $\mathrm{MP}_-$ are success events, and therefore some of the measurement bases may depend on the measurement outcome on the center qubit.
In fact, the measurement bases of the rightmost qubit for $\mathrm{MP}_+$ and $\mathrm{MP}_-$ are different.
Therefore, the probabilistic MEP generation for both $\mathrm{MP}_+$ and $\mathrm{MP}_-$ requires two steps to be implemented.

\begin{proposition}[Probabilistic CZ-gate and MEP generations]\label{prop:list_MBCoperator}
    Let $h(x)$ be the function defined as $h(x)=0$ when $x>0$ and $h(x)=1$ when $x<0$, and $-\pi \leq \theta, \eta \leq \pi$ be the angles such that
    \begin{align}\label{eq:definition_angles}
        \sin(\theta/2) = \frac{\abs{\sin(\phi/2)}}{\sqrt{2}}, \quad 
        \tan(\eta/2) = \frac{{\sgn(\phi)\cos(\phi/2)}}{\sqrt{1+\cos^2(\phi/2)}+1}
    \end{align}
    for $-\pi \leq \phi \leq \pi$.
    Then, up to the global phases, it holds that
    \begin{itemize}
        \item Eqs.~\eqref{eq:success_event} and \eqref{eq:failure_event}:
        \begin{align}
            O^s_+ &\equiv \mathrm{CP}(\phi) \circ_{(0,\phi)} \mathrm{CP}(\phi) = X^-_w(\mathrm{CP}(\phi),\mathrm{CP}(\phi)) = (R_z(\phi/2)\otimes ZR_z(\phi/2))\sin(\phi/2)P_+  \label{eq:CP_CP_success}\\
            O^f_+ &\equiv \mathrm{CP}(\phi) \circ_{(0,\phi+\pi)} \mathrm{CP}(\phi) = X^+_w(\mathrm{CP}(\phi),\mathrm{CP}(\phi)) = (R_z(\phi/2) \otimes R_z(\phi/2))(P_- + \cos(\phi/2)P_+) \label{eq:CP_CP_failure} \\
            O^s_- &\equiv \mathrm{CP}(\phi) \circ_{(0,0)} \mathrm{CP}(-\phi) = X^-_w(\mathrm{CP}(\phi),\mathrm{CP}(-\phi)) = (R_z(\phi/2)\otimes ZR_z(-\phi/2))\sin(\phi/2)P_- \label{eq:CP_ICP_success}\\
            O^f_- &\equiv \mathrm{CP}(\phi) \circ_{(0,\pi)} \mathrm{CP}(-\phi) = X^+_w(\mathrm{CP}(\phi),\mathrm{CP}(-\phi)) = (R_z(\phi/2) \otimes R_z(-\phi/2))(P_+ + \cos(\phi/2)P_-) \label{eq:CP_ICP_failure},
        \end{align}

        \item Probabilistic CZ-gate generation: 
        \begin{align}
            \mathrm{CZ}_\pm &\equiv \mathrm{CP}(\phi) \circ_{(\mp \eta,\phi + \pi/2+\pi h(\pm \phi))} \mathrm{CP}(\phi) \circ_{(0,\phi \pm \theta)} \mathrm{CP}(\phi) \circ_{(\mp \eta,\phi - \pi/2+\pi h(\pm \phi))} \mathrm{CP}(\phi) \notag \\
            & = X^-_w(X^-_w(\mathrm{CP}(\phi), Y^\pm_w(\mathrm{CP}(\phi),\mathrm{CP}(\phi))),\mathrm{CP}(\phi)) = \sqrt{p_{cz}}(R_z((\phi+\pi)/2) \otimes R_z((\phi-\pi)/2))\mathrm{CZ}\label{eq:CZgate_generation}.
        \end{align}
        where 
        \begin{equation}
            p_{cz} = \frac{\sin^4(\phi/2)}{8(1+\cos^2(\phi/2))}.
        \end{equation}
        
        \item Probabilistic MEP generation:
        \begin{align}
            \mathrm{MP}_\pm &\equiv  \mathrm{CP}(\phi) \circ_{(\eta,\phi - \pi/2+\pi h(\phi))} \mathrm{CP}(\phi) \circ_{(0,\phi-\theta)} \mathrm{CP}(\phi) \circ_{(0,\phi+\pi h(\pm 1))}\mathrm{CP}(\phi) \circ_{(0,\phi+\theta)} \mathrm{CP}(\phi) \circ_{(-\eta,\phi + \pi/2+\pi h(\pm \phi))} \mathrm{CP}(\phi) \notag \\
            &= X_w^-(X_w^-(\mathrm{CP}(\phi),X_w^\mp(Y^-_w(\mathrm{CP}(\phi),\mathrm{CP}(\phi)),Y^+_w(\mathrm{CP}(\phi),\mathrm{CP}(\phi)))),\mathrm{CP}(\phi)) \notag \\
            &= \sqrt{p_p}(R_z(\phi/2) \otimes Z R_z(\phi/2))P_\pm, \label{eq:MEP_generation}
        \end{align}
        where 
        \begin{equation}
            p_p=\frac{\sin^6(\phi/2)}{16(1+\cos^2(\phi/2))}.
        \end{equation}
    \end{itemize}
\end{proposition}
\begin{proof}
    We ignore global phases in the proof.
    From Eq.~\eqref{eq:local_invariance} in Proposition~\ref{prop:basis_properties}, all the calculations are reduced to the repeated applications of Corollary~\ref{cor:weighted_PauliY_MBC} or Proposition~\ref{prop:weighted_PauliX_MBC}.
    In particular, Eqs.~\eqref{eq:CP_CP_success}--\eqref{eq:CP_ICP_failure} are obtained from Eqs.~\eqref{eq:CPCP_pauliX+} and \eqref{eq:CPCP_pauliX-} in Proposition~\ref{prop:weighted_PauliX_MBC}.
    To derive Eq.~\eqref{eq:CZgate_generation}, we first use Corollary~\ref{cor:weighted_PauliY_MBC} to obtain
    \begin{equation}\label{eq:calculation_CZ-gate_generation}
        Y^\pm_w(\mathrm{CP}(\phi),\mathrm{CP}(\phi))= \frac{\sin(\phi/2)}{\sqrt{2}} (Z \otimes Z)^{h(\pm \phi)} (R_z((\phi\pm ir)/2) \otimes R_z((\phi\pm ir)/2))\mathrm{CZ},
    \end{equation}
    where $r = \log{\abs{{\sin((\phi +\theta)/2)}/{\sin((\phi -\theta)/2)}}}$ satisfies 
    \begin{equation}\label{eq:calculation_coefficient}
        \mathcal{N}=\cosh(r/2)=\sqrt{1+\cos^2(\phi/2)}
    \end{equation}
    from Eq.~\eqref{eq:def_r*_m*}.
    Then, from 
    \begin{equation}\label{eq:relation_CZgate}
        \mathrm{CZ} = e^{i\pi/4}(R_z(\pi/2) \otimes R_z(\pi/2)) \widetilde{\mathrm{CP}}(\pi)
    \end{equation}
    (where $\mathrm{CZ}$ can also be represented with $\widetilde{\mathrm{CP}}(-\pi)$, but we do not use $\widetilde{\mathrm{CP}}(-\pi)$ to avoid confusion),  
    a measurement basis state for the next two weighted Pauli-X measurements 
    is 
    \begin{equation}\label{eq:measurement_basis_eta}
        \ket{\psi_s(0)} \propto Z^b R_z(\phi + \pi/2)R_z(\mp ir/2)\ket{-},
    \end{equation}
    where $b=0,1$ is determined from the factors of $Z$ in Eq.~\eqref{eq:calculation_CZ-gate_generation} and Eq.~\eqref{eq:calculation_factor} below.
    In particular, for $\ket{(\mp \eta,\phi+\pi/2 + \pi b)}=\ket{\psi_s(0)}$ to hold for $\mathrm{CZ}_\pm$, $\eta$ must satisfy 
    \begin{equation}
        \tan(\eta/2) = \tanh(r/4)= \sgn(r) \sqrt{\frac{\sqrt{1+\cos^2(\phi/2)}-1}{\sqrt{1+\cos^2(\phi/2)}+1}} = \frac{\sgn(\phi)\abs{\cos(\phi/2)}}{\sqrt{1+\cos^2(\phi/2)}+1},
    \end{equation}
    where we have used Eq.~\eqref{eq:calculation_coefficient} in the second equality and $\sgn(r)=\sgn(\phi)\sgn(\theta)$ in the third equality.
    On the other hand, the resulting MBC operator is calculated as
    \begin{equation}
        \mathrm{CZ}_\pm =\frac{\sin(\phi/2)}{\sqrt{2}\mathcal{N}_*}(R_z(\phi/2) \otimes R_z(\phi/2))X^-_w(X^-_w(\widetilde{\mathrm{CP}}(\phi), \widetilde{\mathrm{CP}}(\pi)),\widetilde{\mathrm{CP}}(\phi)) 
    \end{equation}
    where
    \begin{align}
        X^-_w(X^-_w(\widetilde{\mathrm{CP}}(\phi), \widetilde{\mathrm{CP}}(\pi)),\widetilde{\mathrm{CP}}(\phi)) &= \frac{1}{\sqrt{2}}X^-_w((I\otimes Z)\widetilde{\mathrm{EP}}(\phi),\widetilde{\mathrm{CP}}(\phi)) \label{eq:calculation_factor}\\
        &=\frac{\abs{\sin(\phi/2)}}{2}(Z\otimes I)\widetilde{\mathrm{CP}}(\pi)
    \end{align}
    from Eq.~\eqref{eq:CPCP_pauliX-}.

    For the derivation of Eq.~\eqref{eq:MEP_generation}, first we have 
    \begin{align}
        &Y^-_w(\mathrm{CP}(\phi),\mathrm{CP}(\phi))_{AB} Y^+_w(\mathrm{CP}(\phi),\mathrm{CP}(\phi))_{BC} \notag \\
        &= \frac{\sin^2(\phi/2)}{2} (Z^{h(- \phi)} \otimes I \otimes Z^{h(\phi)}) (R_z((\phi - ir+\pi)/2) \otimes R_z(\phi) \otimes R_z((\phi + ir+\pi)/2)) \widetilde{\mathrm{CP}}(\pi)_{AB}\widetilde{\mathrm{CP}}(\pi)_{BC}
    \end{align}
    from Eq.~\eqref{eq:calculation_CZ-gate_generation}, and thereby the measurement basis for the first weighted Pauli-X measurement on the center qubit is identified to be $\{R_z(\phi)\ket{+},R_z(\phi)\ket{-}\}$, where $\ket{\psi_s(\pi)}=R_z(\phi)\ket{+}$ and $\ket{\psi_s(0)}=R_z(\phi)\ket{-}$.
    On the other hand, the measurement bases for the other weighted Pauli-X measurements are the same as Eq.~\eqref{eq:measurement_basis_eta} up to the factor of $Z$. Thus, the MBC operator is calculated as
    \begin{align}
        \mathrm{MP}_\pm 
        &= \frac{\sin^2(\phi/2)}{2\mathcal{N}} (R_z(\phi/2) \otimes R_z(\phi/2)) X_w^-(X_w^-(\widetilde{\mathrm{CP}}(\phi),X_w^\mp(\widetilde{\mathrm{CP}}(\pi),\widetilde{\mathrm{CP}}(\pi))),\widetilde{\mathrm{CP}}(\phi)),
    \end{align}
    where
    \begin{align}
        X_w^-(X_w^-(\widetilde{\mathrm{CP}}(\phi),X_w^\mp(\widetilde{\mathrm{CP}}(\pi),\widetilde{\mathrm{CP}}(\pi))),\widetilde{\mathrm{CP}}(\phi))
        &=  \frac{1}{\sqrt{2}}X_w^-(X_w^-(\widetilde{\mathrm{CP}}(\phi),(I \otimes Z^{h(\mp 1)})\widetilde{\mathrm{EP}}(\pm \pi)),\widetilde{\mathrm{CP}}(\phi)) \\
        &= \frac{1}{2}X_w^-((I \otimes Z^{h(\pm 1)})\widetilde{\mathrm{CP}}(\pm \phi),\widetilde{\mathrm{CP}}(\phi)) \label{eq:calculation_factor2} \\
        &= \frac{\sin(\phi/2)}{2}(I \otimes Z) P_\pm,
    \end{align}
    from Eqs.~\eqref{eq:CPCP_pauliX+} and \eqref{eq:CPCP_pauliX-}.
    Note that the measurement basis of the last weighted Pauli-X measurement in Eq.~\eqref{eq:calculation_factor2} depends on the sign $\pm$, that is, the prior measurement outcome of the first weighted Pauli-X measurement.
    Therefore, the order of these Pauli-X measurements is important for this process to work properly.
\end{proof}

\subsubsection{Near-deterministic generations of MEPs and CZ gates}
Here, we derive the near-deterministic generations of MEP and CZ gate proposed in the manuscript and especially calculate their success probabilities.
To begin with, we briefly review quantum operations and quantum instruments and introduce a notation for efficiently describing all the possible MBC operators.
First, transformations of quantum states conditioned on measurement outcomes are represented as completely positive trace-non-increasing maps $\mathcal{E}$, called quantum operations. 
It is known that $\mathcal{E}$ can be represented with a set of Kraus operators $\{K_i\}$ such that $\sum_i K^\dag_i K_i \leq I$ as $\mathcal{E}(\rho)=\sum_i K_i \rho K_i^\dag$ for any quantum state $\rho$. 
In the current situation, all the considered quantum operations are represented as $\mathcal{E}(\rho) = K \rho K^\dag$ with a single operator $K$.
Therefore, we have represented each quantum operation by writing the operator corresponding to $K$ throughout this paper.
In particular, the success probabilities of the probabilistic CZ-gate generation in Eq.~\eqref{eq:CZgate_generation} and the probabilistic MEP generation in Eq.~\eqref{eq:MEP_generation} are calculated as
\begin{equation}
    \tr(\mathrm{CZ}_\pm \rho \mathrm{CZ}_\pm^\dag) =  p_{cz}, \quad \tr(\mathrm{MP}_+ \rho \mathrm{MP}_+^\dag) + \tr(\mathrm{MP}_- \rho \mathrm{MP}_-^\dag) = p_p\tr(P_+ \rho) + p_p\tr(P_- \rho) = p_p
\end{equation}
for input state $\rho$, respectively.
If we have multiple quantum operations, each of which is represented with $K^{[j]}$, and they are proportional to each other as $K^{[j]} = \sqrt{c_j} K$ for an operator $K$ and coefficients $c_j$, we can combine them into one quantum operation represented with $\sqrt{\sum_i c_i} K$, because the quantum operation corresponding to the case where measurement outcomes are not distinguished becomes $\sum_j K^{[j]} \rho K^{[j]} = (\sum_i c_i) K\rho K^\dag$. We simplify our discussion by combining several possible MBC operators into one.

A quantum instrument is a notion for describing quantum measurements and conditioned quantum operations for all possible measurement outcomes.
A way to represent quantum instrument $\mathcal{I}$ is to  introduce a register system $\mathcal{R}$ for representing all possible measurement outcomes $\{x\}_{x \in X}$. Then, the quantum instrument $\mathcal{I}$ on a system $\mathcal{S}$ is represented as $\mathcal{I}(\rho_\mathcal{S}) = \sum_{x \in X} \mathcal{E}_x(\rho)_\mathcal{S} \otimes \dyad{x}_\mathcal{R}$ for quantum state $\rho$, where each $\mathcal{E}_x$ is a quantum operation such that $\tr(\sum_x \mathcal{E}_x(\rho)) = \tr(\rho)$.
Hereafter, we represent some quantum operations on (system) qubits conditioned on prior measurement outcomes as a single quantum operation on both of the system qubits and register qubits that represent the prior measurement outcomes, without explicitly writing down the register qubits. 
This allows us to combine a broader class of quantum operations into one quantum operation.
For example, let us consider the MBCs of two identical CP gates with the Pauli-Z basis measurement. 
It is calculated from $\mathrm{CP}(\phi)=\dyad{0}\otimes I + e^{i\phi/2}\dyad{1}\otimes R_z(\phi)$ as 
\begin{equation}\label{eq:mbc_pauliZ}
    \mathrm{CP}(\phi) \circ_{0} \mathrm{CP}(\phi) = \frac{1}{\sqrt{2}} I \otimes I, \quad \mathrm{CP}(\phi) \circ_{1} \mathrm{CP}(\phi) = \frac{1}{\sqrt{2}} R_z(\phi) \otimes R_z(\phi),
\end{equation}
up to global phases.
Thus, the resulting MBC operators include the local phase gates only when the corresponding measurement outcome is $Z=-1$. 
Here, we introduce the operator $R_z(\hat{b}^z \phi)$ on a system qubit and an (implicit) register qubit, where $\hat{b}^z=0$ for the measurement outcome of $Z=+1$, and $\hat{b}^z=1$ for the measurement outcome of $Z=-1$.
Then, we can combine both of the MBC operators in Eq.~\eqref{eq:mbc_pauliZ} into $R_z(\hat{b}^z \phi) \otimes R_z(\hat{b}^z \phi)$ and eliminate the need to explicitly write down every MBC operator depending on the measurement outcome.

Here, we further clarify the measurement bases used in the near-deterministic MEP and CZ-gate generations.
We use the notions of 
\begin{align}
    X_n &= \dyad{(\pi,n\phi)} - \dyad{(0,n\phi)} \label{eq:PauliX_1}\\
    X_{\pm_1\pm_2} &= \dyad{(\pi \pm_1 \eta,\phi \pm_2 \pi/2+\pi h(\phi))} \notag \\ &\quad -\dyad{(\pm_1 \eta,\phi \pm_2 \pi/2+\pi h(\phi))}, \label{eq:PauliX_pm}\\
    Y_\pm &= \dyad{(\pi,\phi \pm \theta)} - \dyad{(0,\phi \pm \theta)}, \label{eq:PauliY_pm}
\end{align}
where $\pm_1$ and $\pm_2$ take different signs.
Then, the measurement bases for the probabilistic generations of $\mathrm{CZ}_+$ and $\mathrm{CZ}_-$ in Eq.~\eqref{eq:CZgate_generation} are represented as $(X_{-+},Y_+,X_{--})$ and $(X_{+-},Y_-,X_{++})$, respectively, and these probabilistic generations succeed when all the measurement outcomes are $-1$.
On the other hand, the measurement bases for the probabilistic generations of $\mathrm{MP}_+$ and $\mathrm{MP}_-$ in Eq.~\eqref{eq:MEP_generation} are $(X_{+-},Y_-,X_1,Y_+,X_{-+})$ and $(X_{+-},Y_-,X_1,Y_+,X_{--})$, respectively. 
The success event for $\mathrm{MP}_\pm$ corresponds to the case where $X_1=\mp 1$, and the other measurement outcomes are all $-1$.

\begin{figure}[tbp]
\centering
\includegraphics[scale=1.5, trim=0 0 0 0, clip]{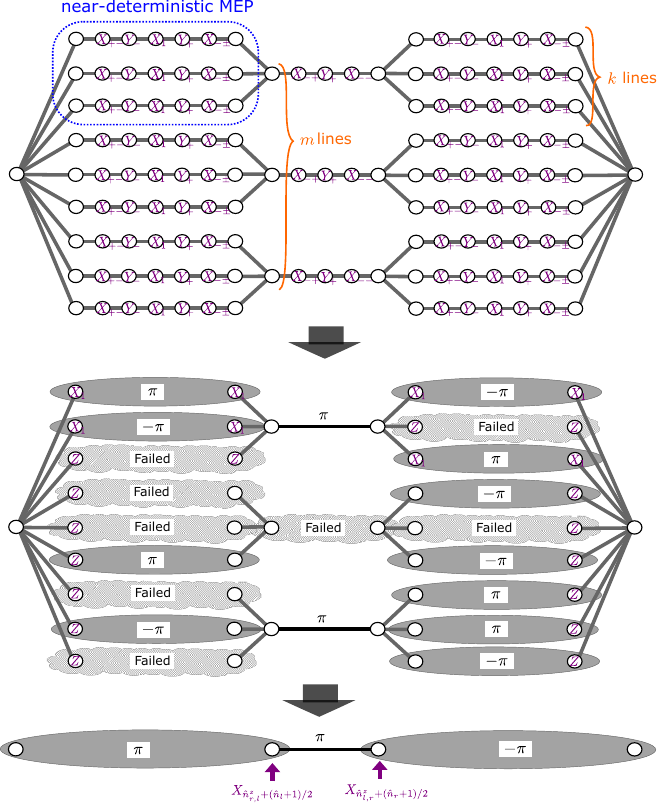}
\caption{Entire scheme of the near-deterministic CZ-gate generation for $k=3$ and $m=3$. In this figure, we suppose that the first and third CZ-gate generations succeeded (that is, their measurement outcomes are $(X_{-+},Y_+,X_{--}) = (-1,-1,-1)$) in the first step and chose the first one. In the second step, we suppose that a pair of $X_1$-basis measurements succeeded (that is, $X_1X_1=+1$ holds) only on the first line in both sides, and the others failed. This example corresponds to the case of $\hat{n}_{l}=\hat{n}_{r}=2$, $\hat{n}^s_{l}=\hat{n}^s_{r}=1$, and $\hat{b}_{p}=1$, while the others depend on the concrete values of the other measurement outcomes.}\label{fig:entire_transformation}
\end{figure}

The entire setup and procedure for the near-deterministic CZ-gate generation is shown in Fig.~\ref{fig:entire_transformation} for $k=3$ and $m=3$, where the region enclosed by the dotted line corresponds to the part of a near-deterministic MEP generation.
For each near-deterministic MEP generation, we apply a pair of weighted Pauli-X measurements to each of the lines in which the probabilistic MEP generation succeeded. 
As shown in the proof of Proposition~\ref{prop:near-deterministic}, the corresponding concrete measurement bases are always $X_1$, and the case of $X_1X_1= +1$ corresponds to the success event in the sense that an MEP is generated again.
For the near-deterministic CZ-gate generation, we perform a pair of weighted Pauli-X measurements in the last step.
Here, the measurement bases can be represented in the form of $X_n$, where 
the value of $n$ depends on the prior measurement outcomes.
As a result, the near-deterministic CZ-gate generation requires four steps in total.

\begin{proposition}[Near-deterministic MEP and CZ-gate generations]\label{prop:near-deterministic}
    The setups and measurement bases for near-deterministic MEP and CZ-gate generations are shown in Fig.~\ref{fig:entire_transformation}.
    \begin{itemize}
        \item Near-deterministic MEP generation: It succeeds when there exists at least one line that satisfies two conditions: (i) the measurement outcomes on the middle five qubits satisfy $(X_{+-},Y_-,X_1,Y_+,X_{-+})=(-1,-1,-1,-1,-1)$ or $(X_{+-},Y_-,X_1,Y_+,X_{--})=(-1,-1,+1,-1,-1)$, and (ii) the measurement outcomes on the leftmost and rightmost qubits satisfy $(X_1,X_1)=(+1,+1)$ or $(-1,-1)$.
        Let $\hat{n}^{z}_{l}$ and $\hat{n}^{z}_{r}$ be the numbers of $Z=-1$ on the $Z$-basis measurements on the leftmost and rightmost qubits, respectively, and $\hat{n}$ and $\hat{n}^{s}$ be the numbers of lines satisfying condition (i) and condition (ii), respectively.
        Then, the corresponding MBC operator when the measurement outcome of the center qubit is $X_1 = \mp 1$ is represented as
        \begin{equation}\label{eq:neardetermistic_MEP}
            \mathrm{MP}^{(k)}_\pm = \sqrt{D_k} \qty{R_z((\hat{n}^{z}_{l} + \hat{n}/2) \phi) \otimes Z^{\hat{n}^{s}} R_z((\hat{n}^{z}_r + \hat{n}/2)\phi)}P_\pm,
        \end{equation}
        where
        \begin{equation}
            D_k = 1-\qty(1-\frac{\sin^8(\phi/2)}{32(1+\cos^2(\phi/2))})^k. \label{eq:Dk_value}
        \end{equation}

        \item Near-deterministic CZ-gate generation: It succeeds when at least one probabilistic CZ-gate generation succeeds, that is, $(X_{-+},Y_+,X_{--})=(-1,-1,-1)$, and the pair of relevant near-deterministic MEP generations also succeed.
        We define $\hat{n}^{z}_{l,l/r}$, $\hat{n}^{z}_{r,l/r}$, $\hat{n}_{l/r}$, and $\hat{n}^{s}_{l/r}$ as $\hat{n}^{z}_{l}$, $\hat{n}^{z}_{r}$, $\hat{n}$, and $\hat{n}^{s}$ defined above for the near-deterministic MEP generation on the left/right-hand sides, respectively. In addition, let $\hat{n}^{z}_{L/R}$ be the number of $Z = -1$ on the $Z$-basis measurements on the leftmost/rightmost qubits in the parts not selected for the near-deterministic MEP generations, and $\hat{b}_{l}$, $\hat{b}_{r}$, and $\hat{b}_p$ be the operators equal to one when $X_{\hat{n}^{z}_{r,l} + (\hat{n}_l+1)/2}=-1$, $X_{\hat{n}^{z}_{l,r} + (\hat{n}_r+1)/2}=-1$, and the pair of generated MEPs, $\mathrm{MP}^{(k)}_+$ or $\mathrm{MP}^{(k)}_-$, are different, respectively, and equal to zero otherwise.
        Then, the  corresponding MBC operator is represented as
        \begin{align}\label{eq:neardetermistic_CZ}
            \mathrm{CZ}^{(k,m)} = D_k \sqrt{E_m} \qty{Z^{\hat{n}'_L}R_z(\hat{n}_{L}\phi+\pi/2) \otimes Z^{\hat{n}'_R} R_z(\hat{n}_{R}\phi-\pi/2)}\mathrm{CZ},
        \end{align}
        where 
        \begin{equation}
            E_m = 1-\qty(1-\frac{\sin^4(\phi/2)}{8(1+\cos^2(\phi/2))})^m ,\label{eq:Em_value}
        \end{equation}
        \begin{equation}
            \hat{n}_{L} = \hat{n}^{z}_{l,l} + \hat{n}^{z}_{L} +  \frac{\hat{n}_{l}}{2}, \quad \hat{n}_{R} = \hat{n}^{z}_{r,r} + \hat{n}^{z}_{R} +  \frac{\hat{n}_{r}}{2}, \quad  \hat{n}'_{L} = \hat{n}^{s}_{l}+\hat{b}_{l}+\hat{b}_p, \quad \hat{n}'_{R} = \hat{n}^{s}_{r}+\hat{b}_{r}+\hat{b}_p.
        \end{equation}
    \end{itemize}
\end{proposition}
\begin{proof}
We ignore global phases in the proof.
First, we calculate the MBC operator generated after the MBC with a CP gate being applied to a successfully generated MEP twice. From Eqs.~\eqref{eq:CPCP_pauliX+} and \eqref{eq:CPCP_pauliX-}, it is calculated as
\begin{equation}
    X^{\pm_1}_w(X^{\pm_2}_w(\mathrm{CP}(\phi), \mathrm{MP}_{\pm_3}),\mathrm{CP}(\phi)) = \sqrt{\frac{p_p}{2}} (R_z(\phi/2)\otimes R_z(\phi/2))X^{\pm_1}_w(X^{\pm_2}_w(\widetilde{\mathrm{CP}}(\phi), (I \otimes Z)\widetilde{\mathrm{EP}}(\pm_3 \pi)),\widetilde{\mathrm{CP}}(\phi)),
\end{equation}
where
\begin{align}
    X^{\pm_1}_w(X^{\pm_2}_w(\widetilde{\mathrm{CP}}(\phi), (I \otimes Z) \widetilde{\mathrm{EP}}(\pm_3 \pi)),\widetilde{\mathrm{CP}}(\phi)) &= \sqrt{2^{-1}}X^{\pm_1}_w((I\otimes Z^{h(\mp_2 1)}) \widetilde{\mathrm{CP}}(\pm_3 \phi),\widetilde{\mathrm{CP}}(\phi)) \\
    &=\begin{dcases}
            \sqrt{2^{-1}}\sin(\phi/2) (I \otimes Z) P_{\pm_3} &  \text{ for $\pm_1 = - $}\\
            \sqrt{2^{-1}} (P_{\mp_3} + \cos(\phi/2) P_{\pm_3}) & \text{ for $\pm_1 = + $}
        \end{dcases}.
\end{align}
where $\pm_1$, $\pm_2$, and $\pm_3$ take different signs.
The sign of $\pm_3$ is determined from the prior measurement outcome on the center qubit as $\pm_3 X_1 =  -1$, while the sign of $\pm_2$ is determined from the measurement outcome on the leftmost qubit as $\pm_2 X_1=  1$.
The sign of $\pm_1$ is determined from the measurement outcome on the rightmost qubit as $\pm_1  \pm_2 X_1=  -1$, that is, it is determined from the measurement outcomes on the leftmost and rightmost qubits as $\pm_1 X_1X_1=  -1$.
As a result, the case of $X_1 X_1 = +1$ corresponds to the success event.
By combining the cases of $\pm_2 = +$ and $\pm_2 = -$, which gives an addition factor of $\sqrt{2}$, we have the four possible MBC operators:
\begin{gather}
    P^s_\pm = \sqrt{\frac{p_p}{2}} (R_z(\phi/2)\otimes ZR_z(\phi/2)) (\sin(\phi/2) P_\pm), \quad 
    P^f_\pm = \sqrt{\frac{p_p}{2}} (R_z(\phi/2)\otimes R_z(\phi/2)) (P_\mp + \cos(\phi/2)P_\pm ).
\end{gather}
In addition, when the initial MEP generation fails, we apply the $Z$-basis measurements to the leftmost and rightmost qubits and obtain the MBC operator $\sqrt{1-p_{p}}R_z(\hat{b}^z_l \phi) \otimes R_z(\hat{b}^z_r \phi)$, as in Eq.~\eqref{eq:mbc_pauliZ}. Here, $\hat{b}^z_l$ and $\hat{b}^z_r$ are equal to zero when $Z=+1$ and equal to one when $Z=-1$ on the leftmost and rightmost qubits, respectively.
The coefficient $\sqrt{1-p_{p}}$ is derived from the fact that the MEP generation succeeds with the probability $p_{p}$, and the operators obtained in the other cases are always in the form of $R_z(\hat{b}^z_l \phi) \otimes R_z(\hat{b}^z_r \phi)$.

Then, we consider parallelizing the above heralded MEP generations, where the number of the lines prepared in parallel is $k$.
Let $n^{s}_+$, $n^{s}_-$, $n^{f}_{+}$, and $n^{f}_{-}$ be the numbers of $P^s_+$, $P^s_-$, $P^f_+$, and $P^f_-$ that have been generated, respectively. 
Then, since $P^s_+ P^s_- = 0$, the case of $n^s_+ > 0$ and $n^s_- > 0$ never happens. Thus, we need to consider three cases: (i) $n^s_+ > 0$ and $n^s_- = 0$, (ii) $n^s_+ = 0$ and $n^s_- > 0$ and (iii) $n^s_+  = n^s_- = 0$, each of which corresponds to
\begin{align}
    &(P^s_+)^{n^{s}_{+}}(P^s_-)^{n^{s}_{-}}(P^f_+)^{n^{f}_{+}}(P^f_-)^{n^{f}_{-}} \notag \\
     &= \qty(\frac{p_p}{2})^{n/2} (R_z(n\phi/2 )\otimes Z^{n^s} R_z(n\phi/2))
    \begin{cases}
        (\sin(\phi/2))^{n^{s}_{+}}(\cos(\phi/2))^{n^{f}_{+}} P_+ & \text{ for case (i)} \\
        (\sin(\phi/2))^{n^{s}_{-}}(\cos(\phi/2))^{n^{f}_{-}} P_- & \text{ for case (ii)} \\
        (\cos(\phi/2))^{n^{f}_{+}}P_+ +  (\cos(\phi/2))^{n^{f}_{-}}P_-& \text{ for case (iii)}
    \end{cases},
    \label{eq:composite_operator}
\end{align}
where $n=n^{s}_{+} + n^{s}_{-} + n^{f}_{+} + n^{f}_{-}$ and $n^s = n^{s}_{+} + n^{s}_{-}$.
The cases (i) and (ii) correspond to the success events of an MEP generation, while the case (iii) corresponds to the failure event.
As in the case of $k=1$, the contributions from the failure events can be represented as $(1-p_p)^{(k-n)/2}R_z(\hat{n}^z_l \phi) \otimes R_z(\hat{n}^z_r \phi)$, where $\hat{n}^z_l$ and $\hat{n}^z_r$ are the sums of $\hat{b}^z_l$ and $\hat{b}^z_r$, respectively, over the $k$ lines.
Therefore, by using the operators $\hat{n}$ and $\hat{n}^s$, the cases (i) and (ii) can be combined into single operators, $\mathrm{MP}^{(k)}_+$ and $\mathrm{MP}^{(k)}_-$, respectively, as in Eq.~\eqref{eq:neardetermistic_MEP}.
Here, $D_k$ is calculated from Eq.~\eqref{eq:composite_operator} as 
\begin{align}
    D_k &= \begin{dcases}
        \sum_{\{n^{s}_{+}, n^{f}_{+}, n^{f}_{-} \mid n^{s}_{+} \neq 0,  n \leq k \}} \frac{k!}{n^{s}_{+}! n^{f}_{+}! n^{f}_{-}! (k-n)!}
        (1-p_p)^{k-n} \qty(\frac{p_p}{2})^n (\sin(\phi/2))^{2n^{s}_{+}}(\cos(\phi/2))^{2n^{f}_{+}} &\text{ for case (i)} \\
        \sum_{\{n^{s}_{-}, n^{f}_{+}, n^{f}_{-} \mid n^{s}_{-} \neq 0,  n \leq k \}} \frac{k!}{n^{s}_{-}! n^{f}_{+}! n^{f}_{-}! (k-n)!}
        (1-p_p)^{k-n} \qty(\frac{p_p}{2})^n (\sin(\phi/2))^{2n^{s}_{-}}(\cos(\phi/2))^{2n^{f}_{-}} &\text{ for case (ii)}
    \end{dcases} \\
    &=\sum_{a+b+c+d=k} \frac{k!}{a!b!c!d!} \qty(\frac{p_p}{2} \sin^2(\phi/2))^a \qty(\frac{p_p}{2}\cos^2(\phi/2))^b \qty(\frac{p_p}{2})^c(1-p_p)^d \notag \\
    & \qquad  - \sum_{b+c+d=k} \frac{k!}{b!c!d!} \qty(\frac{p_p}{2}\cos^2(\phi/2))^b \qty(\frac{p_p}{2})^c(1-p_p)^d \\
    & = \qty(\frac{p_p}{2} \sin^2(\phi/2) + \frac{p_p}{2} \cos^2(\phi/2) + \frac{p_p}{2} + 1-p_p)^k - \qty(\frac{p_p}{2} \cos^2(\phi/2) + \frac{p_p}{2} + 1-p_p)^k \\
    & = 1 - \qty(1-\frac{p_p}{2} \sin^2(\phi/2))^k = 1-\qty(1-\frac{\sin^8(\phi/2)}{32(1+\cos^2(\phi/2))})^k.
\end{align}

Next, we consider the MBCs of $\mathrm{MP}_\pm^{(k)}$ in Eq.~\eqref{eq:neardetermistic_MEP} and $\mathrm{CZ}_+$ in Eq.~\eqref{eq:CZgate_generation}.
Let us represent the operators generated on the near-deterministic generation on the left- and right-hand sides as
\begin{align}
    \mathrm{MP}^{(k)}_{\pm,l} &= \sqrt{D_k} \qty{Z^{\hat{n}^{s}_l} R_z((\hat{n}^{z}_{l,l} + \hat{n}_l/2) \phi) \otimes  R_z((\hat{n}^{z}_{r,l} + \hat{n}_l/2)\phi)}P_\pm, \\
    \mathrm{MP}^{(k)}_{\pm,r} &= \sqrt{D_k} \qty{R_z((\hat{n}^{z}_{l,r} + \hat{n}_r/2) \phi) \otimes Z^{\hat{n}^{s}_r}R_z((\hat{n}^{z}_{r,r} + \hat{n}_r/2)\phi)}P_\pm,
\end{align}
respectively, where we have used $(I\otimes Z) P_\pm = \pm (Z\otimes I) P_\pm$.
From these equations and Eqs.~\eqref{eq:CZgate_generation} and \eqref{eq:relation_CZgate}, 
we have
\begin{align}
    &X_w^{\pm_1}(X_w^{\pm_2}(\mathrm{MP}^{(k)}_{\pm_3,l}, \mathrm{CZ}_+),\mathrm{MP}^{(k)}_{\pm_4,r}) \notag \\
    &= \frac{D_k\sqrt{p_{cz}}}{2} \qty{Z^{\hat{n}^{s}_l}R_z((\hat{n}^{z}_{l,l} + \hat{n}_l/2) \phi) \otimes Z^{\hat{n}^{s}_r}R_z((\hat{n}^{z}_{r,r} + \hat{n}_r/2)\phi)} X_w^{\pm_1}(X_w^{\pm_2}(\widetilde{\mathrm{EP}}(\pm_3 \pi),(Z \otimes I)\widetilde{\mathrm{CP}}(\pi)), \widetilde{\mathrm{EP}}(\pm_4 \pi))
\end{align}
where $\pm_1$, $\pm_2$, $\pm_3$, and $\pm_4$ take different signs, and the measurement bases for the left and right weighted Pauli-X measurements are identified to be $X_{\hat{n}^{z}_{r,l} + (\hat{n}_l+1)/2}$ and $X_{\hat{n}^{z}_{l,r} + (\hat{n}_r+1)/2}$, respectively, although which measurement outcome corresponds to the success event depends on the factors of $Z$.
Then, it is calculated as
\begin{align}
    X_w^{\pm_1}(X_w^{\pm_2}(\widetilde{\mathrm{EP}}(\pm_3 \pi),(Z\otimes I)\widetilde{\mathrm{CP}}(\pi)), \widetilde{\mathrm{EP}}(\pm_4 \pi)) 
    &= \frac{1}{\sqrt{2}}X_w^{\pm_1}((Z^{h(\pm_2 \pm_3 1)} \otimes Z^{h(\pm_3 1)})\widetilde{\mathrm{CP}}( \pi), \widetilde{\mathrm{EP}}(\pm_4 \pi)) \\
    &\quad =\frac{1}{2} (Z^{h(\pm_2 \pm_3 \pm_4 1)} \otimes Z^{h(\pm_1 \pm_4 1)})\widetilde{\mathrm{CP}}(\pi)
\end{align}
from Eqs.~\eqref{eq:CPCP_pauliX+} and \eqref{eq:CPCP_pauliX-}. Therefore, the correspondences between the signs of $\pm_1$ and $\pm_2$ and the measurement outcomes are $\pm_2X_{\hat{n}^{z}_{r,l} + (\hat{n}_l+1)/2}= -1$ and $\pm_1 \pm_3 X_{\hat{n}^{z}_{l,r} + (\hat{n}_r+1)/2} = 1$, that is, $\hat{b}_l=h(\mp_2 1)$ and $\hat{b}_r=h(\pm_1 \pm_3 1)$, respectively.
The signs of $\pm_3$ and $\pm_4$ are determined by the measurement outcomes of $X_1$ on the MEP generations on the left- and right-hand sides, where $\hat{b}_p = h(\pm_3 \pm_4 1)$.
By using the operators $\hat{b}_l$, $\hat{b}_r$, and $\hat{b}_p$, all the $2^4$ cases can be combined to a single operator with an additional factor of $\sqrt{2^4}=4$ as
\begin{equation}
    \mathrm{CZ}^{(k,1)} = D_k\sqrt{p_{cz}} \qty{Z^{\hat{n}^{s}_l+\hat{n}^{x}_l+\hat{n}^{x}_p}R_z((\hat{n}^{z}_{l,l} + \hat{n}_l/2) \phi+\pi/2) \otimes Z^{\hat{n}^{s}_r+\hat{n}^{x}_r+\hat{n}^{x}_p}R_z((\hat{n}^{z}_{r,r} + \hat{n}_r/2)\phi-\pi/2)} \mathrm{CZ}.
\end{equation}

Finally, we consider the near-deterministic CZ-gate generation by preparing $m$ copies of the setup for the above heralded CZ-gate generation.
We can implement the probabilistic CZ-gate and MEP generations on all the copies simultaneously and choose one of the cases in which the CZ-gate generation succeeded.
We continue the procedure for the above heralded CZ-gate generation on the chosen copy, while we apply the $Z$-basis measurements to the other copies.
As a result, the success probability of the CZ-gate generation is replaced from $p_{cz}$ with $E_m = 1-(1-p_{cz})^m$ and the effect of the $Z$-basis measurements can be incorporated as an additional factor of $R_z(\hat{n}^z_L \phi) \otimes R_z(\hat{n}^z_R \phi)$, as in Eq.~\eqref{eq:neardetermistic_CZ}.
\end{proof}

\subsubsection{Overall success probability and its lower bound}
\begin{theorem}
    The WGS shown in Fig.~\ref{fig:entire_graph}, parametrized with $(n,k,m)$, can be transformed to an $n \times n$ cluster state, up to local unitary operations, with the probability of 
    \begin{equation}\label{eq:overall_probability}
        P(n,k,m)= \qty[\qty{1-\qty(1-\frac{\sin^8(\phi/2)}{32(1+\cos^2(\phi/2))})^k}^2\qty{1-\qty{1-\qty(\frac{\sin^4(\phi/2)}{8(1+\cos^2(\phi/2))})}^m}]^{2n(n-1)},
    \end{equation}
    which is lower bounded by $1-\delta$ when $k \geq 64\pi^8 \phi^{-8}\log(8\delta^{-1}n(n-1))$ and $m \geq  16 \pi^4 \phi^{-4}\log(4\delta^{-1}n(n-1))$ for any $\delta > 0$.
\end{theorem}
\begin{proof}
    Since the near-deterministic CZ-gate generation in Proposition~\ref{prop:near-deterministic} succeeds with the probability of $D_k^2E_m$, and there are $2n(n-1)$ perfect edges in an $n \times n$ cluster state, the overall success probability is $P(n,k,m)=(D_k^2E_m)^{2n(n-1)}$. Thus, Eq.~\eqref{eq:overall_probability} is obtained from Eqs.~\eqref{eq:Dk_value} and \eqref{eq:Em_value}.

    To derive a lower bound of $P(n,k,m)$, we use the following inequalities:
    \begin{align}
    \text{(Jordan's inequality) } \qquad &\frac{2}{\pi}x\leq \sin(x) \leq x &&\text{ for } \quad  x\in[0,\pi/2],\label{eq:Jordan_ineq}  \\
    \text{(Bernoulli's inequality) } \qquad & (1+x)^r \geq 1+rx   &&\text{ for } \quad  r \geq 1, \ x \geq -1, \label{eq:Bernoulli_ineq}\\
    &(1+x)^r \leq e^{rx}   &&\text{ for }  \quad  r \geq 0, \ x \geq -1.   \label{eq:exponential_ineq}
    \end{align}
    Then, $D_k$ in Eq.~\eqref{eq:Dk_value} and $E_m$ in Eq.~\eqref{eq:Em_value} are lower bounded as
    \begin{align}
        D_k &\geq 1-\qty(1-\frac{\phi^8}{64 \pi^8})^k \geq 1 - \exp(-\frac{k\phi^8}{64\pi^8}), \label{eq:lowerbound_Dk} \\
        E_m &\geq 1-\qty(1-\frac{\phi^4}{16 \pi^4})^m \geq 1 - \exp(-\frac{m\phi^4}{16\pi^4}), \label{eq:lowerbound_Em} 
    \end{align}
    where we have used Eq.~\eqref{eq:Jordan_ineq} and $\cos(\phi/2)\leq 1$ in the first inequalities and Eq.~\eqref{eq:exponential_ineq} in the second inequalities in both Eqs.~\eqref{eq:lowerbound_Dk} and \eqref{eq:lowerbound_Em}.
    Therefore, the failure probability of the entire protocol is upper bounded as
    \begin{align}
    1-P(n,k,m) &\leq 1 - \qty[1 - \exp(-\frac{k\phi^8}{64\pi^8})]^{4n(n-1)} \qty[1 - \exp(-\frac{m\phi^4}{16\pi^4})]^{2n(n-1)} \\
    &\leq 1- \qty[1-4n(n-1)\exp(-\frac{k\phi^8}{64\pi^8})]\qty[1-2n(n-1)\exp(-\frac{m\phi^4}{16\pi^4})] \\
    &\leq 4n(n-1)\exp(-\frac{k\phi^8}{64\pi^8}) + 2n(n-1)\exp(-\frac{m\phi^4}{16\pi^4}),
\end{align}
where we have used Eq.~\eqref{eq:Bernoulli_ineq} in the second inequality. 
Letting $k\geq 64\pi^8 \phi^{-8}\log(8\delta^{-1}n(n-1))$ and $m \geq 16 \pi^4 \phi^{-4}\log(4\delta^{-1}n(n-1))$, we obtain $1-P(n,k,m) \leq \delta$.
\end{proof}

The number of qubits consumed in the transformation from the WGS to an $n \times n$ cluster state is
\begin{equation}
    N=2n(n-1)(5m + 14km).
\end{equation}
Thus, we can achieve the success probability higher than $1-\delta$ by consuming $N=O(n^2\phi^{-12}\log^2(\delta^{-1}n))$ qubits, that is, the overhead per qubit $N/n^2$ is polynomial in $\phi^{-1}$ and polylogarithmic in $\delta^{-1}$ and $n$.

\subsection{Alternative CZ-gate generation method}\label{subsec:addtional_Pauli}
In this section, we introduce another CZ-gate generation method, in addition to the method of Eq.~\eqref{eq:CZgate_generation} in Proposition~\ref{prop:list_MBCoperator}.
For a 1D (unweighted) cluster state, there are two standard ways to reduce its length; one is to measure a middle qubit in the Pauli-Y basis, and the other is to measure two middle qubits in the Pauli-X basis. 
The method in Proposition~\ref{prop:list_MBCoperator} corresponds to the former, and the method we introduce here corresponds to the latter.
As shown in the manuscript, this section is redundant for proving the universality of the WGS, and, in fact, the method introduced here is less efficient in terms of the success probability and the number of consumed qubits than the method used in the manuscript.
Nevertheless, we expect the alternative CZ-gate method to be useful for other applications because, although the above two methods behave similarly for 1D cluster states, their actions on general graph states are quite different.

\subsubsection{Definitions and properties of weighted Pauli measurements 2}
\begin{figure}[tbp]
\centering
\includegraphics[scale=0.5, trim=0 0 0 0, clip]{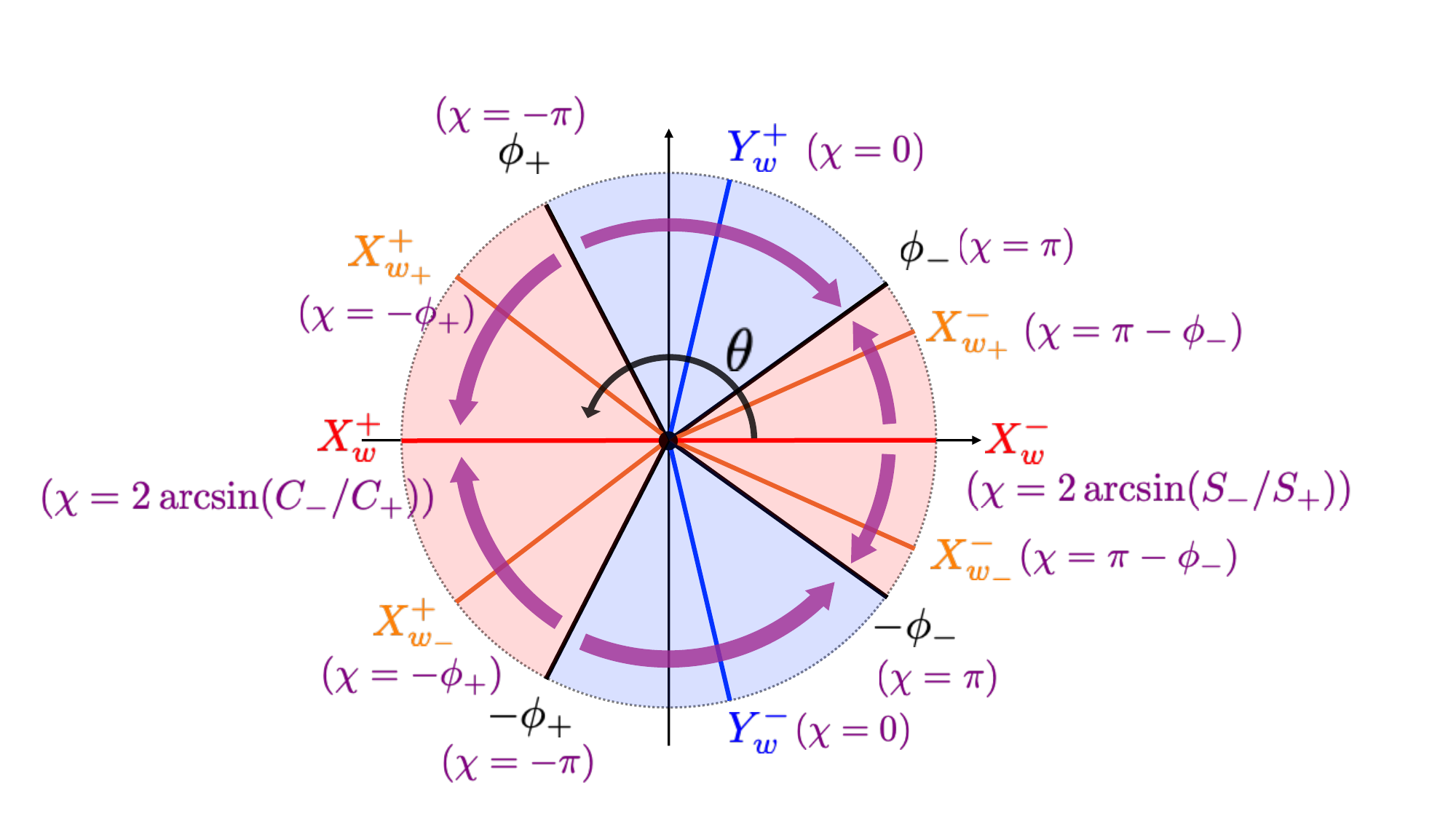}
\caption{Summary of the measurement angles $\theta$ and weights $\chi$ on the weighted Pauli bases for $\phi_1\geq \phi_2 > 0$. The red-shaded and blue-shaded regions correspond to the X-type and Y-type solutions, respectively. The purple arrows indicate the directions in which $\chi$ increases.}\label{fig:weighted_Pauli}
\end{figure}

First, we define additional weighted Pauli measurements in a generalized manner, although we only use a special case for the CZ-gate generation.
While we have defined weighted Pauli-X and Pauli-Y measurements as in Def.~\ref{def:weight_Pauli} as particularly useful measurement bases, we can choose other values of $\theta$ on the measurement basis of $\{\ket{\psi_s(\theta)}, \ket{\psi_f(\theta)}\}$ in Def.~\ref{def:measurement_basis}. 
As shown in Proposition~\ref{prop:fundamental_formula}, the resulting MBC operator takes different forms depending on whether (i) $\abs{\theta}<\phi_l$ or $\phi_u<\abs{\theta}$ or (ii) $\phi_l<\abs{\theta}<\phi_u$ for $\phi_u = \max(\abs{\phi_+}, \abs{\phi_-})$ and $\phi_l = \min(\abs{\phi_+}, \abs{\phi_-})$. 
Here, let us call these cases (i) and (ii) X-type and Y-type solutions, respectively.
The Y-type solutions are complicated to handle, except for the case of weighted Pauli-Y measurements, because the resulting MBC operator includes both an entangling projection and a CZ gate.
On the other hand, the X-type solutions always include only an entangling projection, and their intrinsic entangling power is tunable in the range of Eq.~\eqref{eq:condition_PauliX}.
Here, we introduce an additional family of weighted Pauli-X measurements as the measurement bases in which the weight $\chi$ of the resulting MBC operator in Eq.~\eqref{eq:entangling_power_measure} becomes the addition $\phi_+$ or the difference $\phi_-$ of the weights of input operators, $\phi_1$ and $\phi_2$.
The relation of all the introduced weighted Pauli measurements when $\phi_1 \geq \phi_2>0$ is summarized in Fig.~\ref{fig:weighted_Pauli}.

\begin{definition}[weighted Pauli measurements 2]\label{def:weight_Pauli2}
    $X^{+}_{w_\pm}$-basis and $X^{-}_{w_\pm}$-basis measurements are measurements in the bases of $\{\ket{\psi_s(\pm \theta_{+})}, \ket{\psi_f(\pm \theta_{+})}\}$ and $\{\ket{\psi_s(\pm \theta_{-})}, \ket{\psi_f(\pm \theta_{-})}\}$, respectively, where $0 \leq \theta_+, \theta_- \leq \pi$ are the angles such that
    \begin{equation}\label{eq:def_angle_wpm}
        \sin(\theta_+/2) = \sqrt{S_+ + \abs{S_-}/\sin(\phi_u/2)}, \quad \sin(\theta_-/2) = \sqrt{S_+ - \abs{S_-}/\cos(\phi_l/2)}.
    \end{equation}
    In particular, we use the following notations for the MBC of $O_1$ and $O_2$:
    \begin{equation}
        (O_1 \circ_{\psi_s(\pm \theta_+)} O_2) = X_{w_\pm}^{+} (O_1,O_2), \quad  (O_1 \circ_{\psi_s(\pm \theta_-)} O_2) = X_{w_\pm}^{-} (O_1,O_2).
    \end{equation}
\end{definition}
\begin{proposition}\label{prop:MBC_operator_halfweight}
    The angles $\theta_+$ and $\theta_-$ defined in Def.~\ref{def:weight_Pauli2} always exist, and it holds that
    \begin{align}
        X_{w_\pm}^{+} (\widetilde{\mathrm{CP}}(\phi_1), \widetilde{\mathrm{CP}}(\phi_2)) &= i\sqrt{\frac{\abs{S_-}}{\sin(\phi_u/2)}} (R_z(\pm i(r^+_++r^+_-)/2)\otimes R_z(\pm i(r^+_+-r^+_-)/2))\widetilde{\mathrm{EP}}(-\sgn(S_-)\phi_u) \\
        X_{w_\pm}^{-} (\widetilde{\mathrm{CP}}(\phi_1), \widetilde{\mathrm{CP}}(\phi_2)) &= i\sqrt{\frac{\abs{S_-}}{\cos(\phi_l/2)}} U_{\sigma(1)}U_{\sigma(2)}(R_z(\pm i(r^-_++r^-_-)/2)\otimes R_z(\pm i(r^-_+-r^-_-)/2))\widetilde{\mathrm{EP}}(\sgn(S_-)(\pi-\phi_l))
    \end{align}
    where $r^+_\pm = \log{\abs{{\sin((\phi_\pm +\theta_+)/2)}/{\sin((\phi_\pm -\theta_+)/2)}}}$ and $r^-_\pm = \log{\abs{{\sin((\phi_\pm +\theta_-)/2)}/{\sin((\phi_\pm -\theta_-)/2)}}}$.
\end{proposition}
\begin{proof}
    Substituting $\sin(\chi/2)=-\sgn(S_-)\sin(\phi_u/2)$ and $\sin(\chi/2)=\sgn(S_-)\cos(\phi_l/2)$ into Eq.~\eqref{eq:MBC_angle} for case (i), we obtain the same conditions as Eq.~\eqref{eq:def_angle_wpm} for $\pm \theta_+$ and $\pm \theta_-$, respectively.
    Therefore, the existence of $\theta_+$ and $\theta_-$ is equivalent to the condition of Eq.~\eqref{eq:condition_PauliX} being satisfied. For the case of $\abs{\phi_+} > \abs{\phi_-}$, the conditions for $\theta_+$ and $\theta_-$ are $-\abs{\sin(\phi_+/2)} \leq (C_-/C_+)$ and $(S_-/S_+) \leq \abs{\cos(\phi_-/2)}$, respectively. For the case of $\abs{\phi_-} > \abs{\phi_+}$, the conditions for $\theta_+$ and $\theta_-$ are $(C_-/C_+) \leq \abs{\sin(\phi_-/2)}$ and $-\abs{\cos(\phi_+/2)} \leq (S_-/S_+)$, respectively. In fact, these conditions are always satisfied, which is proven as Lemma~\ref{lemma:sufficient_condition} below.
    Therefore, $X_{w_\pm}^{+} (\widetilde{\mathrm{CP}}(\phi_1), \widetilde{\mathrm{CP}}(\phi_2))$ and $X_{w_\pm}^{-} (\widetilde{\mathrm{CP}}(\phi_1), \widetilde{\mathrm{CP}}(\phi_2))$ are X-type solutions with the weight $\chi$ satisfying $\sin(\chi/2)=-\sgn(S_-)\sin(\phi_u/2)$ and $\sin(\chi/2)=\sgn(S_-)\cos(\phi_l/2)$, respectively. 
    The resulting MBC operators are derived from Proposition~\ref{prop:fundamental_formula}.
\end{proof}

\begin{lemma}\label{lemma:sufficient_condition}
    It holds that 
    \begin{equation}
        -\abs{\cos(\phi_+/2)} \leq (S_-/S_+) \leq \abs{\cos(\phi_-/2)}, \quad -\abs{\sin(\phi_+/2)} \leq (C_-/C_+) \leq \abs{\sin(\phi_-/2)}
    \end{equation}
\end{lemma}
\begin{proof}
It follows from the following decompositions:
\begin{align}
    &R_s \pm \abs{\cos(\phi_\pm/2)} = (\sin^2(\phi_+/2)+\sin^2(\phi_-/2))^{-1} \qty{\sin^2(\phi_+/2)(1 \pm \abs{\cos(\phi_\pm /2)})-\sin^2(\phi_-/2)(1 \mp \abs{\cos(\phi_\pm /2)})} \notag \\
    & \quad =\pm(\sin^2(\phi_+/2)+\sin^2(\phi_-/2))^{-1}(1-\abs{\cos(\phi_\pm/2)})\qty{\sin^2(\phi_\pm/2)\qty(\frac{1 + \abs{\cos(\phi_\pm/2)}}{1 - \abs{\cos(\phi_\pm/2)}}) - \sin^2(\phi_\mp/2)} \notag \\
    & \quad =\pm(\sin^2(\phi_+/2)+\sin^2(\phi_-/2))^{-1}(1-\abs{\cos(\phi_\pm/2)})(1+\abs{\cos(\phi_\pm/2)}+\sin(\phi_\mp/2)) ( 1+\abs{\cos(\phi_\pm/2)}-\sin(\phi_\mp/2)),\\
    &R_c \pm \abs{\sin(\phi_\pm/2)} = (\cos^2(\phi_+/2)+\cos^2(\phi_-/2))^{-1}\qty{\cos^2(\phi_+/2)(1\pm \abs{\sin(\phi_\pm/2)})-\cos^2(\phi_-/2)(1\mp \abs{\sin(\phi_\pm/2)})} \notag \\
    & \quad =\pm(\cos^2(\phi_+/2)+\cos^2(\phi_-/2))^{-1}(1-\abs{\sin(\phi_\pm/2)})\qty{\cos^2(\phi_\pm/2)\qty(\frac{1+\abs{\sin(\phi_\pm/2)}}{1-\abs{\sin(\phi_\pm/2)}})-\cos^2(\phi_\mp/2)} \notag \\
    & \quad =\pm(\cos^2(\phi_+/2)+\cos^2(\phi_-/2))^{-1}(1-\abs{\sin(\phi_\pm/2)})(1+\abs{\sin(\phi_\pm/2)} + \cos(\phi_\mp/2)) (1+\abs{\sin(\phi_\pm/2)} - \cos(\phi_\mp/2)).
\end{align}
\end{proof}

\subsubsection{Alternative CZ-gate generation method and its success probability}
\begin{figure}[tbp]
\centering
\includegraphics[scale=1.3, trim=0 0 0 0, clip]{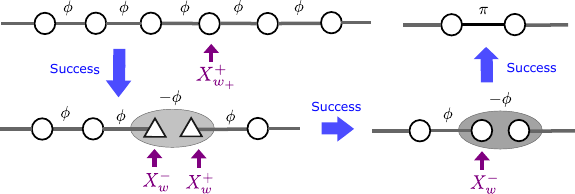}
\caption{Alternative method for probabilistic CZ-gate generation using an $X_{w_+}^+$-basis measurement for $\phi > 0$. The multiple weighted Pauli measurements within one step are defined from the left.}\label{fig:alternative_method}
\end{figure}
Figure~\ref{fig:alternative_method} shows the procedure of the alternative CZ-gate generation method, where we use the $X^+_{w_+}$-basis measurement introduced in Def.~\ref{def:weight_Pauli2} on the MBC of two identical CP gates.
The idea of this CZ-gate generation is to modify a sequence of the weighted Pauli-X measurements in Fig.~\ref{fig:transformation_diagram}.
An obstacle to obtaining a CZ gate with only $X_w^\pm$-basis measurements is that unitary operations are generated only after even numbers of the MBCs, while the weight becomes $\pi$ only after odd numbers of the MBCs, as shown in Corollary~\ref{cor:MBC_operator_weighted_PauliX}.
By using an $X_{w_+}^+$-basis measurement, we can obtain an entangling projection with weight $\phi$ or $-\phi$ from a pair of CP gates with weight $\phi$, enabling us to circumvent the obstacle.
Since an even number of MBCs are required for generating a unitary operation, and two $X_w^\pm$-basis measurements are required after the $X^+_{w_+}$-basis measurement to eliminate the local non-unitary operations, this type of CZ-gate generation needs at least four MBCs in total. As shown below, the success probability $p'_{cz}$ of this CZ-gate generation method is always lower than the success probability $p_{cz}$ of the other CZ-gate generation method.
\begin{proposition}[Alternative probabilistic CZ-gate generation]
    Let $-\pi \leq \theta', \eta' \leq \pi$ be the angles such that
    \begin{equation}
        \sin(\theta'/2) = \sqrt{\frac{\abs{\sin(\phi/2)}(1+\abs{\sin(\phi/2)})}{2}}, \quad \tan(\eta'/2) = \sgn(\phi) \sqrt{\frac{\abs{\sin(\phi/2)}}{2+\abs{\sin(\phi/2)}}}.
    \end{equation}
    Up to the global phases, it holds that
    \begin{align}
        \mathrm{CZ}_+' &\equiv \mathrm{CP}(\phi) \circ_{(0,\phi+\pi h(-\phi))} \mathrm{CP}(\phi) \circ_{(-\eta',\phi)} \mathrm{CP}(\phi) \circ_{(0,\phi+\theta')} \mathrm{CP}(\phi) \circ_{(-\eta',\phi+\pi h(-\phi))} \mathrm{CP}(\phi) \notag \\
        &= X_w^-(\mathrm{CP}(\phi),X_w^+(X_w^-(\mathrm{CP}(\phi), X_{w_+}^+(\mathrm{CP}(\phi),\mathrm{CP}(\phi))),\mathrm{CP}(\phi))) \\
        &= \sqrt{p'_{cz}}(R_z((\phi+\pi)/2) \otimes R_z((\phi-\pi)/2)) \mathrm{CZ}
        \label{eq:alternative_CZgate},
    \end{align}
    where 
    \begin{equation}
        p'_{cz} = \frac{\abs{\sin(\phi/2)}^{5}}{16(1+\abs{\sin(\phi/2)})^2}
    \end{equation}

\end{proposition}
\begin{proof}
    From Proposition~\ref{prop:MBC_operator_halfweight}, we have
    \begin{align}
        X_{w_+}^{+} (\mathrm{CP}(\phi), \mathrm{CP}(\phi)) &= (R_z(\phi/2) \otimes R_z(\phi/2)) X_{w_+}^{+} (\widetilde{\mathrm{CP}}(\phi), \widetilde{\mathrm{CP}}(\phi)) \\
        &= \sqrt{\frac{\abs{\sin(\phi/2)}}{2}} (R_z((\phi +ir')/2)\otimes R_z((\phi+ir')/2))\widetilde{\mathrm{EP}}(-\abs{\phi}),
    \end{align}
    where $r'=\log{\abs{{\sin((\phi +\theta')/2)}/{\sin((\phi -\theta')/2)}}}$ satisfies 
    \begin{equation}\label{eq:alternative_CZgate_amplitude}
        \mathcal{N}'= \cosh(r'/2)= \abs{\frac{\tan(-\abs{\phi}/2)\cos(\phi/2)(1-\cos(\theta'))}{\sin^2(\phi/2)}}=1+\abs{\sin(\phi/2)}
    \end{equation}
    from Eq.~\eqref{eq:MBC_noninitary_amplitude}.
    For $\ket{(-\eta', \phi)}$ being the measurement basis state on the next weighted Pauli measurement, it needs to hold that
    \begin{equation}
        \tan(\eta'/2) = \tanh(r'/4) = \sgn(\phi) \sqrt{\frac{\abs{\sin(\phi/2)}}{2+\abs{\sin(\phi/2)}}},
    \end{equation}
    where we have used Eq.~\eqref{eq:alternative_CZgate_amplitude} and $\sgn(r')=\sgn(\phi)\sgn(\theta')$.
    The resulting MBC operator is 
    \begin{equation}
        \mathrm{CZ}_+' = \frac{\sqrt{\abs{\sin(\phi/2)}}}{\sqrt{2}\mathcal{N}'} (R_z(\phi/2) \otimes R_z(\phi/2)) X_w^-(\widetilde{\mathrm{CP}}(\phi),X_w^+(X_w^-(\widetilde{\mathrm{CP}}(\phi), \widetilde{\mathrm{EP}}(-\abs{\phi})),\widetilde{\mathrm{CP}}(\phi))),
    \end{equation}
    where 
    \begin{align}
        X_w^-(\widetilde{\mathrm{CP}}(\phi),X_w^+(X_w^-(\widetilde{\mathrm{CP}}(\phi), \widetilde{\mathrm{EP}}(-\abs{\phi})),\widetilde{\mathrm{CP}}(\phi))) &= \frac{\sin(\phi/2)}{\sqrt{2}}  X_w^-(\widetilde{\mathrm{CP}}(\phi),X_w^+((Z^{h(-\phi)} \otimes Z^{h(\phi)}) \widetilde{\mathrm{CP}}(\pi),\widetilde{\mathrm{CP}}(\phi))) \\
        &= \frac{\sin(\phi/2)}{2}  X_w^-(\widetilde{\mathrm{CP}}(\phi),(Z^{h(-\phi)} \otimes I)\widetilde{\mathrm{EP}}(-\phi)) 
        \\
        &= \frac{\sin^2(\phi/2)}{2\sqrt{2}}  (Z\otimes I) \widetilde{\mathrm{CP}}(\pi)
    \end{align}
    from Eqs.~\eqref{eq:CPCP_pauliX+} and \eqref{eq:CPCP_pauliX-}.
    Therefore, we obtain Eq.~\eqref{eq:alternative_CZgate} from Eq.~\eqref{eq:relation_CZgate}.
\end{proof}
\bibliography{mbec.bib}
\makeatletter\@input{mbec_main_aux.tex}\makeatother